\documentclass[twocolumn,tighten,trackchanges]{aastex631}
\usepackage{amsmath}
\usepackage{xspace,xcolor}
\usepackage{natbib}
\usepackage{outlines}
\usepackage{savesym}
\savesymbol{tablenum}
\usepackage{siunitx}
\usepackage{comment}
\usepackage{graphicx}
\usepackage{float}
\usepackage{braket}

\newcommand{\beq}{\begin{equation}}
\newcommand{\eeq}{\end{equation}}

\newcommand{\JHU}{Physics and Astronomy Department, Johns Hopkins University, Baltimore, MD 21218, USA}
\newcommand{\ISEF}{ISEF International Fellowship}
\newcommand{\STScI}{Space Telescope Science Institute, Baltimore, MD 21218, USA}
\newcommand{\Caltech}{TAPIR, Mailcode 350-17, California Institute of Technology, Pasadena, CA 91125, USA}
\newcommand{\BurkeCal}{Walter Burke Institute for Theoretical Physics, California Institute of Technology, Pasadena, CA 91125, USA}
\newcommand{\IIT}{Technion - Israel Institute of Technology, Physics department, Haifa Israel 3200002}
\newcommand{\Trinity}{Gonville \& Caius College, Trinity Street, Cambridge, CB2 1TA, UK}
\newcommand{\IACamb}{Institute of Astronomy, University of Cambridge, Madingley Road, Cambridge CB3 0HA, UK}
\newcommand{\IAS}{Institute for Advanced Study, Einstein Drive, Princeton, NJ 08540, USA}

\begin{document}

\title{Mass transfer in eccentric black hole -- neutron star mergers}

\correspondingauthor{Yossef Zenati}
\email{yzenati1@jhu.edu}

\author[0000-0002-0632-8897]{Yossef Zenati}
\altaffiliation{\ISEF}
\affiliation{\JHU}
\affiliation{\STScI}

\author[0000-0002-2728-0132]{Mor Rozner}
\affiliation{\Trinity}
\affiliation{\IACamb}
\affiliation{\IAS}
\affiliation{\IIT}

\author[0000-0002-2995-7717]{Julian~H.~Krolik}
\affiliation{\JHU}

\author[0000-0002-0491-1210]{Elias R. Most}
\affiliation{\Caltech}
\affiliation{\BurkeCal}

\begin{abstract}
Black hole–neutron star (BH/NS) binaries are of interest in many ways: they are intrinsically transient, radiate gravitational waves detectable by LIGO, and may produce $\gamma$-ray bursts.
Although it has long been assumed that their late-stage orbital evolution is driven entirely by gravitational wave emission, we show here that in certain circumstances mass-transfer from the neutron star onto the black hole can both alter the binary's orbital evolution and significantly reduce the neutron star's mass: when the fraction of its mass transferred per orbit is $\gtrsim 10^{-2}$, the neutron star's mass diminishes by order-unity, leading to mergers in which the neutron star mass is exceptionally small. The mass-transfer creates a gas disk around the black hole {\it before} merger that can be comparable in mass to the debris remaining after merger, i.e. $\sim 0.1 M_\odot$.
These processes are most important when the initial neutron star/black hole mass ratio $q$ is in the range $\approx 0.2 - 0.8$, the orbital semimajor axis is $40 \lesssim a_0/r_g \lesssim 300 $ ($r_g \equiv GM_{\rm BH}/c^2$), and the eccentricity is large, $e_0 \gtrsim 0.8$. Systems of this sort may be generated through the dynamical evolution of a triple system, as well as by other means.

\end{abstract}


\section{Introduction} \label{sec:intro}

The first direct detection of gravitational waves \citep{Abbott2016} opened a new era in astrophysics, especially in terms of compact objects. The scientific power of this approach was further amplified when the first binary neutron star merger whose gravitational wave signal was detected \citep{Abbott2017} was also seen in observations across a large part of the electromagnetic (EM) spectrum, from radio frequencies to $\gamma$-rays \citep{Abbott17a,Abbott17b,LIGO_2021c,Abbott2023}.
Since then, LIGO has observed many more double neutron star and black hole/neutron star mergers \citep{Abbott+21_BHNSGW200115,Most+20_GW190814, Zevin+20, Godzieba+21_GW190814, Drozda+22_FMG,Martineau_GW230529}, but no additional EM counterparts. 

The first multi-messenger gravitational wave event, GW170817, prompted intense study of these events, beginning with the formation of the original stars that ultimately evolved to the merging neutron stars and black holes. Whether the system was born with one or more companions (a triple or a binary), or the binary was formed well after the stars by other dynamical processes, there is now a very large literature on the evolutionary tracks that might be followed by progenitor systems of LIGO events \citep{Most+20_GW190814,Zevin+20,Abbott_2020, Abbott+21_BHNSGW200115,GW230529_2024}. In this paper, we will focus on a specific subset of mergers involving neutron stars, those in which the partner is a black hole (henceforth abbreviated BH/NS). 

Moreover, we will, for the most part, further restrict our attention to the final stage of the binary, the period in which its separation diminishes from $\sim 10^2 r_g$ (for black hole gravitational radius $r_g \equiv GM_{\rm BH}/c^2$, with $M_{\rm BH}$ the mass of the black hole) to a distance so close that the two objects merge. Hitherto, it has been widely assumed that when the separation is this small, orbital evolution is dominated by gravitational radiation \citep{Shibata&Uryu06, Etienne+09, Pannarale&Tonita&Rezzolla+11, Kyutoku+13, Foucart+14, Kyutoku+18, Foucart+18, Shibata_Kenta19, Radice+20_AnnRev, Shibata+21, Martineau_GW230529}. This assumption has several immediate consequences. First, everything about the event is determined by the masses of the two objects, the black hole spin parameter, and the spin's orientation relative to the orbital axis. Second, any binary whose lifetime is at least a few times the timescale on which gravitational radiation shrinks the orbit by a factor $\sim O(1)$ becomes very nearly circular. Third, the masses of the neutron star and black hole are unlikely to change until the merger begins.

Despite the general support given to the third consequence, there have been a few studies of classical ``Roche lobe overflow" of matter from the neutron star to the black hole \citep{ClarkEardley1977,Blinnikov+84,Blinnikov+90,Yudin+20,Kramarev+24}. Here, we will show that if the orbit is {\it eccentric} during the later stages of inspiral, mass-loss from the neutron star during pericenter passage can alter the course of orbital evolution and substantially diminish the neutron star's mass. This thought has been previously explored \citep{Davies+05_BHNS}; we return to it now using a much-improved treatment of angular momentum and energy flow in the course of mass-transfer \citep{Hamers&Dosopoulou19}. With this new formalism, we also examine a broader range of $M_{\rm BH}$ and $M_{\rm NS}$, including events that do not necessarily reach the lower bound of neutron star mass. A complementary study focusing on single, very deep pericenter passages was carried out by \citet{East:2011xa}.

Eccentric BH/NS binaries with semimajor axes as small as $\sim 10^2 r_g$ might be created through several channels. If the BH/NS binary is accompanied by a third component that is either a black hole or a neutron star, Kozai-Lidov oscillations can give the inner binary episodes of very high eccentricity (see, e.g., \citealt{Antonini_Perets12,Naoz16,Hamers+21,Shartiat+24}). When this mechanism is active, the mass-transfer could occur during a single high-eccentricity Kozai-Lidov excursion lasting enough orbits to accomplish a sizable total mass-transfer or as a cumulative process spanning many such episodes.

In a dense star cluster, a BS/NH binary orbiting an intermediate-mass black hole could also exhibit Kozai-Lidov oscillations \citep{Hoang+2018}. Dense star clusters provide another path to high-eccentricity in the form of binary-single or binary-binary interactions \citep{XuanZeyuan+24}, or through cluster tides \citep{Hamilton:2019yij}.

Investigating these possibilities also carries other interesting ramifications. We will discuss in detail (Secs. \ref{subsec:Implications} and \ref{sec:Results})
the range of black hole masses for which events of this sort might happen. Here, we use an order of magnitude estimate to define the issue. For mass-transfer to occur, the distance at which the black hole's tidal gravity becomes competitive with the neutron star's self-gravity must be larger than the pericenter distance $r_p$ \citep{Lattimer_Schramm76}. On the other hand, $r_p$ must also be large enough that the black hole does not pass through the neutron star. 
These two conditions taken together demand that during the entire period of mass-transfer, $r_p$ must be not much more than $\simeq 8 r_g (M_\odot/M_{\rm BH})$ if we take the neutron star radius to be $\simeq 12$~km, independent of $M_{\rm NS}$. In other words, this process cannot occur if $M_{\rm BH}$ is more than several $M_\odot$. Thus, searching for BH/NS systems with mass-transfer amounts to a search for black holes in the first mass gap \citep{Abbott_2020, GW230529_2024}.
In addition, this mass-exchange process may lead to mergers in which the neutron star has shallow and potentially sub-solar mass.

\section{Formalism}  \label{sec:BHNS_model}

\subsection{Orbital evolution by gravitational radiation} \label{sec:orb_GW}
For pure gravitational radiation-driven orbital evolution, we take the classic results found by \citet{Peters1964} in the lowest-order post-Newtonian approximation:
\begin{equation}
    \frac{da}{dt} =
    -\frac{64}{5}\mu c \left(a/r_g\right)^{-3}\left(1-e^2 \right)^{-7/2}\left(1 +\frac{73}{24}e^2 + \frac{37}{96}e^4\right)
\label{eq:da_dt_gw}
\end{equation}
and
\begin{equation}
    \frac{de}{dt} =
    -\frac{304}{15}\mu \left(\frac{c}{r_g}\right)e\left(a/r_g\right)^{-4}\left(1-e^2 \right)^{-5/2}\left(1 +\frac{121}{304}e^2\right).
 \label{eq:de_dt_gw} 
\end{equation}
Here $\mu$ is the symmetric mass ratio, $a$ is the orbital semimajor axis, and $e$ is the orbital eccentricity. When it is more convenient, we also use
\begin{equation}
\frac{de}{da} = \frac{de/dt}{da/dt}  =  \frac{19}{12}\frac{e}{a}\left(1-e^2\right)\frac{1 + \frac{121}{304}e^2} {1 + \frac{73}{24}e^2 +\frac{37}{96}e^4}.
    \label{eq:de_da_gw}
\end{equation}

\subsection{Mass-transfer} \label{subsec:MT}

\subsubsection{Onset}

We begin with the criterion for when mass-transfer occurs. In the Introduction, we argued qualitatively that the primary condition to satisfy is for the black hole's tidal gravity to be at least competitive with the star's self-gravity. We follow \citet{Hamers&Dosopoulou19} in making this condition quantitative by defining the quantity
\begin{equation}
    \xi \equiv F_q(1-e), 
    \label{eq:xi}
\end{equation}
where the function $F_q$ is the ratio of the radius of the star's effective Roche lobe to the orbit's semimajor axis. It is given by
\begin{equation}
    F_q = \frac{a}{R_{\rm NS}} \frac{0.49q^{2/3}}{0.6q^{2/3}+\ln \left(1+q^{1/3}\right)},
    \label{eq:Fq}
\end{equation}
where $R_{\rm NS}$ is the neutron star's radius, and $q$ ($\leq 1$) is the binary's mass ratio \citep{Eggleton1983}.
We speak of the ``effective Roche lobe" because true Roche lobes exist only for circular-orbit binaries in which the stars co-rotate with the orbit.
When $\xi > 1$, the effective Roche lobe of the neutron star is larger than the star even at pericenter, so no mass-transfer can take place; mass-transfer begins when $\xi$ drops below unity. Note that $F_q$ is a rather weak function of $q$: it increases from $\approx 0.2$ for $q=0.1$ to $\approx 0.35$ for $q=1$.

To determine $R_{\rm NS}$, we use TOV solutions assuming the SFHo equation of state \citep{Steiner+13_SHFo}. As for all plausible equations of state, these solutions yield a mass-radius relation in which the radius is nearly independent of mass from $M_{\rm NS} \simeq 0.3 M_\odot$ to $M_{\rm NS} \simeq 2.2 M_\odot$; in this case, $R_{\rm NS} \approx 12$~km.

\subsubsection{Magnitude} \label{subsec:Magnitude}

\begin{figure} 
    \centering
    \includegraphics[width=1.1\linewidth]{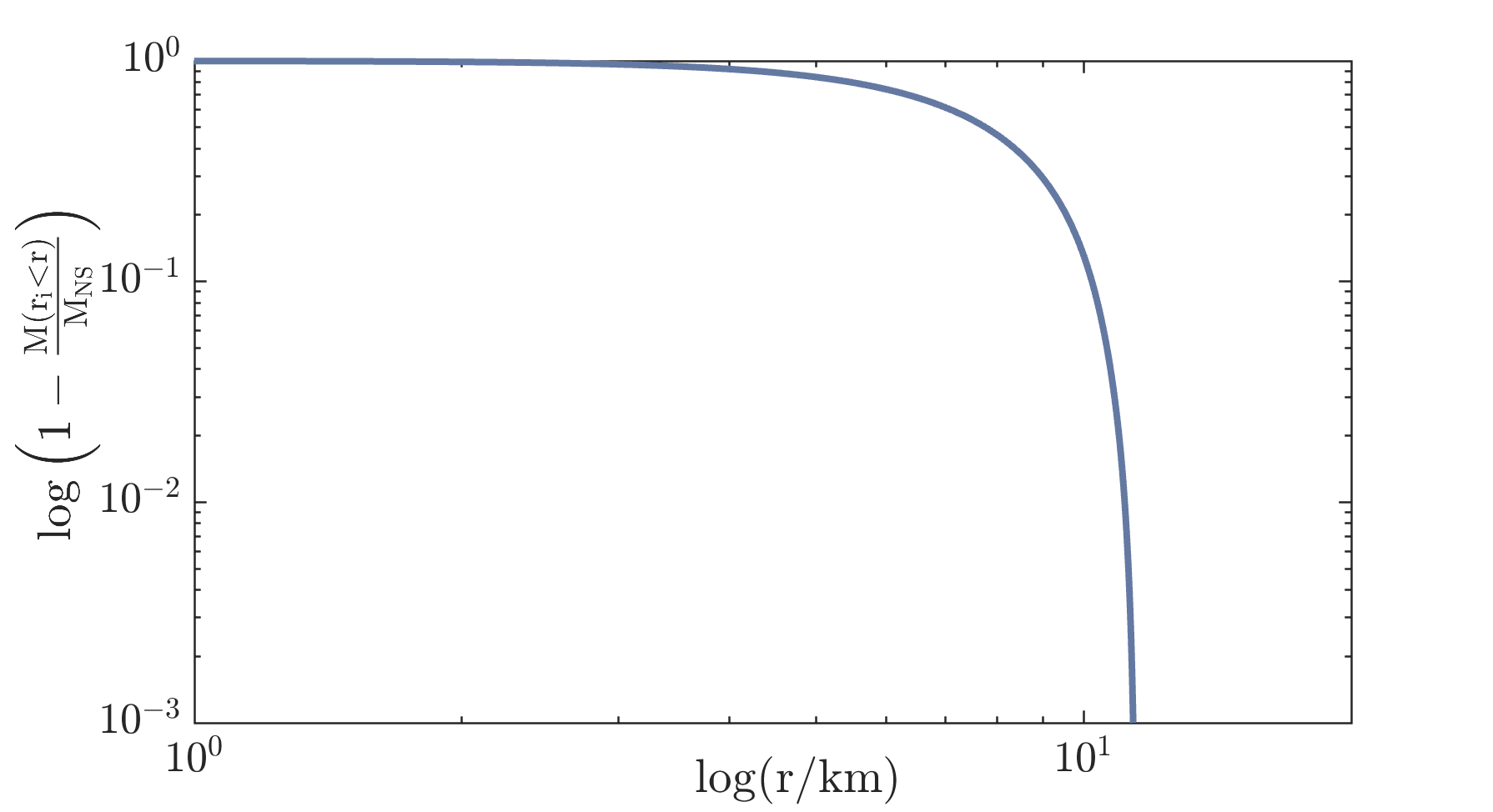}
    \caption{Fraction of a neutron star's mass outside radius $r$. Here, $M_{\rm NS} = 1.35 M_\odot$ and the SFHo equation of state \citep{Steiner+13_SHFo} is assumed; other masses and equations of state produce very similar curves.}
    \label{fig:TOV_1.35M}
\end{figure}

When the orbit is eccentric, most mass-transfer happens when the mass-losing star is near pericenter, and rather little takes place when the star is near apocenter. For the purpose of studying long-term orbital evolution, it is therefore best to work with the fractional mass-loss per orbit. This quantity should grow with greater ``overhang" of the star beyond its effective Roche lobe when it is at pericenter, a quantity given roughly by $1-\xi$. Although main-sequence stars have rather steep density profiles, so that their outer layers contain only a small fraction of their total mass, neutron stars have much shallower density profiles out to very nearly the stellar radius (see Fig.~\ref{fig:TOV_1.35M}). As this figure shows, even for an $\xi$ as large as $\approx 0.9$, the mass fraction beyond the Roche lobe is $\gtrsim 10\%$. Consequently, neutron stars on eccentric orbits are likely to lose a substantially larger fraction of their mass ($\Delta M/M$) in a single orbit than a main-sequence star would.
Note that $\Delta M/M$ refers to the fraction of the instantaneous mass of the neutron star.

\subsection{Orbital evolution for a given $\Delta M/M$} \label{subsec:Orbital_dMM}

\citet{Davies+05_BHNS} assumed that the mass transferred from the neutron star to the black hole carries half the specific angular momentum of the neutron star (they do not state how much of its orbital energy is given to the black hole). By contrast, \citet{Hamers&Dosopoulou19} calculate the angular momentum and energy transferred, accounting for numerous details and their variation with orbital phase: the location in the star from which the mass is taken and where it goes into orbit around the black hole; the velocities of the mass when it leaves the star and enters an orbit around the black hole; and the change in position of the system center-of-mass. Orbit-averaging these quantities, assuming that the instantaneous mass-transfer rate is proportional to the cube of the instantaneous fractional overhang, leads to the following orbital evolution equations in the absence of gravitational radiation:
\begin{eqnarray}\label{eq:da_dt_mdot}
\frac{da}{dt} &=& - (c/r_g)\frac{\Delta M/M}{\pi (a/r_g)^{3/2}f_{\dot M}(e,F_q)} a  \nonumber \\
&\times& \left[(1-q)f_a(e,F_q) + F_q \frac{R_{\rm NS}}{a}g_a(e,F_q) - q \frac{R_A}{a} h_a(e,F_q)\right] \nonumber
\\
\end{eqnarray}
and
\begin{eqnarray}\label{eq:de_dt_mdot}
\frac{de}{dt} &=& - (c/r_g)\frac{\Delta M/M}{\pi (a/r_g)^{3/2}f_{\dot M}(e,F_q)} \nonumber \\ 
&\times& \left[(1-q)f_e(e,F_q)+ F_q \frac{R_{\rm NS}}{a} g_e(e,F_q) - q \frac{R_A}{a} h_e(e,F_q)\right].\nonumber
\\
\end{eqnarray}

The functions $f_{\dot M}(e,F_q)$, $f_{a,e}(e,F_q)$, $g_{a,e}(e,F_q)$, and $h_{a,e}(e,F_q)$ are all given explicitly in Appendix B of \citet{Hamers&Dosopoulou19}; we do not reproduce them here because the expressions are very lengthy. We assume that their accretion delay parameter $\tau = 0$ and $R_A$ is the radius of the accretion disk the transferred mass forms around the black hole. We choose $R_A = 0.3a$ because the Newtonian tidal truncation radius as computed by \citet{Paczynski1977} can be well-fit by $\simeq 0.3a q^{-1/3}$.

To describe the evolution including both mass-transfer and gravitational wave emission, we simply add the contributions to $da/dt$ and $de/dt$. Note, however, that because the mass-transfer part assumes Newtonian dynamics, the orbits themselves do not reflect any relativistic effects.

The preceding equations describe the orbital evolution, but because many of the functions depend on the mass ratio $q$, and mass-transfer changes $q$, its evolution must also be tracked:
\begin{equation}\label{eq:dqdt}
\frac{dq}{dt} = -\frac{c}{r_g}\frac{\Delta M/M}{2\pi (a/r_g)^{3/2}} \left(q^2 + q\right).
\end{equation}
Here we assume that the transfer conserves mass.

\subsection{Implications of the evolution equations} \label{subsec:Implications}

Several qualitative conclusions can be drawn immediately from the forms of these equations. First, as already broached in the Introduction, the $\xi \leq 1$ criterion for mass-transfer places a constraint on the semimajor axis and eccentricity, requiring the pericenter $a(1-e)$ to be smaller than the neutron star's Roche lobe: $a(1-e) \leq R_{\rm NS}/F_q$. For a more or less fixed neutron star radius of $\approx 12$~km, this constraint becomes $a(1-e) \leq 27 (F_q/0.3)^{-1} (M_{\rm BH}/M_\odot)^{-1}r_g$. Second, the concept of mass-transfer via Roche lobe overflow requires that the pericenter $a(1-e) > R_{\rm NS} = 8 (M_{\rm BH}/M_\odot)^{-1} r_g$; otherwise, the black hole would pass through the neutron star. This second constraint is strengthened by the fact that Newtonian dynamics are a good description of orbits only when the separation is $\gtrsim 10 - 15 r_g$. Thus, the parameter space in which mass-transfer from an eccentric orbit is rather limited, and particularly so if $M_{\rm BH}$ is greater than a few solar masses.

A different qualitative conclusion is that the orbital evolution of such a system changes sharply when the mass lost per orbit is great enough that mass-transfer dominates gravitational wave radiation as a driver of orbital evolution. Contrasting the orbital evolution rates in equations~\ref{eq:da_dt_mdot} and \ref{eq:de_dt_mdot} with the gravitational radiation evolution rates in equations~\ref{eq:da_dt_gw} and \ref{eq:de_dt_gw} yields a criterion for this regime change
\begin{equation}
\frac{\Delta M}{M} \gtrsim 2 \times 10^{-3} (a/50r_g)^{-5/2} \mu f_{\dot M},
\end{equation}
where we have dropped the order-unity factor describing the eccentricity-dependence. Thus, the fractional mass-loss per orbit must be significant, but does not need to be a large fraction in order for mass-transfer to influence orbit evolution. However, as the semimajor axis shrinks, the threshold at which mass-transfer drives orbital evolution rises.

Lastly, for mass-transfer to continue, $a(1-e) \leq R_{\rm NS}/F_q$. As the binary's semimajor axis shrinks, its eccentricity does also; note that both gravitational radiation and mass-transfer tend to diminish $a$ and $e$. The question, therefore, arises, ``Does $a$ diminish more rapidly than $1-e$ increases?" This question cannot be answered by estimates; it demands a proper calculation.

\subsection{Validity of the post-Newtonian approximation}  \label{sec:BHNS_PNmethod}

The gravity in the region we are considering---distances $\sim 10 - 100 r_g$ from a black hole, near and inside a neutron star---is strong enough for relativistic corrections to be noticeable.  However, we treat the orbits as being purely Newtonian with the exception of allowance for the binary's evolution by radiation of gravitational waves. Here we explain why we believe our approximations are suitable for this first exploration of mass-transfer from neutron stars to black holes.

There exist a number of formalisms by which relativistic binary orbital evolution can be followed in more-or-less analytic fashion, including the effective-one-body (EOB) approximation and its extension to the inspiral-merger-ringdown multipolar waveform model \citep{Chiaramello_Nagar20,KhalilBuonanno+21,Ramos-Buades+22a, Ramos-Buades+22b}. The primary aim of these methods is to provide a way to generate gravitational waveforms beginning while the binary system is still tracing an almost-closed elliptical orbit, and then continuing seamlessly through the plunge and ringdown stages. Here we are concerned only with the first stage, the almost-closed elliptical orbit. In this regime, the \citet{Peters1964} quadrupole approximation remains reliable, even though it needs modification in the plunge.

Bearing this in mind, our evolutions all stop near the point at which the plunge begins. This point can be determined by comparing the binary's orbital energy to the maximum in the Kerr spacetime's effective potential. If the black hole in the binary has a spin parameter that is either prograde with respect to the orbit and is $\lesssim 0.8$, or is retrograde, these evolutions stop $\sim 50 - 100$ orbits beyond the beginning of the plunge; if the spin parameter is prograde and larger, they stop a similar time before the plunge.  Because the timescale for change in the orbital elements at this stage is several hundred orbits (see Fig.~\ref{fig:da60_de_dt_Mdot}), these offsets from the optimal stopping time make little difference in terms of orbital evolution.  Thus, in this respect, our approximations should be fairly sound.

Other relativistic effects might also affect the orbit, both apsidal and nodal precession. However, both of these precessional motions do not change the size of the orbit; they merely rotate it. They are therefore unlikely to alter mass-transfer effects.

Tidal stresses can also depart from the Newtonian form in this regime. We have not calculated how important the changes are in this context; they should be the focus of the next effort examining this mechanism.

\section{Results} \label{sec:Results}

To answer the question just posed, as well as a great many more, we have integrated the orbital evolution equations for a number of combinations of parameters within the bounds $1M_\odot \leq M_{\rm NS} \leq 2.1 M_\odot$, $2.78 M_\odot \leq M_{\rm BH} \leq 5M_\odot$, initial semimajor axis $a_0 = 50r_g$, $60r_g$, or $100r_g$, and initial eccentricity $e_0$ chosen so that the initial state has $\xi$ very slightly less than 1. Consequently, the initial values of $a$ and $q$ determine the initial value of $e$. In practice, $0.84 \leq e_0 \leq 0.87$ for $a_0 = 50r_g$, $0.88 \leq e_0 \leq 0.90$ for $a_0 = 60 r_g$, and $0.90 \leq e_0 \leq 0.94$ for $a_0 = 100 r_g$ (see Table~\ref{tab:cases}). All evolutions were stopped when the pericenter distance became smaller than $\approx 6 - 8 r_g$, where numerical relativity studies indicate that for the systems we consider the NS will always tidally disrupt, leading to the formation of a remnant debris disk \citep{Foucart:2012nc,Foucart+18}.
Note that the requirement of beginning with $\xi=1$ is, in fact, not restrictive because any system evolving in a smooth fashion from an initial state without mass-transfer must begin mass-transfer when $\xi=1$.

\subsection{Continuity of mass-transfer and orbital evolution} \label{subsec:Continuity_MT}

\begin{figure} 
    \centering
    \includegraphics[width=0.99\linewidth]{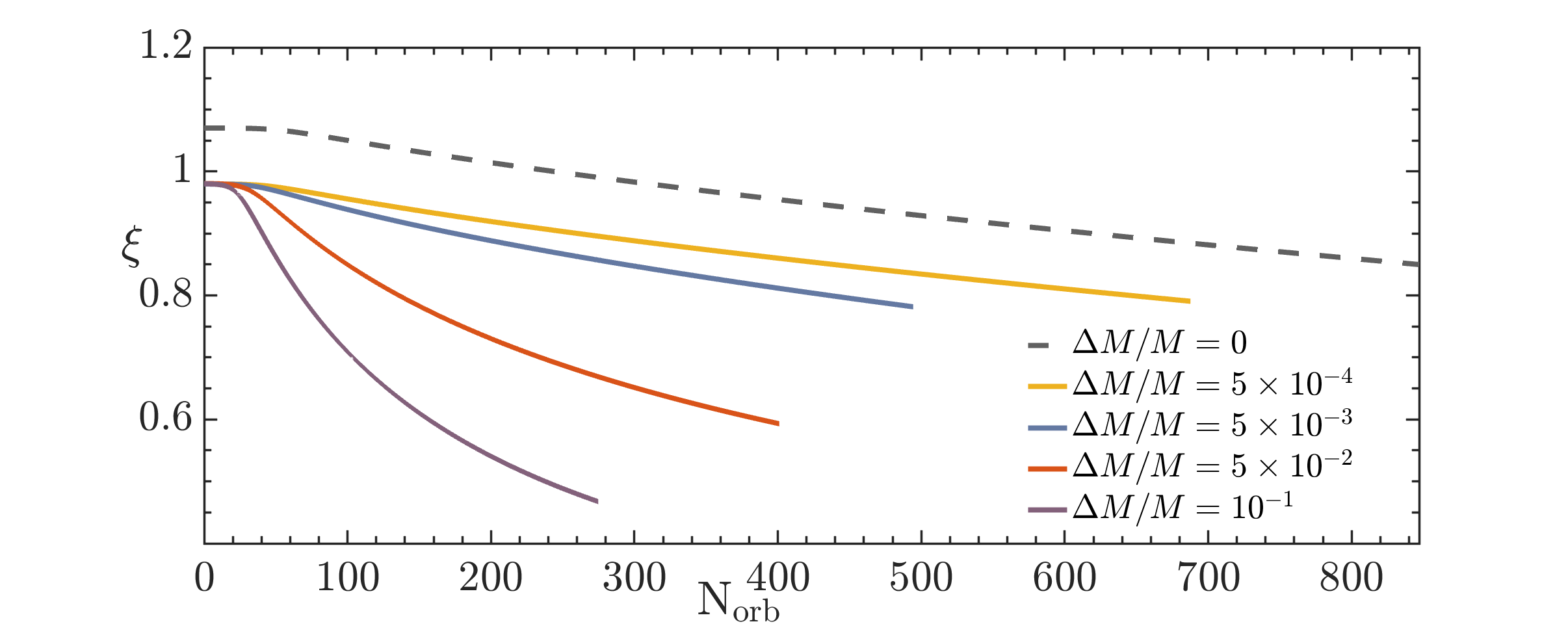}
    \caption{Evolution of the mass-transfer discriminant, $\xi$, as a result of both gravitational radiation and mass-transfer (i.e., eqns.~\ref{eq:da_dt_gw}, \ref{eq:da_dt_mdot}, \ref{eq:de_dt_gw}, \ref{eq:de_dt_mdot} and \ref{eq:dqdt}), plotted as a function of the number of orbits, $N_{\rm orb}$, for each curve. The dashed curve shows the evolution as dictated by gravitational radiation without mass-transfer. The solid curves show the evolution of $\xi$ when both mechanisms contribute. They are distinguished by $\Delta M/M$: $5 \times 10^{-4}$ (yellow), $5 \times 10^{-3}$ (blue), $5 \times 10^{-2}$ (orange), and $10^{-1}$ (purple). The initial mass of the NS is $1.2 M_{\odot}$, and the initial mass of the BH is $3.3 M_\odot$. The initial orbital parameters are $a_0 = 100 r_g$ and $e_0 = 0.917$ for the $\Delta M/M \neq 0$ cases, but a slightly larger value of $a_0$ for $\Delta M/M = 0$.}
    \label{fig:GWMTvsGW}
\end{figure}

Figure~\ref{fig:GWMTvsGW} illustrates how the mass-transfer discriminant $\xi$ evolves in several cases differing only in $\Delta M/M$.
When $\Delta M/M \lesssim 10^{-3}$ or $\xi > 1$, mass-transfer has essentially no effect on the orbital evolution, just as predicted by our qualitative estimate. In this limit, $\xi$ declines gradually over time with a characteristic shape illustrated equally well by the dashed curve in Figure~\ref{fig:GWMTvsGW} (no mass-transfer at all) and the yellow curve (mass-transfer with $\Delta M/M = 5 \times 10^{-4}$, a factor of a few below the critical value).  When $\Delta M/M = 0$, over a span of more than 800 orbits $a$ shrinks from $\simeq 100 r_g$ to $\simeq 40 r_g$ and $e$ falls from $\simeq 0.92$ to $\simeq 0.8$. Importantly, even when $\xi > 1$, so that no mass-transfer can take place, evolution by gravitational wave emission alone leads to a slow decline in $\xi$ so that $\xi$ ultimately falls below unity, permitting mass-transfer.

Because mass-transfer acts in the same sense as gravitational radiation---diminishing both $a$ and $e$---any mass-transfer accelerates the orbital evolution, and mass-transfer at a rate great enough to dominate gravitational radiation causes orbital evolution faster than the rate gravitational radiation can drive. When $\Delta M/M$ is only a factor of several above the threshold (e.g., $5 \times 10^{-3}$), the time from the onset of mass-transfer to the beginning of merger with the black hole is $\lesssim 500$~orbits; only $\simeq 300$~orbits are needed when $\Delta M/M$ is a factor $\sim 20\times$ larger.

Moreover, once $\xi < 1$ and mass-transfer begins, if $\Delta M/M \gtrsim 10^{-3}$, $\xi$ decreases monotonically, so that mass-transfer continues all the way to merger. Although only three cases are shown here, all the other cases we examined behaved in the same way. In other words, once the binary begins mass-transfer, the neutron star continues to lose mass until either the merger takes place or some external effect changes the orbit.

\begin{figure}
    \centering
    \includegraphics[width=0.99\linewidth]{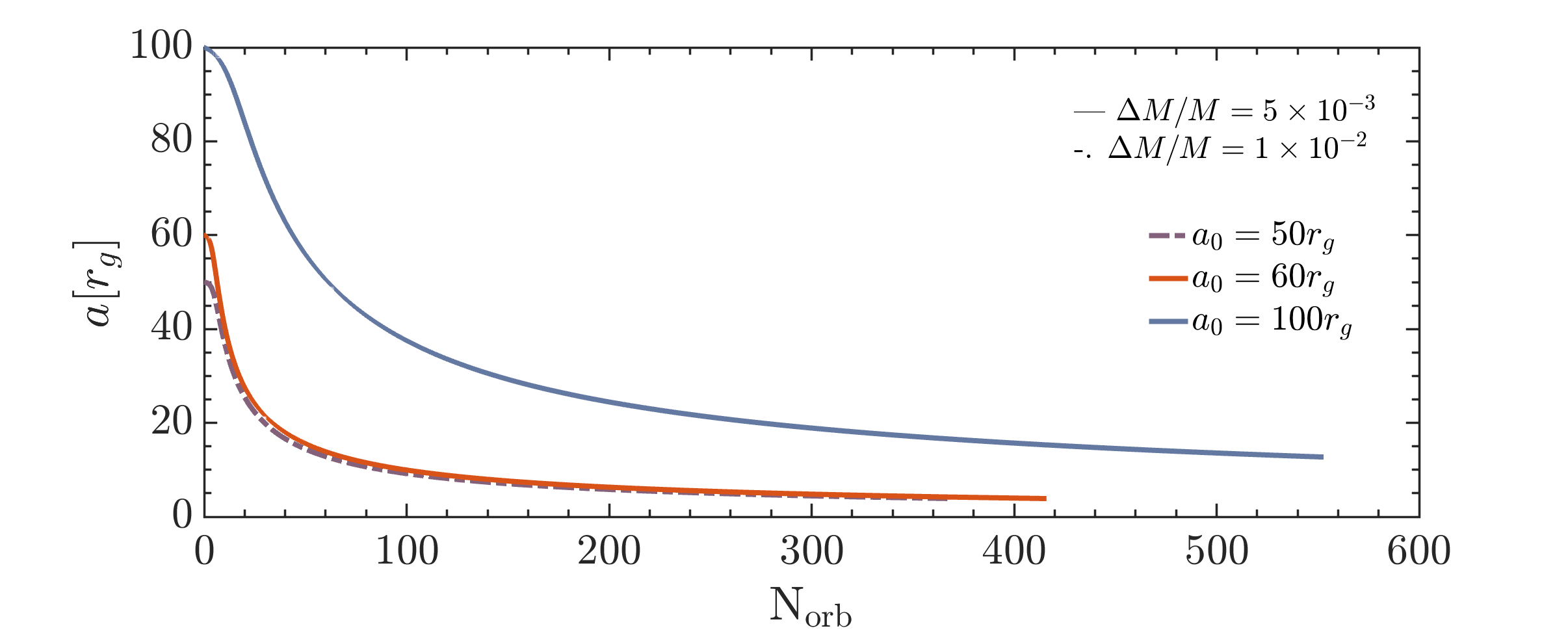}
    \includegraphics[width=0.99\linewidth]{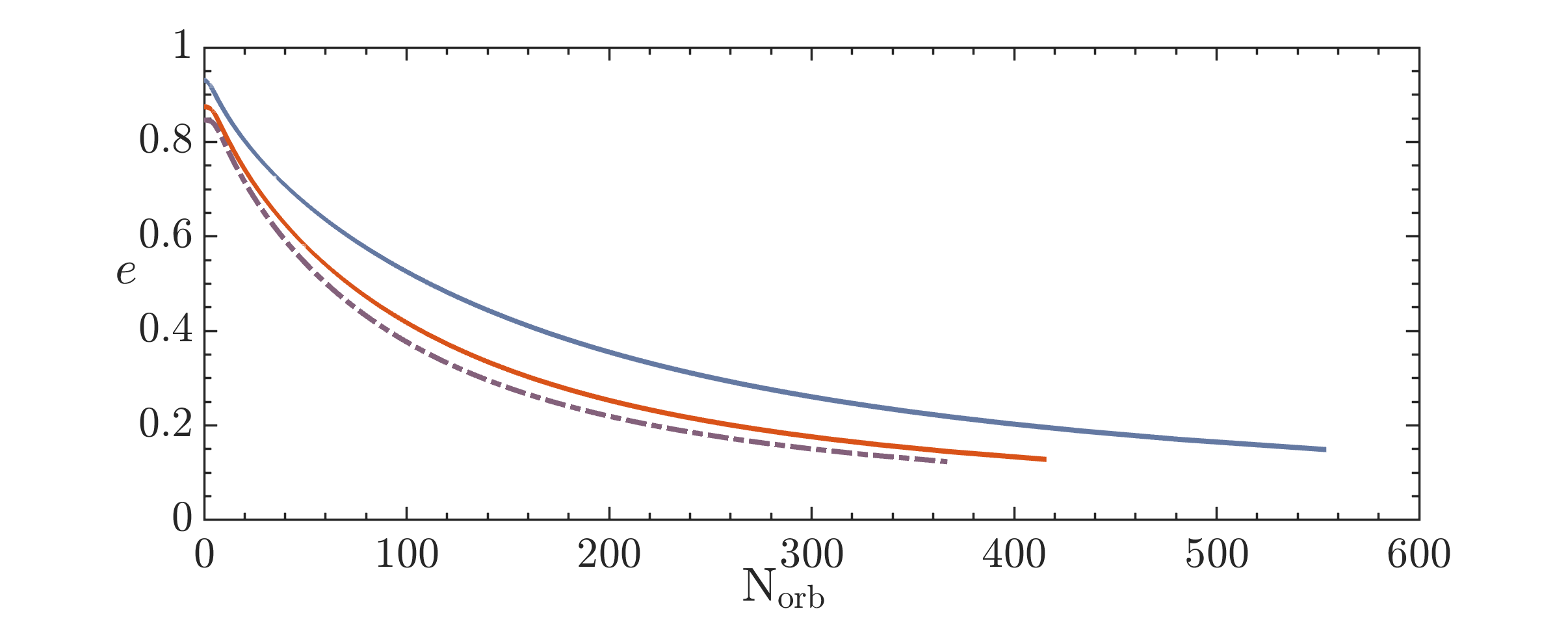}
    \includegraphics[width=0.99\linewidth]{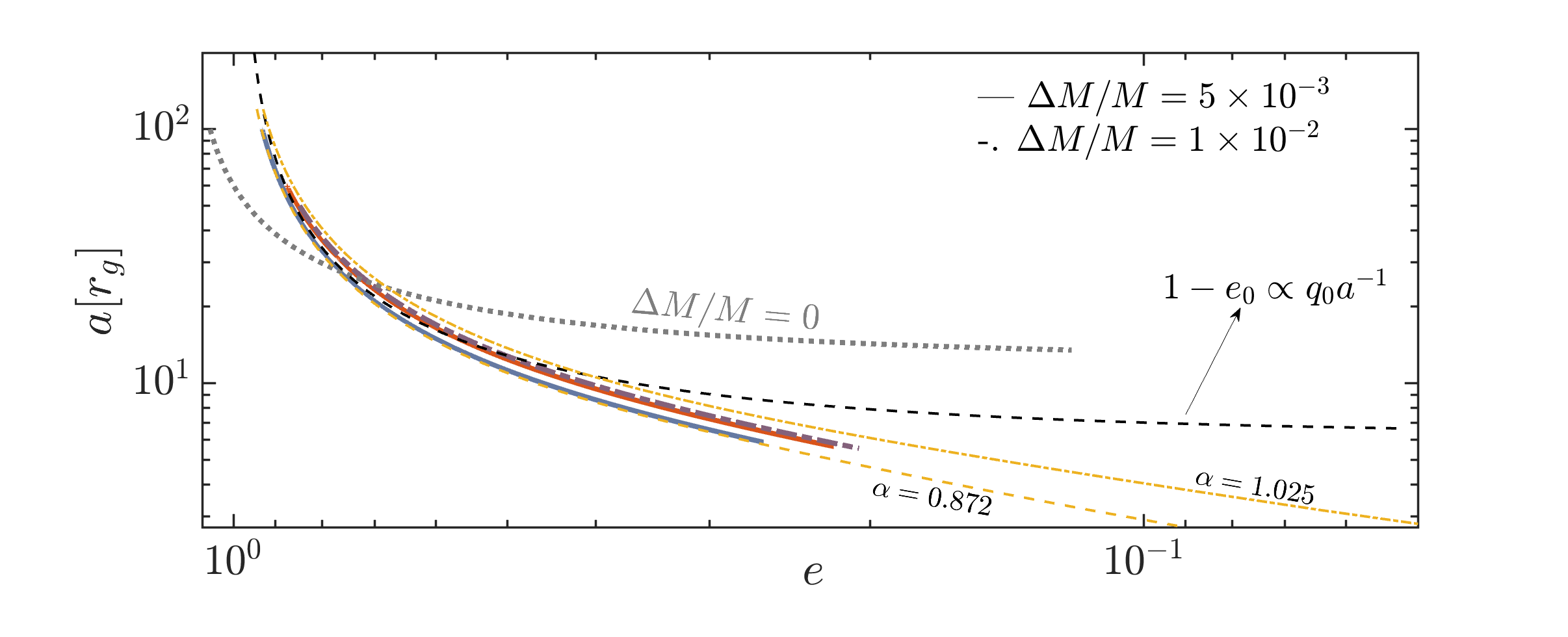}

    \caption{Evolution of semi-major axis (top panel) and eccentricity (middle panel) as functions of the number of orbits. The lower panel shows $a$ as a function of $e$; note that $e$ decreases from left to right. Three different initial semimajor axes are shown, $50r_g$ (purple dashed curves), $60r_g$ (red curves), and $100 r_g$ (blue curves), all having $M_{\rm NS} = 1.2 M_\odot$, $M_{\rm BH} = 3.3 M_\odot$. For the two larger initial semimajor axes, $\Delta M/M = 5 \times 10^{-3}$; for $a_0 = 50 r_g$, $\Delta M/M = 10^{-2}$. The dotted curve in the bottom panel shows $a(e)$ for evolution driven solely by gravitational radiation; the dashed curve shows evolution at fixed $\xi \propto a(1-e)/q_0$. The yellow curves correspond to the relation $a(e) \simeq \beta \exp(\alpha e)$ for two values of the three parameters $(\alpha, \beta, \Delta M/M)$: $(1.02, 6.35, 1\times 10^{-2})$ and $(0.87, 4.41, 5\times 10^{-3})$.  Both $\alpha$ values are close to unity.}
    \label{fig:da60_de_dt_Mdot}
\end{figure}

To gain a more specific sense of how the orbit evolves, three views of three sample histories are shown in Figure~\ref{fig:da60_de_dt_Mdot}. The top and bottom panels ($a$ as a function of $N_{\rm orb}$ and $e$ as a function of $N_{\rm orb}$) tell closely-related stories. 
In both, the contrast between the $a_0 = 100r_g$ and the $a_0=60r_g$ curves, despite their identical values of $\Delta M/M$ and initial $\xi$, shows that the semimajor axis at which mass-transfer begins remains imprinted on the system's orbital evolution throughout its progress toward merger. By contrast, the close similarity of the evolutions for $a_0 = 50r_g$ and $a_0=60r_g$, despite having $\Delta M/M$ values differing by a factor of 2, demonstrates that even though $da/dN_{\rm orb}$ and $de/dN_{\rm orb}$ are explicitly $\propto \Delta M/M$, these dependences can be largely cancelled by the {\it implicit} dependence of $a(N_{\rm orb})$ and $e(N_{\rm orb})$ on $\Delta M/M$ through $q(N_{\rm orb})$ (see eqn.~\ref{eq:dqdt}); terms proportional to both $q$ and $1-q$ appear in the evolution equations for both $a$ and $e$, and the functions $f_{a,e}$, $g_{a,e}$, and $h_{a,e}$ have further implicit dependences on $q$. Such a cancellation is not, however, necessarily a general effect.

Although the mass transfer-driven dependence of $a$ and $e$ on time exhibits interesting parameter-dependences, the relation between $a$ and $e$ is universal (bottom panel of Fig.~\ref{fig:da60_de_dt_Mdot}). All three cases shown in the upper two panels lie on almost exactly the same curve, differing only in the value of $a$ at which they enter it. Initially, their track is decently approximated by the relation $a \propto R_{\rm NS}/(1-e)$ because this is equivalent to $\xi = const$, which is not exactly true for the early stages of orbital evolution, but is also not grossly wrong.  

However, as illustrated in Figure~\ref{fig:GWMTvsGW}, $\xi = const.$ does eventually break down, and this happens sooner when $\Delta M/M$ is larger. From this point onward, $a$ declines more rapidly with respect to $e$ than the $\xi \simeq const.$ approximation would predict.  In fact, the relation between $a$ and $e$ over the entirety of the binary's evolution when it is driven by mass transfer is well described by the function $a(e)/r_g \simeq \beta \exp(\alpha e)$ with $\alpha$ constrained to lie in the range [0.87,1.02] and $\beta \simeq a_0/R_{NS}$. In other words, as $e$ decreases, $a$ shrinks exponentially, and the asymptotic value of $a$ for $e \rightarrow 0$ is $\simeq a_0/R_{\rm NS}$. This follows because
\begin{eqnarray}
(1-q)f_a(e,F_q) + F_q \frac{R_{\rm NS}}{a}g_a(e,F_q) - q \frac{R_A}{a} h_a(e,F_q)  \nonumber
\\
\simeq (1-q)f_e(e,F_q)+ F_q \frac{R_{\rm NS}}{a} g_e(e,F_q) - q \frac{R_A}{a} h_e(e,F_q)  \nonumber
\\
\end{eqnarray}
with remarkable precision despite the variations in $q$, $e$, and $a$ during the binary evolution and the fact that the individual function pairs (e.g., $f_a$ and $f_e$) have similar values, but are not as close as the specific combinations of this equation. Consequently, the ratio of equations \ref{eq:da_dt_mdot} and \ref{eq:de_dt_mdot} simplifies to $da/de \simeq a$.

All of the orbital evolution properties in the mass-transfer dominated regime are quite different from how the evolution proceeds when only gravitational radiation matters. Pure gravitational wave emission causes $e$ to decrease more sharply as a function of $a$ than when mass-transfer dominates. As a result, the orbit becomes almost circular when $a$ has shrunk by a factor of $\sim 2 - 3$, in sharp contrast with the mass-transfer case, in which the eccentricity remains $\gtrsim 0.2$ even when $a$ has shrunk by a factor $\sim 10 -20$. Ultimately, in the limit as $e \rightarrow 0$, gravitational radiation-controlled evolution leads to $a \propto e^{12/19}$ (see eqn.~\ref{eq:de_da_gw}).

\subsection{Total amount and pace of mass exchange} \label{subsec:Mass_excange}

We have already determined that whether gravitational radiation or mass-transfer dominates orbital evolution is largely governed by $\Delta M/M$; when $\Delta M/M \gtrsim 10^{-3}$, mass-transfer plays a role at least comparable to that of gravitational wave emission. Not surprisingly, the fractional mass-loss per orbit is also the critical parameter for determining whether the total mass transferred from the neutron star to the black hole is a sizable fraction of $M_{\rm NS}$. 

\begin{figure}
    \centering
    \includegraphics[width=1.1\linewidth]{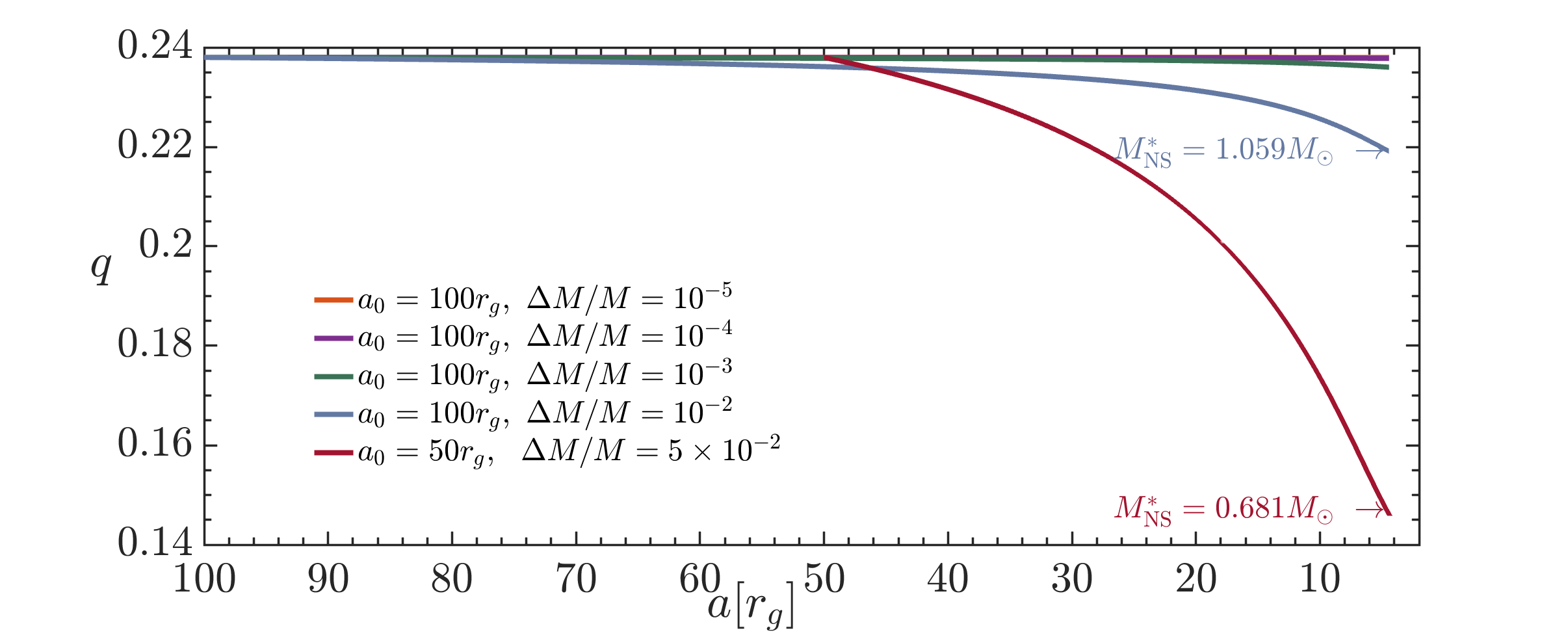}
    \includegraphics[width=1.1\linewidth]{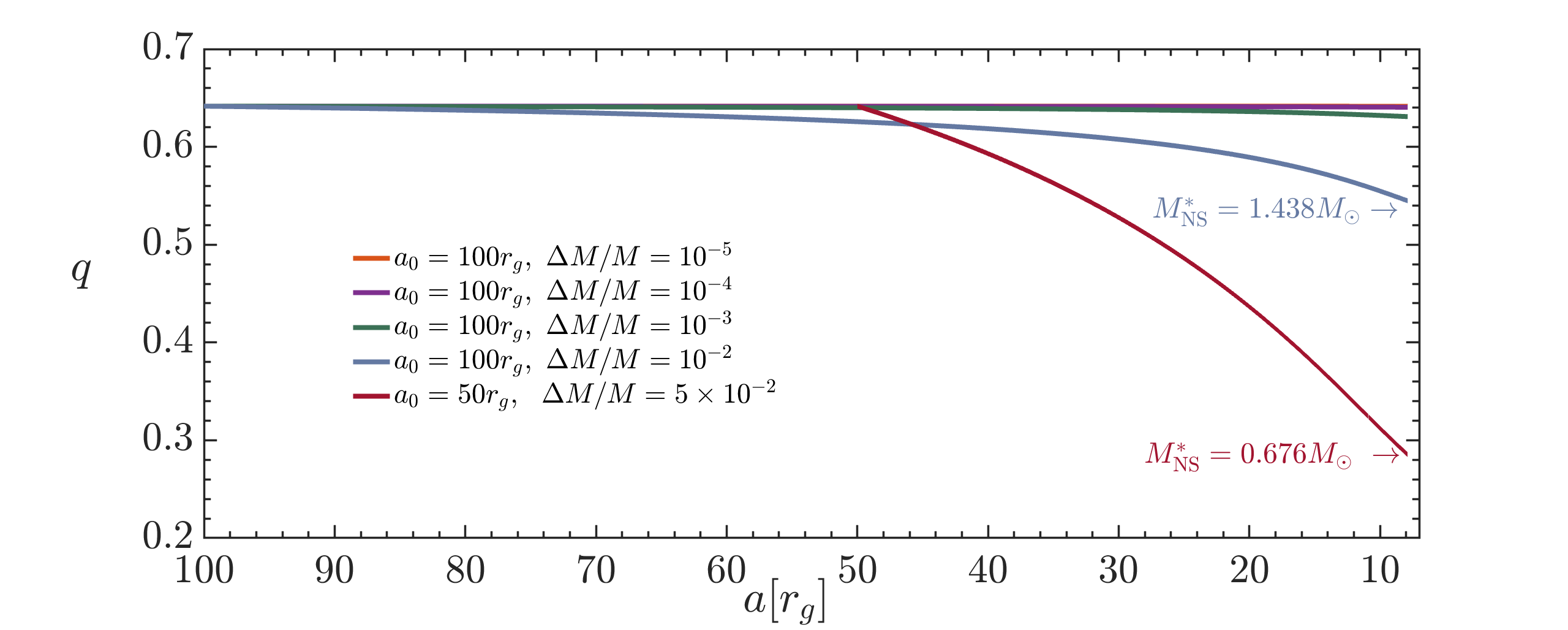}

    \caption{Evolution of the mass ratio as a function of semi-major axis for five different values of $\Delta M/M$. Two pairs of black hole and neutron star mass are portrayed: $M_{\rm BH}= 5.0M_\odot$, $M_{\rm NS} = 1.2M_\odot$ (upper panel); and $M_{\rm BH}= 2.78M_\odot$, $M_{\rm NS} = 1.8M_\odot$ (lower panel).}
    \label{fig:dq_dt_da}
\end{figure}

This fact is demonstrated in Figure~\ref{fig:dq_dt_da}. Whether the initial mass ratio $q=0.24$ (the upper panel of this figure) or $q=0.64$ (the lower panel), the mass lost by the neutron star is at least $\simeq 10 \%$ of its original mass when $\Delta M/M \gtrsim 10^{-2}$. The criterion for substantial total mass-transfer is therefore a factor of 10 more stringent than the one determining which mechanism controls orbital evolution.

The same figure also points out the fact that significant mass-transfer starts when the semimajor axis shrinks to $\lesssim 50 r_g$. As shown in Figure~\ref{fig:da60_de_dt_Mdot}, this is also the evolutionary stage at which the eccentricity drops below $\sim 0.8 \pm$. However, the pace of mass-transfer accelerates once it starts, both in terms of $dM/da$ (as shown in
Fig.~\ref{fig:dq_dt_da}) and even more so in terms of $dM/dt$ because $dM/dt = (dM/da)(da/dt) \propto (dM/da)(a/r_g)^{-3/2}$. Nonetheless, although the precise distribution of mass-loss with semimajor axis depends on parameters, the bulk of the mass-loss occurs when $a > 10r_g$, and, particularly for larger $\Delta M/M$, when $a > 20r_g$ (see also Fig.~\ref{fig:q_Mns_f_a}).

\subsection{Final Neutron star mass} \label{subsec:Final_mass}

\begin{figure}
    \centering
    \includegraphics[width=0.98\linewidth]{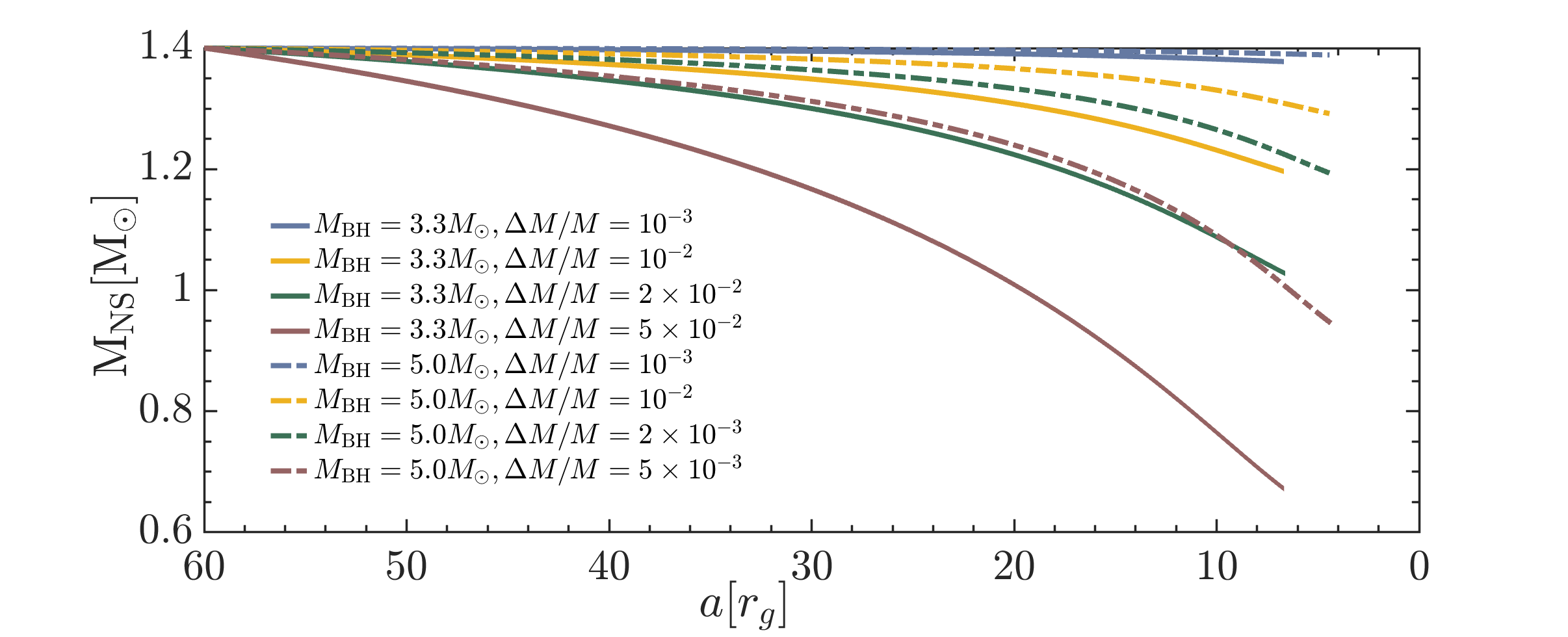}
    \includegraphics[width=1\linewidth]{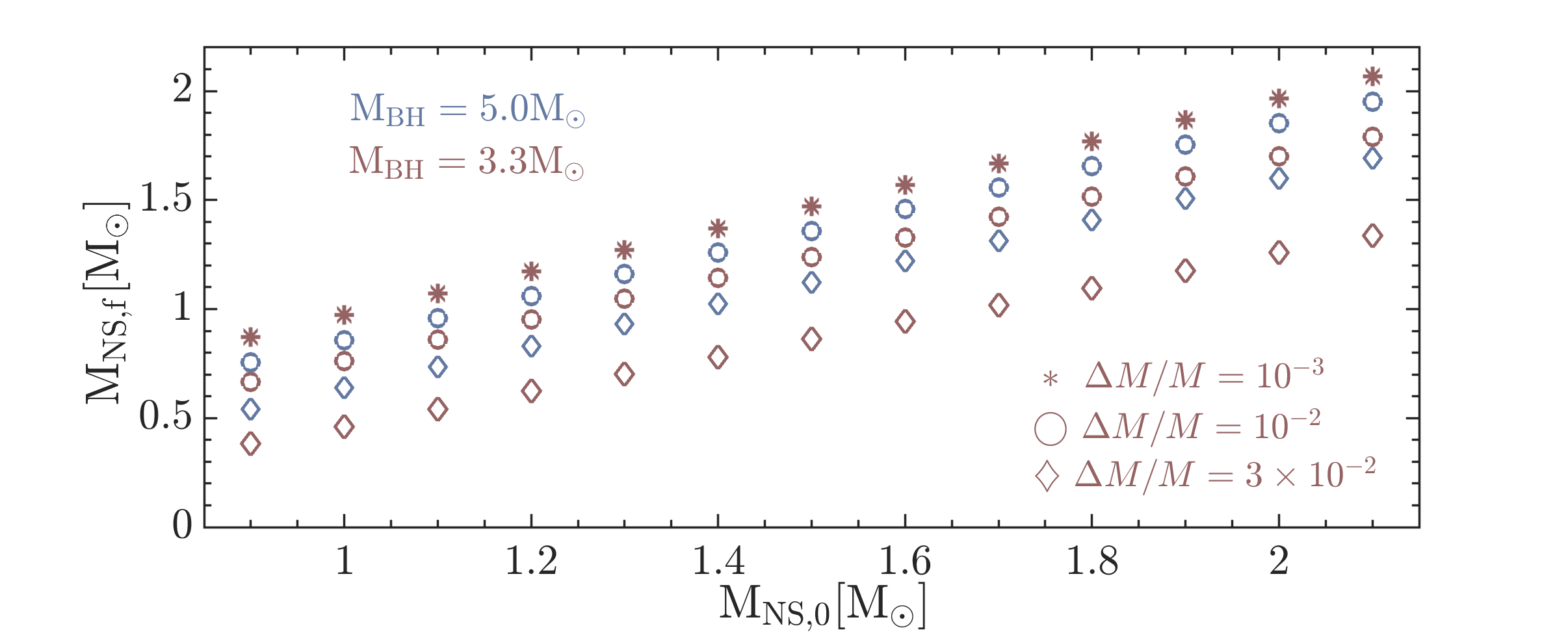}
    \caption{Upper panel: For a single initial neutron star mass ($M_{\rm NS,0} = 1.4 M_{\odot}$), $M_{\rm NS}$ as a function of semimajor axis for four different values of $\Delta M/M$ and two black hole masses ($M_{\rm BH} = 3.3, 5.0 M_\odot$). In order of increasing $\Delta M/M$, the curves have colors blue, red, yellow-green, and purple. The smaller black hole mass is shown with solid curves, the larger with dot-dash curves. Lower panel: Initial neutron star mass $M_{\rm NS,0}$ vs. final mass $M_{\rm NS,f}$ for $\Delta M/M = 10^{-3}$ (stars), $10^{-2}$ (circles), and $3 \times 10^{-2}$ (diamonds) and for $M_{\rm BH} = 3.3M_\odot$ (brown) and $M_{\rm BH} = 5.0 M_\odot$ (blue). When $\Delta M/M = 10^{-3}$, the total mass-loss is so small that the points for both black hole masses are superposed at $M_{\rm NS,f} = M_{\rm NS,0}$.}
    \label{fig:q_Mns_f_a}
\end{figure}

Substantial mass loss from a neutron star leads, naturally enough, to a substantially smaller neutron star mass at the time of the actual merger. How much smaller $M_{\rm NS}$ can be is shown in Figure~\ref{fig:q_Mns_f_a}.

The upper panel of this figure underlines what we have already seen, that the mass-loss increases with $\Delta M/M$ and is substantial when $\Delta M/M \gtrsim 10^{-2}$. However, another dependence is also illustrated by these curves: for fixed $M_{\rm NS,0}$, {\it smaller} $M_{\rm BH}$ promotes {\it greater} mass-transfer. The contrast in total mass-loss between the cases of $M_{\rm BH} = 3.3M_\odot$ and $M_{\rm BH} = 5M_\odot$ is a factor $\sim 2$.

The fruits of a larger sample of parameter values are shown in the lower panel of Figure~\ref{fig:q_Mns_f_a}. A complete tally of all the cases we studied can be found in Table \ref{tab:cases}. As these points show, when $\Delta M/M \gtrsim 10^{-2}$, the neutron star's mass can be reduced by tens of percent or more by the time it reaches merger.

\section{Potential consequences} \label{sec:Summary}

Just as Roche lobe overflow in ordinary stellar binaries leads to the creation of an accretion disk around the mass-receiving star, the mass lost from the neutron star should form such a disk around the black hole; the only requirement is that the mass transferred should have an angular momentum relative to the black hole large enough to place the matter outside the black hole's ISCO. The mass contained in this disk, potentially as much as a few tenths of a Solar mass, could be comparable to the disk that might be formed {\it after} the merger, which for masses in the range treated here, might have a mass $\sim 0.1M_\odot$ \citep{Foucart+18}.

The dynamical history of such a disk is the result of several competing processes. In circular binaries, the tidal truncation radius of a disk around the more massive partner is $\simeq 0.3a q^{-1/3}$; however, when the mass-transfer occurs, these binaries are still moderately eccentric: $e \approx 0.2 - 0.3$, throwing that estimate into some doubt. Pericenter passage could, for a brief time, significantly disturb the disk, both by the stronger tidal gravity and by the impact of additional matter peeled off the neutron star. Moreover, in the semimajor axis range of greatest mass-transfer (several tens of $r_g$), even the circular-orbit truncation scale is quite small: for example, when $a = 20 r_g$ the (Newtonian) circular-orbit truncation radius is only $\sim 8 r_g$.  This is close to the ISCO unless the black hole spins fairly rapidly (spin oblique to the orbital axis could, of course, lead to further complications); this structure echoes that of disks formed after a binary neutron star merger \citep{Zenati+24BNS,Camilletti+2024}. Interestingly, the characteristic duration for the mass-transfer process ($\sim 300$ binary orbits) is equivalent to a time $\lesssim 1$~s, which is close to the expected lifetime of post-merger disks \citep{Shibata_Taniguchi06,Paschalidis+15,Foucart:2012nc}. Thus, in respect to their mass, radial scale, and lifetime, these pre-merger disks resemble post-merger debris disks.

Material removed from a neutron star should already be threaded with a fairly strong magnetic field; if the accretion time for matter in the disk is at least $\sim 10$ orbits within the disk, the magnetorotational instability could amplify it further while also stirring MHD turbulence. Because the dynamics of mass-transfer likely seed the disk with a predominantly toroidal field, poloidal components, the sort necessary to support a jet, may grow slowly \citep{Most:2021ytn}. As the result of mass-transfer beginning $\sim 1$~sec before merger, such jets may appear earlier relative to the gravitational wave signal than jets supported by ordinary debris disks. Unfortunately, it is difficult to make a clear statement of how much earlier they might be launched because uncertainties in the jet launching timescale can be of comparable magnitude \citep{Gottlieb:2023est, Hayashi+24}.

The periodic perturbations at pericenter passage can affect the neutron star as well as the mass-transfer stream and the debris disk. If the crust can re-form as the neutron star goes through apocenter, each orbit's onset of mass transfer will likely trigger a shattering of the neutron star crust, which may potentially lead to electromagnetic transients \citep{Tsang:2011ad,Penner:2011br,Most:2024eig}.
Additionally, strong deformations of the star away from axisymmetry can be induced near black hole passages \citep{East:2011xa}, which may further complicate the mass-loss picture during the final orbits. Such deformations may also excite f-modes at every pericenter fly-by \citep{Chirenti:2016xys,Rosofsky:2018vyg}.

The last complication is that, after $\sim 300$ binary orbits, the remnant neutron star merges with the black hole. What happens to the matter delivered in advance can only be ascertained by a calculation including all these effects; such an effort is far beyond the present paper's scope. The options include everything from quick capture into the (enlarged) black hole to mixture with additional matter drawn from the neutron star during the merger proper.

There are several possibly observable signals from such a disk. Because its physical properties resemble those found in post-merger disks, some of the familiar post-merger phenomenology may be replicated, creation of $\gamma$-ray bursts, for example. Sufficient heat-production within the disk, whether due to dissipation of MHD turbulence or to nuclear reactions, might drive the sort of wind thought to result in kilonova afterglows \citep{Fernandez:2013tya,Fernandez:2016sbf,SiegelMetzger17PRL}. Just as for conventional kilonovae, the neutron-rich composition of the disk is likely to result in r-process nucleosynthesis producing many Lanthanide nuclei \citep{Fernandez&Metzger13,Prego+14,Wanajo+14,Lippuner&Luke15}, whose optical opacity causes the emergent spectrum to be rather red \citep{Kasen13,TankaHotokezaka2013}. However, in the event that only a small amount ($\lesssim 1\%$ of the neutron star's mass) is transferred, the portion of the neutron star from which it was taken is its outermost layers \citep{zenati+23BNS}, where the lepton fraction $Y_e$ is larger. When this is the case, the kilonova might stay blue, although the disk can also neutronize if its density becomes large enough \citep{De:2020jdt,Beloborodov:2002af}.

\begin{center}  \label{Sec:ack}
    \textbf{Acknowledgements}
\end{center}

NASA partially supported this work through grants NNH17ZDA001N and 80NSSC24K0100. YZ and JK were partially supported by NSF grant AST-2009260; in addition, JK received support from NSF grant PHY-2110339. YZ thanks Hagai Perets and Jeremy Schnittman for helpful discussions. ERM acknowledges partial support by the National Science Foundation under grants No.PHY-2309210 and AST-2307394.

\vspace{5mm}
\software{astropy \citep{Astropy+18}, Matplotlib \citep{Hunter07_matplotlib}, and Numpy \citep{2020NumPy-Array}.}

\begin{table*}[!t]
    \centering
    \caption{\label{tab:cases} Final Masses of Neutron star}
    
    \begin{tabular*}{\linewidth}{@{\extracolsep{\stretch{1}}}*{7}{c}}
\toprule
$a_{0}[rg]$&$e_{0}$&$e_{f}$&$\rm M_{BH}[M_\odot]$&$\rm 10^{-3}\times \frac{\Delta M}{M}$&$\rm M_{NS}^{0}[M_\odot]$&$\rm M_{NS}^{f}[M_\odot]$\\
\hline
50&0.842&0.153&3.8&20&1.0&0.633\\
  &0.846&0.192& & &1.1&0.724\\
  &0.851&0.108& & &1.2&0.818\\
  &0.853&0.204& & &1.3&0.909\\
  &0.856&0.219& & &1.4&1.003\\
  &0.859&0.213& & &1.5&1.181\\
  &0.861&0.171& & &1.6&1.181\\
  &0.863&0.201& & &1.7&1.273\\
  &0.865&0.175& & &1.8&1.366\\
  &0.867&0.198& & &1.9&1.453\\
  &0.869&0.227& & &2.0&1.533\\
  &0.871&0.215& & &2.1&1.619\\
  &0.842&0.176&3.8&50&1.0&0.353\\
  &0.842&0.118&3.8&5&1.0&0.885\\
  &0.842&0.222&2.78&5&1.0&0.833\\

\hline
60&0.882&0.138&3.8&20&1.0&0.691\\
  &0.885&0.164& & &1.1&0.784\\
  &0.887&0.111& & &1.2&0.885\\
  &0.891&0.202& & &1.3&0.982\\
  &0.893&0.212& & &1.4&1.071\\
  &0.895&0.208& & &1.5&1.168\\
  &0.896&0.173& & &1.6&1.261\\
  &0.898&0.212& & &1.7&1.337\\
  &0.900&0.178& & &1.8&1.409\\
  &0.902&0.186& & &1.9&1.402\\
  &0.903&0.209& & &2.0&1.621\\
  &0.904&0.207& & &2.1&1.613\\
  &0.881&0.124&3.8&50&1.0&0.402\\
  &0.881&0.117&3.8&5&1.0&0.890\\
  &0.846&0.200&2.78&5&1.0&0.852\\

\hline
100&0.923&0.098&3.8&20&1.0&0.635\\
  &0.925&0.211& & &1.1&0.725\\
  &0.927&0.205& & &1.2&0.824\\
  &0.928&0.182& & &1.3&0.915\\
  &0.930&0.126& & &1.4&0.965\\
  &0.931&0.196& & &1.5&1.102\\
  &0.932&0.105& & &1.6&1.193\\
  &0.934&0.188& & &1.7&1.279\\
  &0.935&0.147& & &1.8&1.376\\
  &0.935&0.192& & &1.9&1.461\\
  &0.936&0.206& & &2.0&1.557\\
  &0.937&0.221& & &2.1&1.639\\
  &0.923&0.209&3.8&5&1.0&0.334\\
  &0.923&0.101&3.8&50&1.0&0.889\\
  &0.902&0.093&2.78&50&1.0&0.827\\
\hline
    \end{tabular*}
\begin{flushleft}
 
\tablecomments{Columns are initial semi-major axis, initial eccentricity, pre-merger eccentricity, initial black hole mass, mass-transfer, initial neutron star mass, and final neutron star mass.}

\end{flushleft}
\end{table*}

\bibliography{refBHNS24}{}

\begin{thebibliography}{}
\expandafter\ifx\csname natexlab\endcsname\relax\def\natexlab#1{#1}\fi
\providecommand{\url}[1]{\href{#1}{#1}}
\providecommand{\dodoi}[1]{doi:~\href{http://doi.org/#1}{\nolinkurl{#1}}}
\providecommand{\doeprint}[1]{\href{http://ascl.net/#1}{\nolinkurl{http://ascl.net/#1}}}
\providecommand{\doarXiv}[1]{\href{https://arxiv.org/abs/#1}{\nolinkurl{https://arxiv.org/abs/#1}}}

\bibitem[{{Abbott} {et~al.}(2016){Abbott}, {Abbott}, {Abbott}, {Abernathy}, {Acernese}, {Ackley}, {Adams}, {Adams}, {Addesso}, {Adhikari}, {Adya}, {Affeldt}, {Agathos}, {Agatsuma}, {Aggarwal}, {Aguiar}, {Aiello}, {Ain}, {Ajith}, {Allen}, {Allocca}, {Altin}, {Anderson}, {Anderson}, {Arai}, {Arain}, {Araya}, {Arceneaux}, {Areeda}, {Arnaud}, {Arun}, {Ascenzi}, {Ashton}, {Ast}, {Aston}, {Astone}, {Aufmuth}, {Aulbert}, {Babak}, {Bacon}, {Bader}, {Baker}, {Baldaccini}, {Ballardin}, {Ballmer}, {Barayoga}, {Barclay}, {Barish}, {Barker}, {Barone}, {Barr}, {Barsotti}, {Barsuglia}, {Barta}, {Bartlett}, {Barton}, {Bartos}, {Bassiri}, {Basti}, {Batch}, {Baune}, {Bavigadda}, {Bazzan}, {Behnke}, {Bejger}, {Belczynski}, {Bell}, {Bell}, {Berger}, {Bergman}, {Bergmann}, {Berry}, {Bersanetti}, {Bertolini}, {Betzwieser}, {Bhagwat}, {Bhandare}, {Bilenko}, {Billingsley}, {Birch}, {Birney}, {Birnholtz}, {Biscans}, {Bisht}, {Bitossi}, {Biwer}, {Bizouard}, {Blackburn}, {Blair}, {Blair}, {Blair}, {Bloemen}, {Bock}, {Bodiya}, {Boer},
  {Bogaert}, {Bogan}, {Bohe}, {Bojtos}, {Bond}, {Bondu}, {Bonnand}, {Boom}, {Bork}, {Boschi}, {Bose}, {Bouffanais}, {Bozzi}, {Bradaschia}, {Brady}, {Braginsky}, {Branchesi}, {Brau}, {Briant}, {Brillet}, {Brinkmann}, {Brisson}, {Brockill}, {Brooks}, {Brown}, {Brown}, {Brown}, {Buchanan}, {Buikema}, {Bulik}, {Bulten}, {Buonanno}, {Buskulic}, {Buy}, {Byer}, {Cabero}, {Cadonati}, {Cagnoli}, {Cahillane}, {Bustillo}, {Callister}, {Calloni}, {Camp}, {Cannon}, {Cao}, {Capano}, {Capocasa}, {Carbognani}, {Caride}, {Casanueva Diaz}, {Casentini}, {Caudill}, {Cavagli{\`a}}, {Cavalier}, {Cavalieri}, {Cella}, {Cepeda}, {Baiardi}, {Cerretani}, {Cesarini}, {Chakraborty}, {Chalermsongsak}, {Chamberlin}, {Chan}, {Chao}, {Charlton}, {Chassande-Mottin}, {Chen}, {Chen}, {Cheng}, {Chincarini}, {Chiummo}, {Cho}, {Cho}, {Chow}, {Christensen}, {Chu}, {Chua}, {Chung}, {Ciani}, {Clara}, {Clark}, {Cleva}, {Coccia}, {Cohadon}, {Colla}, {Collette}, {Cominsky}, {Constancio}, {Conte}, {Conti}, {Cook}, {Corbitt}, {Cornish}, {Corsi},
  {Cortese}, {Costa}, {Coughlin}, {Coughlin}, {Coulon}, {Countryman}, {Couvares}, {Cowan}, {Coward}, {Cowart}, {Coyne}, {Coyne}, {Craig}, {Creighton}, {Creighton}, {Cripe}, {Crowder}, {Cruise}, {Cumming}, {Cunningham}, {Cuoco}, {Dal Canton}, {Danilishin}, {D'Antonio}, {Danzmann}, {Darman}, {Da Silva Costa}, {Dattilo}, {Dave}, {Daveloza}, {Davier}, {Davies}, {Daw}, {Day}, {De}, {DeBra}, {Debreczeni}, {Degallaix}, {De Laurentis}, {Del{\'e}glise}, {Del Pozzo}, {Denker}, {Dent}, {Dereli}, {Dergachev}, {DeRosa}, {De Rosa}, {DeSalvo}, {Dhurandhar}, {D{\'\i}az}, {Di Fiore}, {Di Giovanni}, {Di Lieto}, {Di Pace}, {Di Palma}, {Di Virgilio}, {Dojcinoski}, {Dolique}, {Donovan}, {Dooley}, {Doravari}, {Douglas}, {Downes}, {Drago}, {Drever}, {Driggers}, {Du}, {Ducrot}, {Dwyer}, {Edo}, {Edwards}, {Effler}, {Eggenstein}, {Ehrens}, {Eichholz}, {Eikenberry}, {Engels}, {Essick}, {Etzel}, {Evans}, {Evans}, {Everett}, {Factourovich}, {Fafone}, {Fair}, {Fairhurst}, {Fan}, {Fang}, {Farinon}, {Farr}, {Farr}, {Favata}, {Fays},
  {Fehrmann}, {Fejer}, {Feldbaum}, {Ferrante}, {Ferreira}, {Ferrini}, {Fidecaro}, {Finn}, {Fiori}, {Fiorucci}, {Fisher}, {Flaminio}, {Fletcher}, {Fong}, {Fournier}, {Franco}, {Frasca}, {Frasconi}, {Frede}, {Frei}, {Freise}, {Frey}, {Frey}, {Fricke}, {Fritschel}, {Frolov}, {Fulda}, {Fyffe}, {Gabbard}, {Gair}, {Gammaitoni}, {Gaonkar}, {Garufi}, {Gatto}, {Gaur}, {Gehrels}, {Gemme}, {Gendre}, {Genin}, {Gennai}, {George}, {Gergely}, {Germain}, {Ghosh}, {Ghosh}, {Ghosh}, {Giaime}, {Giardina}, {Giazotto}, {Gill}, {Glaefke}, {Gleason}, {Goetz}, {Goetz}, {Gondan}, {Gonz{\'a}lez}, {Castro}, {Gopakumar}, {Gordon}, {Gorodetsky}, {Gossan}, {Gosselin}, {Gouaty}, {Graef}, {Graff}, {Granata}, {Grant}, {Gras}, {Gray}, {Greco}, {Green}, {Greenhalgh}, {Groot}, {Grote}, {Grunewald}, {Guidi}, {Guo}, {Gupta}, {Gupta}, {Gushwa}, {Gustafson}, {Gustafson}, {Hacker}, {Hall}, {Hall}, {Hammond}, {Haney}, {Hanke}, {Hanks}, {Hanna}, {Hannam}, {Hanson}, {Hardwick}, {Harms}, {Harry}, {Harry}, {Hart}, {Hartman}, {Haster}, {Haughian},
  {Healy}, {Heefner}, {Heidmann}, {Heintze}, {Heinzel}, {Heitmann}, {Hello}, {Hemming}, {Hendry}, {Heng}, {Hennig}, {Heptonstall}, {Heurs}, {Hild}, {Hoak}, {Hodge}, {Hofman}, {Hollitt}, {Holt}, {Holz}, {Hopkins}, {Hosken}, {Hough}, {Houston}, {Howell}, {Hu}, {Huang}, {Huerta}, {Huet}, {Hughey}, {Husa}, {Huttner}, {Huynh-Dinh}, {Idrisy}, {Indik}, {Ingram}, {Inta}, {Isa}, {Isac}, {Isi}, {Islas}, {Isogai}, {Iyer}, {Izumi}, {Jacobson}, {Jacqmin}, {Jang}, {Jani}, {Jaranowski}, {Jawahar}, {Jim{\'e}nez-Forteza}, {Johnson}, {Johnson-McDaniel}, {Jones}, {Jones}, {Jonker}, {Ju}, {Haris}, {Kalaghatgi}, {Kalogera}, {Kandhasamy}, {Kang}, {Kanner}, {Karki}, {Kasprzack}, {Katsavounidis}, {Katzman}, {Kaufer}, {Kaur}, {Kawabe}, {Kawazoe}, {K{\'e}f{\'e}lian}, {Kehl}, {Keitel}, {Kelley}, {Kells}, {Kennedy}, {Keppel}, {Key}, {Khalaidovski}, {Khalili}, {Khan}, {Khan}, {Khan}, {Khazanov}, {Kijbunchoo}, {Kim}, {Kim}, {Kim}, {Kim}, {Kim}, {Kim}, {King}, {King}, {Kinzel}, {Kissel}, {Kleybolte}, {Klimenko}, {Koehlenbeck}, {Kokeyama},
  {Koley}, {Kondrashov}, {Kontos}, {Koranda}, {Korobko}, {Korth}, {Kowalska}, {Kozak}, {Kringel}, {Krishnan}, {Kr{\'o}lak}, {Krueger}, {Kuehn}, {Kumar}, {Kumar}, {Kuo}, {Kutynia}, {Kwee}, {Lackey}, {Landry}, {Lange}, {Lantz}, {Lasky}, {Lazzarini}, {Lazzaro}, {Leaci}, {Leavey}, {Lebigot}, {Lee}, {Lee}, {Lee}, {Lee}, {Lenon}, {Leonardi}, {Leong}, {Leroy}, {Letendre}, {Levin}, {Levine}, {Li}, {Libson}, {Littenberg}, {Lockerbie}, {Logue}, {Lombardi}, {London}, {Lord}, {Lorenzini}, {Loriette}, {Lormand}, {Losurdo}, {Lough}, {Lousto}, {Lovelace}, {L{\"u}ck}, {Lundgren}, {Luo}, {Lynch}, {Ma}, {MacDonald}, {Machenschalk}, {MacInnis}, {Macleod}, {Maga{\~n}a-Sandoval}, {Magee}, {Mageswaran}, {Majorana}, {Maksimovic}, {Malvezzi}, {Man}, {Mandel}, {Mandic}, {Mangano}, {Mansell}, {Manske}, {Mantovani}, {Marchesoni}, {Marion}, {M{\'a}rka}, {M{\'a}rka}, {Markosyan}, {Maros}, {Martelli}, {Martellini}, {Martin}, {Martin}, {Martynov}, {Marx}, {Mason}, {Masserot}, {Massinger}, {Masso-Reid}, {Matichard}, {Matone}, {Mavalvala},
  {Mazumder}, {Mazzolo}, {McCarthy}, {McClelland}, {McCormick}, {McGuire}, {McIntyre}, {McIver}, {McManus}, {McWilliams}, {Meacher}, {Meadors}, {Meidam}, {Melatos}, {Mendell}, {Mendoza-Gandara}, {Mercer}, {Merilh}, {Merzougui}, {Meshkov}, {Messenger}, {Messick}, {Meyers}, {Mezzani}, {Miao}, {Michel}, {Middleton}, {Mikhailov}, {Milano}, {Miller}, {Millhouse}, {Minenkov}, {Ming}, {Mirshekari}, {Mishra}, {Mitra}, {Mitrofanov}, {Mitselmakher}, {Mittleman}, {Moggi}, {Mohan}, {Mohapatra}, {Montani}, {Moore}, {Moore}, {Moraru}, {Moreno}, {Morriss}, {Mossavi}, {Mours}, {Mow-Lowry}, {Mueller}, {Mueller}, {Muir}, {Mukherjee}, {Mukherjee}, {Mukherjee}, {Mukund}, {Mullavey}, {Munch}, {Murphy}, {Murray}, {Mytidis}, {Nardecchia}, {Naticchioni}, {Nayak}, {Necula}, {Nedkova}, {Nelemans}, {Neri}, {Neunzert}, {Newton}, {Nguyen}, {Nielsen}, {Nissanke}, {Nitz}, {Nocera}, {Nolting}, {Normandin}, {Nuttall}, {Oberling}, {Ochsner}, {O'Dell}, {Oelker}, {Ogin}, {Oh}, {Oh}, {Ohme}, {Oliver}, {Oppermann}, {Oram}, {O'Reilly},
  {O'Shaughnessy}, {Ott}, {Ottaway}, {Ottens}, {Overmier}, {Owen}, {Pai}, {Pai}, {Palamos}, {Palashov}, {Palomba}, {Pal-Singh}, {Pan}, {Pan}, {Pankow}, {Pannarale}, {Pant}, {Paoletti}, {Paoli}, {Papa}, {Paris}, {Parker}, {Pascucci}, {Pasqualetti}, {Passaquieti}, {Passuello}, {Patricelli}, {Patrick}, {Pearlstone}, {Pedraza}, {Pedurand}, {Pekowsky}, {Pele}, {Penn}, {Perreca}, {Pfeiffer}, {Phelps}, {Piccinni}, {Pichot}, {Pickenpack}, {Piergiovanni}, {Pierro}, {Pillant}, {Pinard}, {Pinto}, {Pitkin}, {Poeld}, {Poggiani}, {Popolizio}, {Post}, {Powell}, {Prasad}, {Predoi}, {Premachandra}, {Prestegard}, {Price}, {Prijatelj}, {Principe}, {Privitera}, {Prix}, {Prodi}, {Prokhorov}, {Puncken}, {Punturo}, {Puppo}, {P{\"u}rrer}, {Qi}, {Qin}, {Quetschke}, {Quintero}, {Quitzow-James}, {Raab}, {Rabeling}, {Radkins}, {Raffai}, {Raja}, {Rakhmanov}, {Ramet}, {Rapagnani}, {Raymond}, {Razzano}, {Re}, {Read}, {Reed}, {Regimbau}, {Rei}, {Reid}, {Reitze}, {Rew}, {Reyes}, {Ricci}, {Riles}, {Robertson}, {Robie}, {Robinet}, {Rocchi},
  {Rolland}, {Rollins}, {Roma}, {Romano}, {Romano}, {Romanov}, {Romie}, {Rosi{\'n}ska}, {Rowan}, {R{\"u}diger}, {Ruggi}, {Ryan}, {Sachdev}, {Sadecki}, {Sadeghian}, {Salconi}, {Saleem}, {Salemi}, {Samajdar}, {Sammut}, {Sampson}, {Sanchez}, {Sandberg}, {Sandeen}, {Sanders}, {Sanders}, {Sassolas}, {Sathyaprakash}, {Saulson}, {Sauter}, {Savage}, {Sawadsky}, {Schale}, {Schilling}, {Schmidt}, {Schmidt}, {Schnabel}, {Schofield}, {Sch{\"o}nbeck}, {Schreiber}, {Schuette}, {Schutz}, {Scott}, {Scott}, {Sellers}, {Sengupta}, {Sentenac}, {Sequino}, {Sergeev}, {Serna}, {Setyawati}, {Sevigny}, {Shaddock}, {Shaffer}, {Shah}, {Shahriar}, {Shaltev}, {Shao}, {Shapiro}, {Shawhan}, {Sheperd}, {Shoemaker}, {Shoemaker}, {Siellez}, {Siemens}, {Sigg}, {Silva}, {Simakov}, {Singer}, {Singer}, {Singh}, {Singh}, {Singhal}, {Sintes}, {Slagmolen}, {Smith}, {Smith}, {Smith}, {Smith}, {Son}, {Sorazu}, {Sorrentino}, {Souradeep}, {Srivastava}, {Staley}, {Steinke}, {Steinlechner}, {Steinlechner}, {Steinmeyer}, {Stephens}, {Stevenson}, {Stone},
  {Strain}, {Straniero}, {Stratta}, {Strauss}, {Strigin}, {Sturani}, {Stuver}, {Summerscales}, {Sun}, {Sutton}, {Swinkels}, {Szczepa{\'n}czyk}, {Tacca}, {Talukder}, {Tanner}, {T{\'a}pai}, {Tarabrin}, {Taracchini}, {Taylor}, {Theeg}, {Thirugnanasambandam}, {Thomas}, {Thomas}, {Thomas}, {Thorne}, {Thorne}, {Thrane}, {Tiwari}, {Tiwari}, {Tokmakov}, {Tomlinson}, {Tonelli}, {Torres}, {Torrie}, {T{\"o}yr{\"a}}, {Travasso}, {Traylor}, {Trifir{\`o}}, {Tringali}, {Trozzo}, {Tse}, {Turconi}, {Tuyenbayev}, {Ugolini}, {Unnikrishnan}, {Urban}, {Usman}, {Vahlbruch}, {Vajente}, {Valdes}, {Vallisneri}, {van Bakel}, {van Beuzekom}, {van den Brand}, {Van Den Broeck}, {Vander-Hyde}, {van der Schaaf}, {van Heijningen}, {van Veggel}, {Vardaro}, {Vass}, {Vas{\'u}th}, {Vaulin}, {Vecchio}, {Vedovato}, {Veitch}, {Veitch}, {Venkateswara}, {Verkindt}, {Vetrano}, {Vicer{\'e}}, {Vinciguerra}, {Vine}, {Vinet}, {Vitale}, {Vo}, {Vocca}, {Vorvick}, {Voss}, {Vousden}, {Vyatchanin}, {Wade}, {Wade}, {Wade}, {Waldman}, {Walker}, {Wallace},
  {Walsh}, {Wang}, {Wang}, {Wang}, {Wang}, {Wang}, {Ward}, {Ward}, {Warner}, {Was}, {Weaver}, {Wei}, {Weinert}, {Weinstein}, {Weiss}, {Welborn}, {Wen}, {We{\ss}els}, {Westphal}, {Wette}, {Whelan}, {Whitcomb}, {White}, {Whiting}, {Wiesner}, {Wilkinson}, {Willems}, {Williams}, {Williams}, {Williamson}, {Willis}, {Willke}, {Wimmer}, {Winkelmann}, {Winkler}, {Wipf}, {Wiseman}, {Wittel}, {Woan}, {Worden}, {Wright}, {Wu}, {Yablon}, {Yakushin}, {Yam}, {Yamamoto}, {Yancey}, {Yap}, {Yu}, {Yvert}, {Zadro{\.Z}ny}, {Zangrando}, {Zanolin}, {Zendri}, {Zevin}, {Zhang}, {Zhang}, {Zhang}, {Zhang}, {Zhao}, {Zhou}, {Zhou}, {Zhu}, {Zucker}, {Zuraw}, {Zweizig}, {LIGO Scientific Collaboration}, \& {Virgo Collaboration}}]{Abbott2016}
{Abbott}, B.~P., {Abbott}, R., {Abbott}, T.~D., {et~al.} 2016, \prl, 116, 061102, \dodoi{10.1103/PhysRevLett.116.061102}

\bibitem[{{Abbott} {et~al.}(2017{\natexlab{a}}){Abbott}, {Abbott}, {Abbott}, {Acernese}, {Ackley}, {Adams}, {Adams}, {Addesso}, {Adhikari}, {Adya}, {Affeldt}, {Afrough}, {Agarwal}, {Agathos}, {Agatsuma}, {Aggarwal}, {Aguiar}, {Aiello}, {Ain}, {Ajith}, {Allen}, {Allen}, {Allocca}, {Altin}, {Amato}, {Ananyeva}, {Anderson}, {Anderson}, {Angelova}, {Antier}, {Appert}, {Arai}, {Araya}, {Areeda}, {Arnaud}, {Arun}, {Ascenzi}, {Ashton}, {Ast}, {Aston}, {Astone}, {Atallah}, {Aufmuth}, {Aulbert}, {AultONeal}, {Austin}, {Avila-Alvarez}, {Babak}, {Bacon}, {Bader}, {Bae}, {Bailes}, {Baker}, {Baldaccini}, {Ballardin}, {Ballmer}, {Banagiri}, {Barayoga}, {Barclay}, {Barish}, {Barker}, {Barkett}, {Barone}, {Barr}, {Barsotti}, {Barsuglia}, {Barta}, {Barthelmy}, {Bartlett}, {Bartos}, {Bassiri}, {Basti}, {Batch}, {Bawaj}, {Bayley}, {Bazzan}, {B{\'e}csy}, {Beer}, {Bejger}, {Belahcene}, {Bell}, {Berger}, {Bergmann}, {Bernuzzi}, {Bero}, {Berry}, {Bersanetti}, {Bertolini}, {Betzwieser}, {Bhagwat}, {Bhandare}, {Bilenko},
  {Billingsley}, {Billman}, {Birch}, {Birney}, {Birnholtz}, {Biscans}, {Biscoveanu}, {Bisht}, {Bitossi}, {Biwer}, {Bizouard}, {Blackburn}, {Blackman}, {Blair}, {Blair}, {Blair}, {Bloemen}, {Bock}, {Bode}, {Boer}, {Bogaert}, {Bohe}, {Bondu}, {Bonilla}, {Bonnand}, {Boom}, {Bork}, {Boschi}, {Bose}, {Bossie}, {Bouffanais}, {Bozzi}, {Bradaschia}, {Brady}, {Branchesi}, {Brau}, {Briant}, {Brillet}, {Brinkmann}, {Brisson}, {Brockill}, {Broida}, {Brooks}, {Brown}, {Brown}, {Brunett}, {Buchanan}, {Buikema}, {Bulik}, {Bulten}, {Buonanno}, {Buskulic}, {Buy}, {Byer}, {Cabero}, {Cadonati}, {Cagnoli}, {Cahillane}, {Calder{\'o}n Bustillo}, {Callister}, {Calloni}, {Camp}, {Canepa}, {Canizares}, {Cannon}, {Cao}, {Cao}, {Capano}, {Capocasa}, {Carbognani}, {Caride}, {Carney}, {Carullo}, {Casanueva Diaz}, {Casentini}, {Caudill}, {Cavagli{\`a}}, {Cavalier}, {Cavalieri}, {Cella}, {Cepeda}, {Cerd{\'a}-Dur{\'a}n}, {Cerretani}, {Cesarini}, {Chamberlin}, {Chan}, {Chao}, {Charlton}, {Chase}, {Chassande-Mottin}, {Chatterjee},
  {Chatziioannou}, {Cheeseboro}, {Chen}, {Chen}, {Chen}, {Cheng}, {Chia}, {Chincarini}, {Chiummo}, {Chmiel}, {Cho}, {Cho}, {Chow}, {Christensen}, {Chu}, {Chua}, {Chua}, {Chung}, {Chung}, {Ciani}, {Ciolfi}, {Cirelli}, {Cirone}, {Clara}, {Clark}, {Clearwater}, {Cleva}, {Cocchieri}, {Coccia}, {Cohadon}, {Cohen}, {Colla}, {Collette}, {Cominsky}, {Constancio}, {Conti}, {Cooper}, {Corban}, {Corbitt}, {Cordero-Carri{\'o}n}, {Corley}, {Cornish}, {Corsi}, {Cortese}, {Costa}, {Coughlin}, {Coughlin}, {Coulon}, {Countryman}, {Couvares}, {Covas}, {Cowan}, {Coward}, {Cowart}, {Coyne}, {Coyne}, {Creighton}, {Creighton}, {Cripe}, {Crowder}, {Cullen}, {Cumming}, {Cunningham}, {Cuoco}, {Dal Canton}, {D{\'a}lya}, {Danilishin}, {D'Antonio}, {Danzmann}, {Dasgupta}, {Da Silva Costa}, {Dattilo}, {Dave}, {Davier}, {Davis}, {Daw}, {Day}, {De}, {DeBra}, {Degallaix}, {De Laurentis}, {Del{\'e}glise}, {Del Pozzo}, {Demos}, {Denker}, {Dent}, {De Pietri}, {Dergachev}, {De Rosa}, {DeRosa}, {De Rossi}, {DeSalvo}, {de Varona}, {Devenson},
  {Dhurandhar}, {D{\'\i}az}, {Dietrich}, {Di Fiore}, {Di Giovanni}, {Di Girolamo}, {Di Lieto}, {Di Pace}, {Di Palma}, {Di Renzo}, {Doctor}, {Dolique}, {Donovan}, {Dooley}, {Doravari}, {Dorrington}, {Douglas}, {Dovale {\'A}lvarez}, {Downes}, {Drago}, {Dreissigacker}, {Driggers}, {Du}, {Ducrot}, {Dudi}, {Dupej}, {Dwyer}, {Edo}, {Edwards}, {Effler}, {Eggenstein}, {Ehrens}, {Eichholz}, {Eikenberry}, {Eisenstein}, {Essick}, {Estevez}, {Etienne}, {Etzel}, {Evans}, {Evans}, {Factourovich}, {Fafone}, {Fair}, {Fairhurst}, {Fan}, {Farinon}, {Farr}, {Farr}, {Fauchon-Jones}, {Favata}, {Fays}, {Fee}, {Fehrmann}, {Feicht}, {Fejer}, {Fernandez-Galiana}, {Ferrante}, {Ferreira}, {Ferrini}, {Fidecaro}, {Finstad}, {Fiori}, {Fiorucci}, {Fishbach}, {Fisher}, {Fitz-Axen}, {Flaminio}, {Fletcher}, {Fong}, {Font}, {Forsyth}, {Forsyth}, {Fournier}, {Frasca}, {Frasconi}, {Frei}, {Freise}, {Frey}, {Frey}, {Fries}, {Fritschel}, {Frolov}, {Fulda}, {Fyffe}, {Gabbard}, {Gadre}, {Gaebel}, {Gair}, {Gammaitoni}, {Ganija}, {Gaonkar},
  {Garcia-Quiros}, {Garufi}, {Gateley}, {Gaudio}, {Gaur}, {Gayathri}, {Gehrels}, {Gemme}, {Genin}, {Gennai}, {George}, {George}, {Gergely}, {Germain}, {Ghonge}, {Ghosh}, {Ghosh}, {Ghosh}, {Giaime}, {Giardina}, {Giazotto}, {Gill}, {Glover}, {Goetz}, {Goetz}, {Gomes}, {Goncharov}, {Gonz{\'a}lez}, {Gonzalez Castro}, {Gopakumar}, {Gorodetsky}, {Gossan}, {Gosselin}, {Gouaty}, {Grado}, {Graef}, {Granata}, {Grant}, {Gras}, {Gray}, {Greco}, {Green}, {Gretarsson}, {Groot}, {Grote}, {Grunewald}, {Gruning}, {Guidi}, {Guo}, {Gupta}, {Gupta}, {Gushwa}, {Gustafson}, {Gustafson}, {Halim}, {Hall}, {Hall}, {Hamilton}, {Hammond}, {Haney}, {Hanke}, {Hanks}, {Hanna}, {Hannam}, {Hannuksela}, {Hanson}, {Hardwick}, {Harms}, {Harry}, {Harry}, {Hart}, {Haster}, {Haughian}, {Healy}, {Heidmann}, {Heintze}, {Heitmann}, {Hello}, {Hemming}, {Hendry}, {Heng}, {Hennig}, {Heptonstall}, {Heurs}, {Hild}, {Hinderer}, {Ho}, {Hoak}, {Hofman}, {Holt}, {Holz}, {Hopkins}, {Horst}, {Hough}, {Houston}, {Howell}, {Hreibi}, {Hu}, {Huerta}, {Huet},
  {Hughey}, {Husa}, {Huttner}, {Huynh-Dinh}, {Indik}, {Inta}, {Intini}, {Isa}, {Isac}, {Isi}, {Iyer}, {Izumi}, {Jacqmin}, {Jani}, {Jaranowski}, {Jawahar}, {Jim{\'e}nez-Forteza}, {Johnson}, {Johnson-McDaniel}, {Jones}, {Jones}, {Jonker}, {Ju}, {Junker}, {Kalaghatgi}, {Kalogera}, {Kamai}, {Kandhasamy}, {Kang}, {Kanner}, {Kapadia}, {Karki}, {Karvinen}, {Kasprzack}, {Kastaun}, {Katolik}, {Katsavounidis}, {Katzman}, {Kaufer}, {Kawabe}, {K{\'e}f{\'e}lian}, {Keitel}, {Kemball}, {Kennedy}, {Kent}, {Key}, {Khalili}, {Khan}, {Khan}, {Khan}, {Khazanov}, {Kijbunchoo}, {Kim}, {Kim}, {Kim}, {Kim}, {Kim}, {Kim}, {Kimbrell}, {King}, {King}, {Kinley-Hanlon}, {Kirchhoff}, {Kissel}, {Kleybolte}, {Klimenko}, {Knowles}, {Koch}, {Koehlenbeck}, {Koley}, {Kondrashov}, {Kontos}, {Korobko}, {Korth}, {Kowalska}, {Kozak}, {Kr{\"a}mer}, {Kringel}, {Krishnan}, {Kr{\'o}lak}, {Kuehn}, {Kumar}, {Kumar}, {Kumar}, {Kuo}, {Kutynia}, {Kwang}, {Lackey}, {Lai}, {Landry}, {Lang}, {Lange}, {Lantz}, {Lanza}, {Larson}, {Lartaux-Vollard}, {Lasky},
  {Laxen}, {Lazzarini}, {Lazzaro}, {Leaci}, {Leavey}, {Lee}, {Lee}, {Lee}, {Lee}, {Lee}, {Lehmann}, {Lenon}, {Leon}, {Leonardi}, {Leroy}, {Letendre}, {Levin}, {Li}, {Linker}, {Littenberg}, {Liu}, {Liu}, {Lo}, {Lockerbie}, {London}, {Lord}, {Lorenzini}, {Loriette}, {Lormand}, {Losurdo}, {Lough}, {Lousto}, {Lovelace}, {L{\"u}ck}, {Lumaca}, {Lundgren}, {Lynch}, {Ma}, {Macas}, {Macfoy}, {Machenschalk}, {MacInnis}, {Macleod}, {Maga{\~n}a Hernandez}, {Maga{\~n}a-Sandoval}, {Maga{\~n}a Zertuche}, {Magee}, {Majorana}, {Maksimovic}, {Man}, {Mandic}, {Mangano}, {Mansell}, {Manske}, {Mantovani}, {Marchesoni}, {Marion}, {M{\'a}rka}, {M{\'a}rka}, {Markakis}, {Markosyan}, {Markowitz}, {Maros}, {Marquina}, {Marsh}, {Martelli}, {Martellini}, {Martin}, {Martin}, {Martynov}, {Marx}, {Mason}, {Massera}, {Masserot}, {Massinger}, {Masso-Reid}, {Mastrogiovanni}, {Matas}, {Matichard}, {Matone}, {Mavalvala}, {Mazumder}, {McCarthy}, {McClelland}, {McCormick}, {McCuller}, {McGuire}, {McIntyre}, {McIver}, {McManus}, {McNeill}, {McRae},
  {McWilliams}, {Meacher}, {Meadors}, {Mehmet}, {Meidam}, {Mejuto-Villa}, {Melatos}, {Mendell}, {Mercer}, {Merilh}, {Merzougui}, {Meshkov}, {Messenger}, {Messick}, {Metzdorff}, {Meyers}, {Miao}, {Michel}, {Middleton}, {Mikhailov}, {Milano}, {Miller}, {Miller}, {Miller}, {Millhouse}, {Milovich-Goff}, {Minazzoli}, {Minenkov}, {Ming}, {Mishra}, {Mitra}, {Mitrofanov}, {Mitselmakher}, {Mittleman}, {Moffa}, {Moggi}, {Mogushi}, {Mohan}, {Mohapatra}, {Molina}, {Montani}, {Moore}, {Moraru}, {Moreno}, {Morisaki}, {Morriss}, {Mours}, {Mow-Lowry}, {Mueller}, {Muir}, {Mukherjee}, {Mukherjee}, {Mukherjee}, {Mukund}, {Mullavey}, {Munch}, {Mu{\~n}iz}, {Muratore}, {Murray}, {Nagar}, {Napier}, {Nardecchia}, {Naticchioni}, {Nayak}, {Neilson}, {Nelemans}, {Nelson}, {Nery}, {Neunzert}, {Nevin}, {Newport}, {Newton}, {Ng}, {Nguyen}, {Nguyen}, {Nichols}, {Nielsen}, {Nissanke}, {Nitz}, {Noack}, {Nocera}, {Nolting}, {North}, {Nuttall}, {Oberling}, {O'Dea}, {Ogin}, {Oh}, {Oh}, {Ohme}, {Okada}, {Oliver}, {Oppermann}, {Oram}, {O'Reilly},
  {Ormiston}, {Ortega}, {O'Shaughnessy}, {Ossokine}, {Ottaway}, {Overmier}, {Owen}, {Pace}, {Page}, {Page}, {Pai}, {Pai}, {Palamos}, {Palashov}, {Palomba}, {Pal-Singh}, {Pan}, {Pan}, {Pang}, {Pang}, {Pankow}, {Pannarale}, {Pant}, {Paoletti}, {Paoli}, {Papa}, {Parida}, {Parker}, {Pascucci}, {Pasqualetti}, {Passaquieti}, {Passuello}, {Patil}, {Patricelli}, {Pearlstone}, {Pedraza}, {Pedurand}, {Pekowsky}, {Pele}, {Penn}, {Perez}, {Perreca}, {Perri}, {Pfeiffer}, {Phelps}, {Piccinni}, {Pichot}, {Piergiovanni}, {Pierro}, {Pillant}, {Pinard}, {Pinto}, {Pirello}, {Pitkin}, {Poe}, {Poggiani}, {Popolizio}, {Porter}, {Post}, {Powell}, {Prasad}, {Pratt}, {Pratten}, {Predoi}, {Prestegard}, {Prijatelj}, {Principe}, {Privitera}, {Prix}, {Prodi}, {Prokhorov}, {Puncken}, {Punturo}, {Puppo}, {P{\"u}rrer}, {Qi}, {Quetschke}, {Quintero}, {Quitzow-James}, {Raab}, {Rabeling}, {Radkins}, {Raffai}, {Raja}, {Rajan}, {Rajbhandari}, {Rakhmanov}, {Ramirez}, {Ramos-Buades}, {Rapagnani}, {Raymond}, {Razzano}, {Read}, {Regimbau}, {Rei},
  {Reid}, {Reitze}, {Ren}, {Reyes}, {Ricci}, {Ricker}, {Rieger}, {Riles}, {Rizzo}, {Robertson}, {Robie}, {Robinet}, {Rocchi}, {Rolland}, {Rollins}, {Roma}, {Romano}, {Romano}, {Romel}, {Romie}, {Rosi{\'n}ska}, {Ross}, {Rowan}, {R{\"u}diger}, {Ruggi}, {Rutins}, {Ryan}, {Sachdev}, {Sadecki}, {Sadeghian}, {Sakellariadou}, {Salconi}, {Saleem}, {Salemi}, {Samajdar}, {Sammut}, {Sampson}, {Sanchez}, {Sanchez}, {Sanchis-Gual}, {Sandberg}, {Sanders}, {Sassolas}, {Sathyaprakash}, {Saulson}, {Sauter}, {Savage}, {Sawadsky}, {Schale}, {Scheel}, {Scheuer}, {Schmidt}, {Schmidt}, {Schnabel}, {Schofield}, {Sch{\"o}nbeck}, {Schreiber}, {Schuette}, {Schulte}, {Schutz}, {Schwalbe}, {Scott}, {Scott}, {Seidel}, {Sellers}, {Sengupta}, {Sentenac}, {Sequino}, {Sergeev}, {Shaddock}, {Shaffer}, {Shah}, {Shahriar}, {Shaner}, {Shao}, {Shapiro}, {Shawhan}, {Sheperd}, {Shoemaker}, {Shoemaker}, {Siellez}, {Siemens}, {Sieniawska}, {Sigg}, {Silva}, {Singer}, {Singh}, {Singhal}, {Sintes}, {Slagmolen}, {Smith}, {Smith}, {Smith}, {Somala},
  {Son}, {Sonnenberg}, {Sorazu}, {Sorrentino}, {Souradeep}, {Spencer}, {Srivastava}, {Staats}, {Staley}, {Steinke}, {Steinlechner}, {Steinlechner}, {Steinmeyer}, {Stevenson}, {Stone}, {Stops}, {Strain}, {Stratta}, {Strigin}, {Strunk}, {Sturani}, {Stuver}, {Summerscales}, {Sun}, {Sunil}, {Suresh}, {Sutton}, {Swinkels}, {Szczepa{\'n}czyk}, {Tacca}, {Tait}, {Talbot}, {Talukder}, {Tanner}, {T{\'a}pai}, {Taracchini}, {Tasson}, {Taylor}, {Taylor}, {Tewari}, {Theeg}, {Thies}, {Thomas}, {Thomas}, {Thomas}, {Thorne}, {Thorne}, {Thrane}, {Tiwari}, {Tiwari}, {Tokmakov}, {Toland}, {Tonelli}, {Tornasi}, {Torres-Forn{\'e}}, {Torrie}, {T{\"o}yr{\"a}}, {Travasso}, {Traylor}, {Trinastic}, {Tringali}, {Trozzo}, {Tsang}, {Tse}, {Tso}, {Tsukada}, {Tsuna}, {Tuyenbayev}, {Ueno}, {Ugolini}, {Unnikrishnan}, {Urban}, {Usman}, {Vahlbruch}, {Vajente}, {Valdes}, {Vallisneri}, {van Bakel}, {van Beuzekom}, {van den Brand}, {Van Den Broeck}, {Vander-Hyde}, {van der Schaaf}, {van Heijningen}, {van Veggel}, {Vardaro}, {Varma}, {Vass},
  {Vas{\'u}th}, {Vecchio}, {Vedovato}, {Veitch}, {Veitch}, {Venkateswara}, {Venugopalan}, {Verkindt}, {Vetrano}, {Vicer{\'e}}, {Viets}, {Vinciguerra}, {Vine}, {Vinet}, {Vitale}, {Vo}, {Vocca}, {Vorvick}, {Vyatchanin}, {Wade}, {Wade}, {Wade}, {Walet}, {Walker}, {Wallace}, {Walsh}, {Wang}, {Wang}, {Wang}, {Wang}, {Wang}, {Ward}, {Warner}, {Was}, {Watchi}, {Weaver}, {Wei}, {Weinert}, {Weinstein}, {Weiss}, {Wen}, {Wessel}, {We{\ss}els}, {Westerweck}, {Westphal}, {Wette}, {Whelan}, {Whitcomb}, {Whiting}, {Whittle}, {Wilken}, {Williams}, {Williams}, {Williamson}, {Willis}, {Willke}, {Wimmer}, {Winkler}, {Wipf}, {Wittel}, {Woan}, {Woehler}, {Wofford}, {Wong}, {Worden}, {Wright}, {Wu}, {Wysocki}, {Xiao}, {Yamamoto}, {Yancey}, {Yang}, {Yap}, {Yazback}, {Yu}, {Yu}, {Yvert}, {Zadro{\.Z}ny}, {Zanolin}, {Zelenova}, {Zendri}, {Zevin}, {Zhang}, {Zhang}, {Zhang}, {Zhang}, {Zhao}, {Zhou}, {Zhou}, {Zhu}, {Zhu}, {Zimmerman}, {Zucker}, {Zweizig}, {LIGO Scientific Collaboration}, \& {Virgo Collaboration}}]{Abbott2017}
---. 2017{\natexlab{a}}, \prl, 119, 161101, \dodoi{10.1103/PhysRevLett.119.161101}

\bibitem[{{Abbott} {et~al.}(2017{\natexlab{b}}){Abbott}, {Abbott}, {Abbott}, {Acernese}, {Ackley}, {Adams}, {Adams}, {Addesso}, {Adhikari}, {Adya}, \& et~al.}]{Abbott17a}
---. 2017{\natexlab{b}}, \apjl, 848, L12, \dodoi{10.3847/2041-8213/aa91c9}

\bibitem[{{Abbott} {et~al.}(2017{\natexlab{c}}){Abbott}, {Abbott}, {Abbott}, {Acernese}, {Ackley}, {Adams}, {Adams}, {Addesso}, {Adhikari}, {Adya}, \& et~al.}]{Abbott17b}
---. 2017{\natexlab{c}}, \nat, 551, 85, \dodoi{10.1038/nature24471}

\bibitem[{Abbott {et~al.}(2020)Abbott, Abbott, Abraham, Acernese, Ackley, Adams, Adhikari, Adya, Affeldt, Agathos, Agatsuma, Aggarwal, Aguiar, Aich, Aiello, Ain, Ajith, Akcay, Allen, Allocca, Altin, Amato, Anand, Ananyeva, Anderson, Anderson, Angelova, Ansoldi, Antier, Appert, Arai, Araya, Areeda, Arène, Arnaud, Aronson, Arun, Asali, Ascenzi, Ashton, Aston, Astone, Aubin, Aufmuth, AultONeal, Austin, Avendano, Babak, Bacon, Badaracco, Bader, Bae, Baer, Baird, Baldaccini, Ballardin, Ballmer, Bals, Balsamo, Baltus, Banagiri, Bankar, Bankar, Barayoga, Barbieri, Barish, Barker, Barkett, Barneo, Barone, Barr, Barsotti, Barsuglia, Barta, Bartlett, Bartos, Bassiri, Basti, Bawaj, Bayley, Bazzan, Bécsy, Bejger, Belahcene, Bell, Beniwal, Benjamin, Benkel, Bentley, Bergamin, Berger, Bergmann, Bernuzzi, Berry, Bersanetti, Bertolini, Betzwieser, Bhandare, Bhandari, Bidler, Biggs, Bilenko, Billingsley, Birney, Birnholtz, Biscans, Bischi, Biscoveanu, Bisht, Bissenbayeva, Bitossi, Bizouard, Blackburn, Blackman, Blair,
  Blair, Blair, Bobba, Bode, Boer, Boetzel, Bogaert, Bondu, Bonilla, Bonnand, Booker, Boom, Bork, Boschi, Bose, Bossilkov, Bosveld, Bouffanais, Bozzi, Bradaschia, Brady, Bramley, Branchesi, Brau, Breschi, Briant, Briggs, Brighenti, Brillet, Brinkmann, Brito, Brockill, Brooks, Brooks, Brown, Brunett, Bruno, Bruntz, Buikema, Bulik, Bulten, Buonanno, Buskulic, Byer, Cabero, Cadonati, Cagnoli, Cahillane, Bustillo, Callaghan, Callister, Calloni, Camp, Canepa, Cannon, Cao, Cao, Carapella, Carbognani, Caride, Carney, Carullo, Diaz, Casentini, Castañeda, Caudill, Cavaglià, Cavalier, Cavalieri, Cella, Cerdá-Durán, Cesarini, Chaibi, Chakravarti, Chan, Chan, Chao, Charlton, Chase, Chassande-Mottin, Chatterjee, Chaturvedi, Chatziioannou, Chen, Chen, Chen, Cheng, Cheong, Chia, Chiadini, Chierici, Chincarini, Chiummo, Cho, Cho, Cho, Christensen, Chu, Chua, Chung, Chung, Ciani, Ciecielag, Cieślar, Ciobanu, Ciolfi, Cipriano, Cirone, Clara, Clark, Clearwater, Clesse, Cleva, Coccia, Cohadon, Cohen, Colleoni, Collette,
  Collins, Colpi, Constancio, Conti, Cooper, Corban, Corbitt, Cordero-Carrión, Corezzi, Corley, Cornish, Corre, Corsi, Cortese, Costa, Cotesta, Coughlin, Coughlin, Coulon, Countryman, Couvares, Covas, Coward, Cowart, Coyne, Coyne, Creighton, Creighton, Cripe, Croquette, Crowder, Cudell, Cullen, Cumming, Cummings, Cunningham, Cuoco, Curylo, Canton, Dálya, Dana, Daneshgaran-Bajastani, D’Angelo, Danilishin, D’Antonio, Danzmann, Darsow-Fromm, Dasgupta, Datrier, Dattilo, Dave, Davier, Davies, Davis, Daw, DeBra, Deenadayalan, Degallaix, Laurentis, Deléglise, Delfavero, Lillo, Pozzo, DeMarchi, D’Emilio, Demos, Dent, Pietri, Rosa, Rossi, DeSalvo, de~Varona, Dhurandhar, Díaz, Diaz-Ortiz, Dietrich, Fiore, Fronzo, Giorgio, Giovanni, Giovanni, Girolamo, Lieto, Ding, Pace, Palma, Renzo, Divakarla, Dmitriev, Doctor, Donovan, Dooley, Doravari, Dorrington, Downes, Drago, Driggers, Du, Ducoin, Dupej, Durante, D’Urso, Dwyer, Easter, Eddolls, Edelman, Edo, Edy, Effler, Ehrens, Eichholz, Eikenberry, Eisenmann,
  Eisenstein, Ejlli, Errico, Essick, Estelles, Estevez, Etienne, Etzel, Evans, Evans, Ewing, Fafone, Fairhurst, Fan, Farinon, Farr, Farr, Fauchon-Jones, Favata, Fays, Fazio, Feicht, Fejer, Feng, Fenyvesi, Ferguson, Fernandez-Galiana, Ferrante, Ferreira, Ferreira, Fidecaro, Fiori, Fiorucci, Fishbach, Fisher, Fittipaldi, Fitz-Axen, Fiumara, Flaminio, Floden, Flynn, Fong, Font, Forsyth, Fournier, Frasca, Frasconi, Frei, Freise, Frey, Frey, Fritschel, Frolov, Fronzè, Fulda, Fyffe, Gabbard, Gadre, Gaebel, Gair, Galaudage, Ganapathy, Ganguly, Gaonkar, García-Quirós, Garufi, Gateley, Gaudio, Gayathri, Gemme, Genin, Gennai, George, George, Gergely, Ghonge, Ghosh, Ghosh, Ghosh, Giacomazzo, Giaime, Giardina, Gibson, Gier, Gill, Glanzer, Gniesmer, Godwin, Goetz, Goetz, Gohlke, Goncharov, González, Gopakumar, Gossan, Gosselin, Gouaty, Grace, Grado, Granata, Grant, Gras, Grassia, Gray, Gray, Greco, Green, Green, Gretarsson, Griggs, Grignani, Grimaldi, Grimm, Grote, Grunewald, Gruning, Guidi, Guimaraes, Guixé, Gulati,
  Guo, Gupta, Gupta, Gupta, Gustafson, Gustafson, Haegel, Halim, Hall, Hamilton, Hammond, Haney, Hanke, Hanks, Hanna, Hannam, Hannuksela, Hansen, Hanson, Harder, Hardwick, Haris, Harms, Harry, Harry, Hasskew, Haster, Haughian, Hayes, Healy, Heidmann, Heintze, Heinze, Heitmann, Hellman, Hello, Hemming, Hendry, Heng, Hennes, Hennig, Heurs, Hild, Hinderer, Hoback, Hochheim, Hofgard, Hofman, Holgado, Holland, Holt, Holz, Hopkins, Horst, Hough, Howell, Hoy, Huang, Hübner, Huerta, Huet, Hughey, Hui, Husa, Huttner, Huxford, Huynh-Dinh, Idzkowski, Iess, Inchauspe, Ingram, Intini, Isac, Isi, Iyer, Jacqmin, Jadhav, Jadhav, James, Jani, Janthalur, Jaranowski, Jariwala, Jaume, Jenkins, Jiang, Johns, Johnson-McDaniel, Jones, Jones, Jones, Jones, Jones, Jonker, Ju, Junker, Kalaghatgi, Kalogera, Kamai, Kandhasamy, Kang, Kanner, Kapadia, Karki, Kashyap, Kasprzack, Kastaun, Katsanevas, Katsavounidis, Katzman, Kaufer, Kawabe, Kéfélian, Keitel, Keivani, Kennedy, Key, Khadka, Khalili, Khan, Khan, Khan, Khazanov, Khetan,
  Khursheed, Kijbunchoo, Kim, Kim, Kim, Kim, Kim, Kim, Kim, Kimball, King, Kinley-Hanlon, Kirchhoff, Kissel, Kleybolte, Klimenko, Knowles, Knyazev, Koch, Koehlenbeck, Koekoek, Koley, Kondrashov, Kontos, Koper, Korobko, Korth, Kovalam, Kozak, Kringel, Krishnendu, Królak, Krupinski, Kuehn, Kumar, Kumar, Kumar, Kumar, Kumar, Kuo, Kutynia, Lackey, Laghi, Lalande, Lam, Lamberts, Landry, Landry, Lane, Lang, Lange, Lantz, Lanza, Rosa, Lartaux-Vollard, Lasky, Laxen, Lazzarini, Lazzaro, Leaci, Leavey, Lecoeuche, Lee, Lee, Lee, Lee, Lee, Lehmann, Leroy, Letendre, Levin, Li, Li, li, Li, Li, Linde, Linker, Linley, Littenberg, Liu, Liu, Llorens-Monteagudo, Lo, Lockwood, London, Longo, Lorenzini, Loriette, Lormand, Losurdo, Lough, Lousto, Lovelace, Lück, Lumaca, Lundgren, Ma, Macas, Macfoy, MacInnis, Macleod, MacMillan, Macquet, Hernandez, Magaña-Sandoval, Magee, Majorana, Maksimovic, Malik, Man, Mandic, Mangano, Mansell, Manske, Mantovani, Mapelli, Marchesoni, Marion, Márka, Márka, Markakis, Markosyan, Markowitz,
  Maros, Marquina, Marsat, Martelli, Martin, Martin, Martinez, Martynov, Masalehdan, Mason, Massera, Masserot, Massinger, Masso-Reid, Mastrogiovanni, Matas, Matichard, Mavalvala, Maynard, McCann, McCarthy, McClelland, McCormick, McCuller, McGuire, McIsaac, McIver, McManus, McRae, McWilliams, Meacher, Meadors, Mehmet, Mehta, Villa, Melatos, Mendell, Mercer, Mereni, Merfeld, Merilh, Merritt, Merzougui, Meshkov, Messenger, Messick, Metzdorff, Meyers, Meylahn, Mhaske, Miani, Miao, Michaloliakos, Michel, Middleton, Milano, Miller, Millhouse, Mills, Milotti, Milovich-Goff, Minazzoli, Minenkov, Mishkin, Mishra, Mistry, Mitra, Mitrofanov, Mitselmakher, Mittleman, Mo, Mogushi, Mohapatra, Mohite, Molina-Ruiz, Mondin, Montani, Moore, Moraru, Morawski, Moreno, Morisaki, Mours, Mow-Lowry, Mozzon, Muciaccia, Mukherjee, Mukherjee, Mukherjee, Mukherjee, Mukund, Mullavey, Munch, Muñiz, Murray, Nagar, Nardecchia, Naticchioni, Nayak, Neil, Neilson, Nelemans, Nelson, Nery, Neunzert, Ng, Ng, Nguyen, Nguyen, Nichols, Nichols,
  Nissanke, Nocera, Noh, North, Nothard, Nuttall, Oberling, O’Brien, Oganesyan, Ogin, Oh, Oh, Ohme, Ohta, Okada, Oliver, Olivetto, Oppermann, Oram, O’Reilly, Ormiston, Ortega, O’Shaughnessy, Ossokine, Osthelder, Ottaway, Overmier, Owen, Pace, Pagano, Page, Pagliaroli, Pai, Pai, Palamos, Palashov, Palomba, Pan, Panda, Pang, Pankow, Pannarale, Pant, Paoletti, Paoli, Parida, Parker, Pascucci, Pasqualetti, Passaquieti, Passuello, Patricelli, Payne, Pearlstone, Pechsiri, Pedersen, Pedraza, Pele, Penn, Perego, Perez, Périgois, Perreca, Perriès, Petermann, Pfeiffer, Phelps, Phukon, Piccinni, Pichot, Piendibene, Piergiovanni, Pierro, Pillant, Pinard, Pinto, Piotrzkowski, Pirello, Pitkin, Plastino, Poggiani, Pong, Ponrathnam, Popolizio, Porter, Powell, Prajapati, Prasai, Prasanna, Pratten, Prestegard, Principe, Prodi, Prokhorov, Punturo, Puppo, Pürrer, Qi, Quetschke, Quinonez, Raab, Raaijmakers, Radkins, Radulesco, Raffai, Rafferty, Raja, Rajan, Rajbhandari, Rakhmanov, Ramirez, Ramos-Buades, Rana, Rao,
  Rapagnani, Raymond, Razzano, Read, Regimbau, Rei, Reid, Reitze, Rettegno, Ricci, Richardson, Richardson, Ricker, Riemenschneider, Riles, Rizzo, Robertson, Robinet, Rocchi, Rodriguez-Soto, Rolland, Rollins, Roma, Romanelli, Romano, Romel, Romero-Shaw, Romie, Rose, Rose, Rose, Rosińska, Rosofsky, Ross, Rowan, Rowlinson, Roy, Roy, Roy, Ruggi, Rutins, Ryan, Sachdev, Sadecki, Sakellariadou, Salafia, Salconi, Saleem, Salemi, Samajdar, Sanchez, Sanchez, Sanchis-Gual, Sanders, Santiago, Santos, Sarin, Sassolas, Sathyaprakash, Sauter, Savage, Savant, Sawant, Sayah, Schaetzl, Schale, Scheel, Scheuer, Schmidt, Schnabel, Schofield, Schönbeck, Schreiber, Schulte, Schutz, Schwarm, Schwartz, Scott, Scott, Seidel, Sellers, Sengupta, Sennett, Sentenac, Sequino, Sergeev, Setyawati, Shaddock, Shaffer, Shahriar, Sharma, Sharma, Shawhan, Shen, Shikauchi, Shink, Shoemaker, Shoemaker, Shukla, ShyamSundar, Siellez, Sieniawska, Sigg, Singer, Singh, Singh, Singha, Singhal, Sintes, Sipala, Skliris, Slagmolen, Slaven-Blair, Smetana,
  Smith, Smith, Somala, Son, Soni, Sorazu, Sordini, Sorrentino, Souradeep, Sowell, Spencer, Spera, Srivastava, Srivastava, Staats, Stachie, Standke, Steer, Steinhoff, Steinke, Steinlechner, Steinlechner, Steinmeyer, Stevenson, Stocks, Stops, Stover, Strain, Stratta, Strunk, Sturani, Stuver, Sudhagar, Sudhir, Summerscales, Sun, Sunil, Sur, Suresh, Sutton, Swinkels, Szczepańczyk, Tacca, Tait, Talbot, Tanasijczuk, Tanner, Tao, Tápai, Tapia, Martin, Tasson, Taylor, Tenorio, Terkowski, Thirugnanasambandam, Thomas, Thomas, Thompson, Thondapu, Thorne, Thrane, Tinsman, Saravanan, Tiwari, Tiwari, Tiwari, Toland, Tonelli, Tornasi, Torres-Forné, Torrie, e~Melo, Töyrä, Trail, Travasso, Traylor, Tringali, Tripathee, Trovato, Trudeau, Tsang, Tse, Tso, Tsukada, Tsuna, Tsutsui, Turconi, Ubhi, Ueno, Ugolini, Unnikrishnan, Urban, Usman, Utina, Vahlbruch, Vajente, Valdes, Valentini, van Bakel, van Beuzekom, van~den Brand, Broeck, Vander-Hyde, van~der Schaaf, Heijningen, van Veggel, Vardaro, Varma, Vass, Vasúth, Vecchio,
  Vedovato, Veitch, Veitch, Venkateswara, Venugopalan, Verkindt, Veske, Vetrano, Viceré, Viets, Vinciguerra, Vine, Vinet, Vitale, Vivanco, Vo, Vocca, Vorvick, Vyatchanin, Wade, Wade, Wade, Walet, Walker, Wallace, Wallace, Walsh, Wang, Wang, Wang, Ward, Warden, Warner, Was, Watchi, Weaver, Wei, Weinert, Weinstein, Weiss, Wellmann, Wen, Weßels, Westhouse, Wette, Whelan, Whiting, Whittle, Wilken, Williams, Willis, Willke, Winkler, Wipf, Wittel, Woan, Woehler, Wofford, Wong, Wright, Wu, Wysocki, Xiao, Yamamoto, Yang, Yang, Yang, Yap, Yazback, Yeeles, Yu, Yu, Yuen, Zadrożny, Zadrożny, Zanolin, Zelenova, Zendri, Zevin, Zhang, Zhang, Zhang, Zhao, Zhao, Zhou, Zhou, Zhu, Zimmerman, Zucker, Zweizig, Collaboration, \& Collaboration}]{Abbott_2020}
Abbott, R., Abbott, T.~D., Abraham, S., {et~al.} 2020, The Astrophysical Journal Letters, 896, L44

\bibitem[{{Abbott} {et~al.}(2021){Abbott}, {Abbott}, {Abraham}, {Acernese}, {Ackley}, {Adams}, {Adams}, {Adhikari}, {Adya}, {Affeldt}, {Agarwal}, {Agathos}, {Agatsuma}, {Aggarwal}, {Aguiar}, {Aiello}, {Ain}, {Ajith}, {Akutsu}, {Aleman}, {Allen}, {Allocca}, {Altin}, {Amato}, {Anand}, {Ananyeva}, {Anderson}, {Anderson}, {Ando}, {Angelova}, {Ansoldi}, {Antelis}, {Antier}, {Appert}, {Arai}, {Arai}, {Arai}, {Araki}, {Araya}, {Araya}, {Areeda}, {Ar{\`e}ne}, {Aritomi}, {Arnaud}, {Aronson}, {Arun}, {Asada}, {Asali}, {Ashton}, {Aso}, {Aston}, {Astone}, {Aubin}, {Aufmuth}, {Aultoneal}, {Austin}, {Babak}, {Badaracco}, {Bader}, {Bae}, {Bae}, {Baer}, {Bagnasco}, {Bai}, {Baiotti}, {Baird}, {Bajpai}, {Ball}, {Ballardin}, {Ballmer}, {Bals}, {Balsamo}, {Baltus}, {Banagiri}, {Bankar}, {Bankar}, {Barayoga}, {Barbieri}, {Barish}, {Barker}, {Barneo}, {Barone}, {Barr}, {Barsotti}, {Barsuglia}, {Barta}, {Bartlett}, {Barton}, {Bartos}, {Bassiri}, {Basti}, {Bawaj}, {Bayley}, {Baylor}, {Bazzan}, {B{\'e}csy}, {Bedakihale}, {Bejger},
  {Belahcene}, {Benedetto}, {Beniwal}, {Benjamin}, {Benkel}, {Bennett}, {Bentley}, {Benyaala}, {Bergamin}, {Berger}, {Bernuzzi}, {Berry}, {Bersanetti}, {Bertolini}, {Betzwieser}, {Bhandare}, {Bhandari}, {Bhattacharjee}, {Bhaumik}, {Bidler}, {Bilenko}, {Billingsley}, {Birney}, {Birnholtz}, {Biscans}, {Bischi}, {Biscoveanu}, {Bisht}, {Biswas}, {Bitossi}, {Bizouard}, {Blackburn}, {Blackman}, {Blair}, {Blair}, {Blair}, {Bobba}, {Bode}, {Boer}, {Bogaert}, {Boldrini}, {Bondu}, {Bonilla}, {Bonnand}, {Booker}, {Boom}, {Bork}, {Boschi}, {Bose}, {Bose}, {Bossilkov}, {Boudart}, {Bouffanais}, {Bozzi}, {Bradaschia}, {Brady}, {Bramley}, {Branch}, {Branchesi}, {Brau}, {Breschi}, {Briant}, {Briggs}, {Brillet}, {Brinkmann}, {Brockill}, {Brooks}, {Brooks}, {Brown}, {Brunett}, {Bruno}, {Bruntz}, {Bryant}, {Buikema}, {Bulik}, {Bulten}, {Buonanno}, {Buscicchio}, {Buskulic}, {Byer}, {Cadonati}, {Caesar}, {Cagnoli}, {Cahillane}, {Cain}, {Calder{\'o}n Bustillo}, {Callaghan}, {Callister}, {Calloni}, {Camp}, {Canepa},
  {Cannavacciuolo}, {Cannon}, {Cao}, {Cao}, {Cao}, {Capocasa}, {Capote}, {Carapella}, {Carbognani}, {Carlin}, {Carney}, {Carpinelli}, {Carullo}, {Carver}, {Casanueva Diaz}, {Casentini}, {Castaldi}, {Caudill}, {Cavagli{\`a}}, {Cavalier}, {Cavalieri}, {Cella}, {Cerd{\'a}-Dur{\'a}n}, {Cesarini}, {Chaibi}, {Chakravarti}, {Champion}, {Chan}, {Chan}, {Chan}, {Chan}, {Chandra}, {Chanial}, {Chao}, {Charlton}, {Chase}, {Chassande-Mottin}, {Chatterjee}, {Chaturvedi}, {Chatziioannou}, {Chen}, {Chen}, {Chen}, {Chen}, {Chen}, {Chen}, {Chen}, {Chen}, {Chen}, {Cheng}, {Cheong}, {Cheung}, {Chia}, {Chiadini}, {Chiang}, {Chierici}, {Chincarini}, {Chiofalo}, {Chiummo}, {Cho}, {Cho}, {Choate}, {Choudhary}, {Choudhary}, {Christensen}, {Chu}, {Chu}, {Chu}, {Chua}, {Chung}, {Ciani}, {Ciecielag}, {Cie{\'s}lar}, {Cifaldi}, {Ciobanu}, {Ciolfi}, {Cipriano}, {Cirone}, {Clara}, {Clark}, {Clark}, {Clarke}, {Clearwater}, {Clesse}, {Cleva}, {Coccia}, {Cohadon}, {Cohen}, {Cohen}, {Colleoni}, {Collette}, {Colpi}, {Compton}, {Constancio},
  {Conti}, {Cooper}, {Corban}, {Corbitt}, {Cordero-Carri{\'o}n}, {Corezzi}, {Corley}, {Cornish}, {Corre}, {Corsi}, {Cortese}, {Costa}, {Cotesta}, {Coughlin}, {Coughlin}, {Coulon}, {Countryman}, {Cousins}, {Couvares}, {Covas}, {Coward}, {Cowart}, {Coyne}, {Coyne}, {Creighton}, {Creighton}, {Criswell}, {Croquette}, {Crowder}, {Cudell}, {Cullen}, {Cumming}, {Cummings}, {Cuoco}, {Cury{\l}o}, {Dal Canton}, {D{\'a}lya}, {Dana}, {Daneshgaranbajastani}, {D'Angelo}, {Danilishin}, {D'Antonio}, {Danzmann}, {Darsow-Fromm}, {Dasgupta}, {Datrier}, {Dattilo}, {Dave}, {Davier}, {Davies}, {Davis}, {Daw}, {Dean}, {Debra}, {Deenadayalan}, {Degallaix}, {de Laurentis}, {Del{\'e}glise}, {Del Favero}, {de Lillo}, {de Lillo}, {Del Pozzo}, {Demarchi}, {de Matteis}, {D'Emilio}, {Demos}, {Dent}, {Depasse}, {de Pietri}, {De Rosa}, {de Rossi}, {Desalvo}, {de Simone}, {Dhurandhar}, {D{\'\i}az}, {Diaz-Ortiz}, {Didio}, {Dietrich}, {di Fiore}, {di Fronzo}, {di Giorgio}, {di Giovanni}, {di Girolamo}, {di Lieto}, {Ding}, {di Pace}, {di Palma},
  {di Renzo}, {Divakarla}, {Dmitriev}, {Doctor}, {D'Onofrio}, {Donovan}, {Dooley}, {Doravari}, {Dorrington}, {Drago}, {Driggers}, {Drori}, {Du}, {Ducoin}, {Dupej}, {Durante}, {D'Urso}, {Duverne}, {Dwyer}, {Easter}, {Ebersold}, {Eddolls}, {Edelman}, {Edo}, {Edy}, {Effler}, {Eguchi}, {Eichholz}, {Eikenberry}, {Eisenmann}, {Eisenstein}, {Ejlli}, {Enomoto}, {Errico}, {Essick}, {Estell{\'e}s}, {Estevez}, {Etienne}, {Etzel}, {Evans}, {Evans}, {Ewing}, {Fafone}, {Fair}, {Fairhurst}, {Fan}, {Farah}, {Farinon}, {Farr}, {Farr}, {Farrow}, {Fauchon-Jones}, {Favata}, {Fays}, {Fazio}, {Feicht}, {Fejer}, {Feng}, {Fenyvesi}, {Ferguson}, {Fernandez-Galiana}, {Ferrante}, {Ferreira}, {Fidecaro}, {Figura}, {Fiori}, {Fishbach}, {Fisher}, {Fittipaldi}, {Fiumara}, {Flaminio}, {Floden}, {Flynn}, {Fong}, {Font}, {Fornal}, {Forsyth}, {Franke}, {Frasca}, {Frasconi}, {Frederick}, {Frei}, {Freise}, {Frey}, {Fritschel}, {Frolov}, {Fronz{\'e}}, {Fujii}, {Fujikawa}, {Fukunaga}, {Fukushima}, {Fulda}, {Fyffe}, {Gabbard}, {Gadre}, {Gaebel},
  {Gair}, {Gais}, {Galaudage}, {Gamba}, {Ganapathy}, {Ganguly}, {Gao}, {Gaonkar}, {Garaventa}, {Garc{\'\i}a-N{\'u}{\~n}ez}, {Garc{\'\i}a-Quir{\'o}s}, {Garufi}, {Gateley}, {Gaudio}, {Gayathri}, {Ge}, {Gemme}, {Gennai}, {George}, {Gergely}, {Gewecke}, {Ghonge}, {Ghosh}, {Ghosh}, {Ghosh}, {Ghosh}, {Ghosh}, {Giacomazzo}, {Giacoppo}, {Giaime}, {Giardina}, {Gibson}, {Gier}, {Giesler}, {Giri}, {Gissi}, {Glanzer}, {Gleckl}, {Godwin}, {Goetz}, {Goetz}, {Gohlke}, {Goncharov}, {Gonz{\'a}lez}, {Gopakumar}, {Gosselin}, {Gouaty}, {Grace}, {Grado}, {Granata}, {Granata}, {Grant}, {Gras}, {Grassia}, {Gray}, {Gray}, {Greco}, {Green}, {Green}, {Gretarsson}, {Gretarsson}, {Griffith}, {Griffiths}, {Griggs}, {Grignani}, {Grimaldi}, {Grimes}, {Grimm}, {Grote}, {Grunewald}, {Gruning}, {Guerrero}, {Guidi}, {Guimaraes}, {Guix{\'e}}, {Gulati}, {Guo}, {Guo}, {Gupta}, {Gupta}, {Gupta}, {Gustafson}, {Gustafson}, {Guzman}, {Ha}, {Haegel}, {Hagiwara}, {Haino}, {Halim}, {Hall}, {Hamilton}, {Hammond}, {Han}, {Haney}, {Hanks}, {Hanna},
  {Hannam}, {Hannuksela}, {Hansen}, {Hansen}, {Hanson}, {Harder}, {Hardwick}, {Haris}, {Harms}, {Harry}, {Harry}, {Hartwig}, {Hasegawa}, {Haskell}, {Hasskew}, {Haster}, {Hattori}, {Haughian}, {Hayakawa}, {Hayama}, {Hayes}, {Healy}, {Heidmann}, {Heintze}, {Heinze}, {Heinzel}, {Heitmann}, {Hellman}, {Hello}, {Helmling-Cornell}, {Hemming}, {Hendry}, {Heng}, {Hennes}, {Hennig}, {Hennig}, {Hernandez Vivanco}, {Heurs}, {Hild}, {Hill}, {Himemoto}, {Hinderer}, {Hines}, {Hiranuma}, {Hirata}, {Hirose}, {Ho}, {Hochheim}, {Hofman}, {Hohmann}, {Holgado}, {Holland}, {Hollows}, {Holmes}, {Holt}, {Holz}, {Hong}, {Hopkins}, {Hough}, {Howell}, {Hoy}, {Hoyland}, {Hreibi}, {Hsieh}, {Hsu}, {Huang}, {Huang}, {Huang}, {Huang}, {Huang}, {Huang}, {H{\"u}bner}, {Huddart}, {Huerta}, {Hughey}, {Hui}, {Hui}, {Husa}, {Huttner}, {Huxford}, {Huynh-Dinh}, {Ide}, {Idzkowski}, {Iess}, {Ikenoue}, {Imam}, {Inayoshi}, {Inchauspe}, {Ingram}, {Inoue}, {Intini}, {Ioka}, {Isi}, {Isleif}, {Ito}, {Itoh}, {Iyer}, {Izumi}, {Jaberianhamedan}, {Jacqmin},
  {Jadhav}, {Jadhav}, {James}, {Jan}, {Jani}, {Janssens}, {Janthalur}, {Jaranowski}, {Jariwala}, {Jaume}, {Jenkins}, {Jeon}, {Jeunon}, {Jia}, {Jiang}, {Jin}, {Johns}, {Jones}, {Jones}, {Jones}, {Jones}, {Jones}, {Jonker}, {Ju}, {Jung}, {Jung}, {Junker}, {Kaihotsu}, {Kajita}, {Kakizaki}, {Kalaghatgi}, {Kalogera}, {Kamai}, {Kamiizumi}, {Kanda}, {Kandhasamy}, {Kang}, {Kanner}, {Kao}, {Kapadia}, {Kapasi}, {Karat}, {Karathanasis}, {Karki}, {Kashyap}, {Kasprzack}, {Kastaun}, {Katsanevas}, {Katsavounidis}, {Katzman}, {Kaur}, {Kawabe}, {Kawaguchi}, {Kawai}, {Kawasaki}, {K{\'e}f{\'e}lian}, {Keitel}, {Key}, {Khadka}, {Khalili}, {Khan}, {Khan}, {Khazanov}, {Khetan}, {Khursheed}, {Kijbunchoo}, {Kim}, {Kim}, {Kim}, {Kim}, {Kim}, {Kim}, {Kimball}, {Kimura}, {King}, {Kinley-Hanlon}, {Kirchhoff}, {Kissel}, {Kita}, {Kitazawa}, {Kleybolte}, {Klimenko}, {Knee}, {Knowles}, {Knyazev}, {Koch}, {Koekoek}, {Kojima}, {Kokeyama}, {Koley}, {Kolitsidou}, {Kolstein}, {Komori}, {Kondrashov}, {Kong}, {Kontos}, {Koper}, {Korobko}, {Kotake},
  {Kovalam}, {Kozak}, {Kozakai}, {Kozu}, {Kringel}, {Krishnendu}, {Kr{\'o}lak}, {Kuehn}, {Kuei}, {Kumar}, {Kumar}, {Kumar}, {Kumar}, {Kume}, {Kuns}, {Kuo}, {Kuo}, {Kuromiya}, {Kuroyanagi}, {Kusayanagi}, {Kwak}, {Kwang}, {Laghi}, {Lalande}, {Lam}, {Lamberts}, {Landry}, {Landry}, {Lane}, {Lang}, {Lange}, {Lantz}, {La Rosa}, {Lartaux-Vollard}, {Lasky}, {Laxen}, {Lazzarini}, {Lazzaro}, {Leaci}, {Leavey}, {Lecoeuche}, {Lee}, {Lee}, {Lee}, {Lee}, {Lee}, {Lee}, {Lehmann}, {Lema{\^\i}tre}, {Leon}, {Leonardi}, {Leroy}, {Letendre}, {Levin}, {Leviton}, {Li}, {Li}, {Li}, {Li}, {Li}, {Li}, {Lin}, {Lin}, {Lin}, {Lin}, {Lin}, {Linde}, {Linker}, {Linley}, {Littenberg}, {Liu}, {Liu}, {Liu}, {Liu}, {Llorens-Monteagudo}, {Lo}, {Lockwood}, {Lollie}, {London}, {Longo}, {Lopez}, {Lorenzini}, {Loriette}, {Lormand}, {Losurdo}, {Lough}, {Lousto}, {Lovelace}, {L{\"u}ck}, {Lumaca}, {Lundgren}, {Luo}, {Macas}, {Macinnis}, {MacLeod}, {MacMillan}, {Macquet}, {Maga{\~n}a Hernandez}, {Maga{\~n}a-Sandoval}, {Magazz{\`u}}, {Magee},
  {Maggiore}, {Majorana}, {Makarem}, {Maksimovic}, {Maliakal}, {Malik}, {Man}, {Mandic}, {Mangano}, {Mango}, {Mansell}, {Manske}, {Mantovani}, {Mapelli}, {Marchesoni}, {Marchio}, {Marion}, {Mark}, {M{\'a}rka}, {M{\'a}rka}, {Markakis}, {Markosyan}, {Markowitz}, {Maros}, {Marquina}, {Marsat}, {Martelli}, {Martin}, {Martin}, {Martinez}, {Martinez}, {Martinovic}, {Martynov}, {Marx}, {Masalehdan}, {Mason}, {Massera}, {Masserot}, {Massinger}, {Masso-Reid}, {Mastrogiovanni}, {Matas}, {Mateu-Lucena}, {Matichard}, {Matiushechkina}, {Mavalvala}, {McCann}, {McCarthy}, {McClelland}, {McClincy}, {McCormick}, {McCuller}, {McGhee}, {McGuire}, {McIsaac}, {McIver}, {McManus}, {McRae}, {McWilliams}, {Meacher}, {Mehmet}, {Mehta}, {Melatos}, {Melchor}, {Mendell}, {Menendez-Vazquez}, {Menoni}, {Mercer}, {Mereni}, {Merfeld}, {Merilh}, {Merritt}, {Merzougui}, {Meshkov}, {Messenger}, {Messick}, {Meyers}, {Meylahn}, {Mhaske}, {Miani}, {Miao}, {Michaloliakos}, {Michel}, {Michimura}, {Middleton}, {Milano}, {Miller}, {Millhouse},
  {Mills}, {Milotti}, {Milovich-Goff}, {Minazzoli}, {Minenkov}, {Mio}, {Mir}, {Mishkin}, {Mishra}, {Mishra}, {Mistry}, {Mitra}, {Mitrofanov}, {Mitselmakher}, {Mittleman}, {Miyakawa}, {Miyamoto}, {Miyazaki}, {Miyo}, {Miyoki}, {Mo}, {Mogushi}, {Mohapatra}, {Mohite}, {Molina}, {Molina-Ruiz}, {Mondin}, {Montani}, {Moore}, {Moraru}, {Morawski}, {More}, {Moreno}, {Moreno}, {Mori}, {Morisaki}, {Moriwaki}, {Mours}, {Mow-Lowry}, {Mozzon}, {Muciaccia}, {Mukherjee}, {Mukherjee}, {Mukherjee}, {Mukherjee}, {Mukund}, {Mullavey}, {Munch}, {Mu{\~n}iz}, {Murray}, {Musenich}, {Nadji}, {Nagano}, {Nagano}, {Nagar}, {Nakamura}, {Nakano}, {Nakano}, {Nakashima}, {Nakayama}, {Nardecchia}, {Narikawa}, {Naticchioni}, {Nayak}, {Nayak}, {Negishi}, {Neil}, {Neilson}, {Nelemans}, {Nelson}, {Nery}, {Neunzert}, {Ng}, {Ng}, {Nguyen}, {Nguyen}, {Nguyen}, {Nguyen Quynh}, {Ni}, {Nichols}, {Nishizawa}, {Nissanke}, {Nocera}, {Noh}, {Norman}, {North}, {Nozaki}, {Nuttall}, {Oberling}, {O'Brien}, {Obuchi}, {O'Dell}, {Ogaki}, {Oganesyan}, {Oh}, {Oh},
  {Oh}, {Ohashi}, {Ohishi}, {Ohkawa}, {Ohme}, {Ohta}, {Okada}, {Okutani}, {Okutomi}, {Olivetto}, {Oohara}, {Ooi}, {Oram}, {O'Reilly}, {Ormiston}, {Ormsby}, {Ortega}, {O'Shaughnessy}, {O'Shea}, {Oshino}, {Ossokine}, {Osthelder}, {Otabe}, {Ottaway}, {Overmier}, {Pace}, {Pagano}, {Page}, {Pagliaroli}, {Pai}, {Pai}, {Palamos}, {Palashov}, {Palomba}, {Pan}, {Panda}, {Pang}, {Pang}, {Pankow}, {Pannarale}, {Pant}, {Paoletti}, {Paoli}, {Paolone}, {Parisi}, {Park}, {Parker}, {Pascucci}, {Pasqualetti}, {Passaquieti}, {Passuello}, {Patel}, {Patricelli}, {Payne}, {Pechsiri}, {Pedraza}, {Pegoraro}, {Pele}, {Pe{\~n}a Arellano}, {Penn}, {Perego}, {Pereira}, {Pereira}, {Perez}, {P{\'e}rigois}, {Perreca}, {Perri{\`e}s}, {Petermann}, {Petterson}, {Pfeiffer}, {Pham}, {Phukon}, {Piccinni}, {Pichot}, {Piendibene}, {Piergiovanni}, {Pierini}, {Pierro}, {Pillant}, {Pilo}, {Pinard}, {Pinto}, {Piotrzkowski}, {Piotrzkowski}, {Pirello}, {Pitkin}, {Placidi}, {Plastino}, {Pluchar}, {Poggiani}, {Polini}, {Pong}, {Ponrathnam}, {Popolizio},
  {Porter}, {Powell}, {Pracchia}, {Pradier}, {Prajapati}, {Prasai}, {Prasanna}, {Pratten}, {Prestegard}, {Principe}, {Prodi}, {Prokhorov}, {Prosposito}, {Prudenzi}, {Puecher}, {Punturo}, {Puosi}, {Puppo}, {P{\"u}rrer}, {Qi}, {Quetschke}, {Quinonez}, {Quitzow-James}, {Raab}, {Raaijmakers}, {Radkins}, {Radulesco}, {Raffai}, {Rail}, {Raja}, {Rajan}, {Ramirez}, {Ramirez}, {Ramos-Buades}, {Rana}, {Rapagnani}, {Rapol}, {Ratto}, {Ray}, {Raymond}, {Raza}, {Razzano}, {Read}, {Rees}, {Regimbau}, {Rei}, {Reid}, {Reitze}, {Relton}, {Rettegno}, {Ricci}, {Richardson}, {Richardson}, {Richardson}, {Ricker}, {Riemenschneider}, {Riles}, {Rizzo}, {Robertson}, {Robie}, {Robinet}, {Rocchi}, {Rocha}, {Rodriguez}, {Rodriguez-Soto}, {Rolland}, {Rollins}, {Roma}, {Romanelli}, {Romano}, {Romel}, {Romero}, {Romero-Shaw}, {Romie}, {Rose}, {Rosi{\'n}ska}, {Rosofsky}, {Ross}, {Rowan}, {Rowlinson}, {Roy}, {Roy}, {Rozza}, {Ruggi}, {Ryan}, {Sachdev}, {Sadecki}, {Sadiq}, {Sago}, {Saito}, {Saito}, {Sakai}, {Sakai}, {Sakellariadou}, {Sakuno},
  {Salafia}, {Salconi}, {Saleem}, {Salemi}, {Samajdar}, {Sanchez}, {Sanchez}, {Sanchez}, {Sanchis-Gual}, {Sanders}, {Sanuy}, {Saravanan}, {Sarin}, {Sassolas}, {Satari}, {Sathyaprakash}, {Sato}, {Sato}, {Sauter}, {Savage}, {Savant}, {Sawada}, {Sawant}, {Sawant}, {Sayah}, {Schaetzl}, {Scheel}, {Scheuer}, {Schindler-Tyka}, {Schmidt}, {Schnabel}, {Schneewind}, {Schofield}, {Sch{\"o}nbeck}, {Schulte}, {Schutz}, {Schwartz}, {Scott}, {Scott}, {Seglar-Arroyo}, {Seidel}, {Sekiguchi}, {Sekiguchi}, {Sellers}, {Sengupta}, {Sennett}, {Sentenac}, {Seo}, {Sequino}, {Sergeev}, {Setyawati}, {Shaffer}, {Shahriar}, {Shams}, {Shao}, {Sharifi}, {Sharma}, {Sharma}, {Shawhan}, {Shcheblanov}, {Shen}, {Shibagaki}, {Shikauchi}, {Shimizu}, {Shimoda}, {Shimode}, {Shink}, {Shinkai}, {Shishido}, {Shoda}, {Shoemaker}, {Shoemaker}, {Shukla}, {Shyamsundar}, {Sieniawska}, {Sigg}, {Singer}, {Singh}, {Singh}, {Singha}, {Sintes}, {Sipala}, {Skliris}, {Slagmolen}, {Slaven-Blair}, {Smetana}, {Smith}, {Smith}, {Somala}, {Somiya}, {Son}, {Soni},
  {Soni}, {Sorazu}, {Sordini}, {Sorrentino}, {Sorrentino}, {Sotani}, {Soulard}, {Souradeep}, {Sowell}, {Spagnuolo}, {Spencer}, {Spera}, {Srivastava}, {Srivastava}, {Staats}, {Stachie}, {Steer}, {Steinlechner}, {Steinlechner}, {Stops}, {Stevenson}, {Stover}, {Strain}, {Strang}, {Stratta}, {Strunk}, {Sturani}, {Stuver}, {S{\"u}dbeck}, {Sudhagar}, {Sudhir}, {Sugimoto}, {Suh}, {Summerscales}, {Sun}, {Sun}, {Sunil}, {Sur}, {Suresh}, {Sutton}, {Suzuki}, {Suzuki}, {Swinkels}, {Szczepa{\'n}czyk}, {Szewczyk}, {Tacca}, {Tagoshi}, {Tait}, {Takahashi}, {Takahashi}, {Takamori}, {Takano}, {Takeda}, {Takeda}, {Talbot}, {Tanaka}, {Tanaka}, {Tanaka}, {Tanaka}, {Tanaka}, {Tanasijczuk}, {Tanioka}, {Tanner}, {Tao}, {Tapia}, {Tapia San Martin}, {Tasson}, {Telada}, {Tenorio}, {Terkowski}, {Test}, {Thirugnanasambandam}, {Thomas}, {Thomas}, {Thompson}, {Thondapu}, {Thorne}, {Thrane}, {Tiwari}, {Tiwari}, {Tiwari}, {Toland}, {Tolley}, {Tomaru}, {Tomigami}, {Tomura}, {Tonelli}, {Torres-Forn{\'e}}, {Torrie}, {Tosta E Melo},
  {T{\"o}yr{\"a}}, {Trapananti}, {Travasso}, {Traylor}, {Tringali}, {Tripathee}, {Troiano}, {Trovato}, {Trozzo}, {Trudeau}, {Tsai}, {Tsai}, {Tsang}, {Tsang}, {Tsao}, {Tse}, {Tso}, {Tsubono}, {Tsuchida}, {Tsukada}, {Tsuna}, {Tsutsui}, {Tsuzuki}, {Turconi}, {Tuyenbayev}, {Ubhi}, {Uchikata}, {Uchiyama}, {Udall}, {Ueda}, {Uehara}, {Ueno}, {Ueshima}, {Ugolini}, {Unnikrishnan}, {Uraguchi}, {Urban}, {Ushiba}, {Usman}, {Utina}, {Vahlbruch}, {Vajente}, {Vajpeyi}, {Valdes}, {Valentini}, {Valsan}, {van Bakel}, {van Beuzekom}, {van den Brand}, {van den Broeck}, {Vander-Hyde}, {van der Schaaf}, {van Heijningen}, {Vanosky}, {van Putten}, {Vardaro}, {Vargas}, {Varma}, {Vas{\'u}th}, {Vecchio}, {Vedovato}, {Veitch}, {Veitch}, {Venkateswara}, {Venneberg}, {Venugopalan}, {Verkindt}, {Verma}, {Veske}, {Vetrano}, {Vicer{\'e}}, {Viets}, {Villa-Ortega}, {Vinet}, {Vitale}, {Vo}, {Vocca}, {von Reis}, {von Wrangel}, {Vorvick}, {Vyatchanin}, {Wade}, {Wade}, {Wagner}, {Walet}, {Walker}, {Wallace}, {Wallace}, {Walsh}, {Wang}, {Wang},
  {Wang}, {Ward}, {Warner}, {Was}, {Washimi}, {Washington}, {Watchi}, {Weaver}, {Wei}, {Weinert}, {Weinstein}, {Weiss}, {Weller}, {Wellmann}, {Wen}, {We{\ss}els}, {Westhouse}, {Wette}, {Whelan}, {White}, {Whiting}, {Whittle}, {Wilken}, {Williams}, {Williams}, {Williamson}, {Willis}, {Willke}, {Wilson}, {Winkler}, {Wipf}, {Wlodarczyk}, {Woan}, {Woehler}, {Wofford}, {Wong}, {Wu}, {Wu}, {Wu}, {Wu}, {Wysocki}, {Xiao}, {Xu}, {Yamada}, {Yamamoto}, {Yamamoto}, {Yamamoto}, {Yamamoto}, {Yamashita}, {Yamazaki}, {Yang}, {Yang}, {Yang}, {Yang}, {Yang}, {Yap}, {Yeeles}, {Yelikar}, {Ying}, {Yokogawa}, {Yokoyama}, {Yokozawa}, {Yoon}, {Yoshioka}, {Yu}, {Yu}, {Yuzurihara}, {Zadro{\.z}ny}, {Zanolin}, {Zappa}, {Zeidler}, {Zelenova}, {Zendri}, {Zevin}, {Zhan}, {Zhang}, {Zhang}, {Zhang}, {Zhang}, {Zhang}, {Zhao}, {Zhao}, {Zhao}, {Zhao}, {Zhou}, {Zhu}, {Zhu}, {Zimmerman}, {Zlochower}, {Zucker}, {Zweizig}, {Ligo Scientific Collaboration}, {VIRGO Collaboration}, \& {KAGRA Collaboration}}]{Abbott+21_BHNSGW200115}
{Abbott}, R., {Abbott}, T.~D., {Abraham}, S., {et~al.} 2021, \apjl, 915, L5, \dodoi{10.3847/2041-8213/ac082e}

\bibitem[{Abbott {et~al.}(2023{\natexlab{a}})Abbott, Abbott, Acernese, Ackley, Adams, Adhikari, Adhikari, Adya, Affeldt, Agarwal, Agathos, Agatsuma, Aggarwal, Aguiar, Aiello, Ain, Ajith, Akcay, Akutsu, Albanesi, Allocca, Altin, Amato, Anand, Anand, Ananyeva, Anderson, Anderson, Ando, Andrade, Andres, Andri\ifmmode~\acute{c}\else \'{c}\fi{}, Angelova, Ansoldi, Antelis, Antier, Appert, Arai, Arai, Arai, Araki, Araya, Araya, Areeda, Ar\`ene, Aritomi, Arnaud, Arogeti, Aronson, Arun, Asada, Asali, Ashton, Aso, Assiduo, Aston, Astone, Aubin, Austin, Babak, Badaracco, Bader, Badger, Bae, Bae, Baer, Bagnasco, Bai, Baiotti, Baird, Bajpai, Ball, Ballardin, Ballmer, Balsamo, Baltus, Banagiri, Bankar, Barayoga, Barbieri, Barish, Barker, Barneo, Barone, Barr, Barsotti, Barsuglia, Barta, Bartlett, Barton, Bartos, Bassiri, Basti, Bawaj, Bayley, Baylor, Bazzan, B\'ecsy, Bedakihale, Bejger, Belahcene, Benedetto, Beniwal, Bennett, Bentley, BenYaala, Bergamin, Berger, Bernuzzi, Berry, Bersanetti, Bertolini, Betzwieser,
  Beveridge, Bhandare, Bhardwaj, Bhattacharjee, Bhaumik, Bilenko, Billingsley, Bini, Birney, Birnholtz, Biscans, Bischi, Biscoveanu, Bisht, Biswas, Bitossi, Bizouard, Blackburn, Blair, Blair, Blair, Bobba, Bode, Boer, Bogaert, Boldrini, Bonavena, Bondu, Bonilla, Bonnand, Booker, Boom, Bork, Boschi, Bose, Bose, Bossilkov, Boudart, Bouffanais, Bozzi, Bradaschia, Brady, Bramley, Branch, Branchesi, Brandt, Brau, Breschi, Briant, Briggs, Brillet, Brinkmann, Brockill, Brooks, Brooks, Brown, Brunett, Bruno, Bruntz, Bryant, Bulik, Bulten, Buonanno, Buscicchio, Buskulic, Buy, Byer, Davies, Cadonati, Cagnoli, Cahillane, Bustillo, Callaghan, Callister, Calloni, Cameron, Camp, Canepa, Canevarolo, Cannavacciuolo, Cannon, Cao, Cao, Capocasa, Capote, Carapella, Carbognani, Carlin, Carney, Carpinelli, Carrillo, Carullo, Carver, Diaz, Casentini, Castaldi, Caudill, Cavagli\`a, Cavalier, Cavalieri, Ceasar, Cella, Cerd\'a-Dur\'an, Cesarini, Chaibi, Chakravarti, Subrahmanya, Champion, Chan, Chan, Chan, Chan, Chan, Chandra,
  Chanial, Chao, Chapman-Bird, Charlton, Chase, Chassande-Mottin, Chatterjee, Chatterjee, Chatterjee, Chaturvedi, Chaty, Chatziioannou, Chen, Chen, Chen, Chen, Chen, Chen, Chen, Chen, Cheng, Cheong, Cheung, Chia, Chiadini, Chiang, Chiarini, Chierici, Chincarini, Chiofalo, Chiummo, Cho, Cho, Choudhary, Choudhary, Christensen, Chu, Chu, Chu, Chua, Chung, Ciani, Ciecielag, Cie\ifmmode~\acute{s}\else \'{s}\fi{}lar, Cifaldi, Ciobanu, Ciolfi, Cipriano, Cirone, Clara, Clark, Clark, Clarke, Clearwater, Clesse, Cleva, Coccia, Codazzo, Cohadon, Cohen, Cohen, Colleoni, Collette, Colombo, Colpi, Compton, Constancio, Conti, Cooper, Corban, Corbitt, Cordero-Carri\'on, Corezzi, Corley, Cornish, Corre, Corsi, Cortese, Costa, Cotesta, Coughlin, Coulon, Countryman, Cousins, Couvares, Coward, Cowart, Coyne, Coyne, Creighton, Creighton, Criswell, Croquette, Crowder, Cudell, Cullen, Cumming, Cummings, Cunningham, Cuoco, Cury\l{}o, Dabadie, Canton, Dall'Osso, D\'alya, Dana, DaneshgaranBajastani, D'Angelo, Danila, Danilishin,
  D'Antonio, Danzmann, Darsow-Fromm, Dasgupta, Datrier, Dattilo, Dave, Davier, Davis, Davis, Daw, de~Alarc\'on, Dean, DeBra, Deenadayalan, Degallaix, De~Laurentis, Del\'eglise, Del~Favero, De~Lillo, De~Lillo, Del~Pozzo, DeMarchi, De~Matteis, D'Emilio, Demos, Dent, Depasse, De~Pietri, De~Rosa, De~Rossi, DeSalvo, De~Simone, Dhurandhar, D\'{\i}az, Diaz-Ortiz, Didio, Dietrich, Di~Fiore, Di~Fronzo, Di~Giorgio, Di~Giovanni, Di~Giovanni, Di~Girolamo, Di~Lieto, Ding, Di~Pace, Di~Palma, Di~Renzo, Divakarla, Dmitriev, Doctor, D'Onofrio, Donovan, Dooley, Doravari, Dorrington, Drago, Driggers, Drori, Ducoin, Dupej, Durante, D'Urso, Duverne, Dwyer, Eassa, Easter, Ebersold, Eckhardt, Eddolls, Edelman, Edo, Edy, Effler, Eguchi, Eichholz, Eikenberry, Eisenmann, Eisenstein, Ejlli, Engelby, Enomoto, Errico, Essick, Estell\'es, Estevez, Etienne, Etzel, Evans, Evans, Ewing, Fafone, Fair, Fairhurst, Farah, Farinon, Farr, Farr, Farrow, Fauchon-Jones, Favaro, Favata, Fays, Fazio, Feicht, Fejer, Fenyvesi, Ferguson,
  Fernandez-Galiana, Ferrante, Ferreira, Fidecaro, Figura, Fiori, Fishbach, Fisher, Fittipaldi, Fiumara, Flaminio, Floden, Fong, Font, Fornal, Forsyth, Franke, Frasca, Frasconi, Frederick, Freed, Frei, Freise, Frey, Fritschel, Frolov, Fronz\'e, Fujii, Fujikawa, Fukunaga, Fukushima, Fulda, Fyffe, Gabbard, Gabella, Gadre, Gair, Gais, Galaudage, Gamba, Ganapathy, Ganguly, Gao, Gaonkar, Garaventa, Garc\'{\i}a, Garc\'{\i}a-N\'u\~nez, Garc\'{\i}a-Quir\'os, Garufi, Gateley, Gaudio, Gayathri, Ge, Gemme, Gennai, George, George, Gerberding, Gergely, Gewecke, Ghonge, Ghosh, Ghosh, Ghosh, Ghosh, Giacomazzo, Giacoppo, Giaime, Giardina, Gibson, Gier, Giesler, Giri, Gissi, Glanzer, Gleckl, Godwin, Goetz, Goetz, Gohlke, Golomb, Goncharov, Gonz\'alez, Gopakumar, Gosselin, Gouaty, Gould, Grace, Grado, Granata, Granata, Grant, Gras, Grassia, Gray, Gray, Greco, Green, Green, Gretarsson, Gretarsson, Griffith, Griffiths, Griggs, Grignani, Grimaldi, Grimm, Grote, Grunewald, Gruning, Guerra, Guidi, Guimaraes, Guix\'e, Gulati, Guo,
  Guo, Gupta, Gupta, Gupta, Gustafson, Gustafson, Guzman, Ha, Haegel, Hagiwara, Haino, Halim, Hall, Hamilton, Hammond, Han, Haney, Hanks, Hanna, Hannam, Hannuksela, Hansen, Hansen, Hanson, Harder, Hardwick, Haris, Harms, Harry, Harry, Hartwig, Hasegawa, Haskell, Hasskew, Haster, Hattori, Haughian, Hayakawa, Hayama, Hayes, Healy, Heidmann, Heidt, Heintze, Heinze, Heinzel, Heitmann, Hellman, Hello, Helmling-Cornell, Hemming, Hendry, Heng, Hennes, Hennig, Hennig, Hernandez, Hernandez~Vivanco, Heurs, Hild, Hill, Himemoto, Hines, Hiranuma, Hirata, Hirose, Hochheim, Hofman, Hohmann, Holcomb, Holland, Holley-Bockelmann, Hollows, Holmes, Holt, Holz, Hong, Hopkins, Hough, Hourihane, Howell, Hoy, Hoyland, Hreibi, Hsieh, Hsu, Huang, Huang, Huang, Huang, Huang, Huang, H\"ubner, Huddart, Hughey, Hui, Hui, Husa, Huttner, Huxford, Huynh-Dinh, Ide, Idzkowski, Iess, Ikenoue, Imam, Inayoshi, Ingram, Inoue, Ioka, Isi, Isleif, Ito, Itoh, Iyer, Izumi, JaberianHamedan, Jacqmin, Jadhav, Jadhav, James, Jan, Jani, Janquart, Janssens,
  Janthalur, Jaranowski, Jariwala, Jaume, Jenkins, Jenner, Jeon, Jeunon, Jia, Jin, Johns, Johnson-McDaniel, Jones, Jones, Jones, Jones, Jones, Jonker, Ju, Jung, Jung, Junker, Juste, Kaihotsu, Kajita, Kakizaki, Kalaghatgi, Kalogera, Kamai, Kamiizumi, Kanda, Kandhasamy, Kang, Kanner, Kao, Kapadia, Kapasi, Karat, Karathanasis, Karki, Kashyap, Kasprzack, Kastaun, Katsanevas, Katsavounidis, Katzman, Kaur, Kawabe, Kawaguchi, Kawai, Kawasaki, K\'ef\'elian, Keitel, Key, Khadka, Khalili, Khan, Khazanov, Khetan, Khursheed, Kijbunchoo, Kim, Kim, Kim, Kim, Kim, Kim, Kimball, Kimura, Kinley-Hanlon, Kirchhoff, Kissel, Kita, Kitazawa, Kleybolte, Klimenko, Knee, Knowles, Knyazev, Koch, Koekoek, Kojima, Kokeyama, Koley, Kolitsidou, Kolstein, Komori, Kondrashov, Kong, Kontos, Koper, Korobko, Kotake, Kovalam, Kozak, Kozakai, Kozu, Kringel, Krishnendu, Kr\'olak, Kuehn, Kuei, Kuijer, Kulkarni, Kumar, Kumar, Kumar, Kumar, Kume, Kuns, Kuo, Kuo, Kuromiya, Kuroyanagi, Kusayanagi, Kuwahara, Kwak, Lagabbe, Laghi, Lalande, Lam,
  Lamberts, Landry, Lane, Lang, Lange, Lantz, La~Rosa, Lartaux-Vollard, Lasky, Laxen, Lazzarini, Lazzaro, Leaci, Leavey, Lecoeuche, Lee, Lee, Lee, Lee, Lee, Lee, Lehmann, Lema\^{\i}tre, Leonardi, Leroy, Letendre, Levesque, Levin, Leviton, Leyde, Li, Li, Li, Li, Li, Li, Lin, Lin, Lin, Lin, Lin, Linde, Linker, Linley, Littenberg, Liu, Liu, Liu, Liu, Llamas, Llorens-Monteagudo, Lo, Lockwood, Loh, London, Longo, Lopez, Portilla, Lorenzini, Loriette, Lormand, Losurdo, Lott, Lough, Lousto, Lovelace, Lucaccioni, L\"uck, Lumaca, Lundgren, Luo, Lynam, Macas, MacInnis, Macleod, MacMillan, Macquet, Hernandez, Magazz\`u, Magee, Maggiore, Magnozzi, Mahesh, Majorana, Makarem, Maksimovic, Maliakal, Malik, Man, Mandic, Mangano, Mango, Mansell, Manske, Mantovani, Mapelli, Marchesoni, Marchio, Marion, Mark, M\'arka, M\'arka, Markakis, Markosyan, Markowitz, Maros, Marquina, Marsat, Martelli, Martin, Martin, Martinez, Martinez, Martinez, Martinovic, Martynov, Marx, Masalehdan, Mason, Massera, Masserot, Massinger, Masso-Reid,
  Mastrogiovanni, Matas, Mateu-Lucena, Matichard, Matiushechkina, Mavalvala, McCann, McCarthy, McClelland, McClincy, McCormick, McCuller, McGhee, McGuire, McIsaac, McIver, McRae, McWilliams, Meacher, Mehmet, Mehta, Meijer, Melatos, Melchor, Mendell, Menendez-Vazquez, Menoni, Mercer, Mereni, Merfeld, Merilh, Merritt, Merzougui, Meshkov, Messenger, Messick, Meyers, Meylahn, Mhaske, Miani, Miao, Michaloliakos, Michel, Michimura, Middleton, Milano, Miller, Miller, Miller, Millhouse, Mills, Milotti, Minazzoli, Minenkov, Mio, Mir, Miravet-Ten\'es, Mishra, Mishra, Mistry, Mitra, Mitrofanov, Mitselmakher, Mittleman, Miyakawa, Miyamoto, Miyazaki, Miyo, Miyoki, Mo, Modafferi, Moguel, Mogushi, Mohapatra, Mohite, Molina, Molina-Ruiz, Mondin, Montani, Moore, Moraru, Morawski, More, Moreno, Moreno, Mori, Morisaki, Moriwaki, Morr\'as, Mours, Mow-Lowry, Mozzon, Muciaccia, Mukherjee, Mukherjee, Mukherjee, Mukherjee, Mukherjee, Mukund, Mullavey, Munch, Mu\~niz, Murray, Musenich, Muusse, Nadji, Nagano, Nagano, Nagar, Nakamura,
  Nakano, Nakano, Nakashima, Nakayama, Napolano, Nardecchia, Narikawa, Naticchioni, Nayak, Nayak, Negishi, Neil, Neilson, Nelemans, Nelson, Nery, Neubauer, Neunzert, Ng, Ng, Nguyen, Nguyen, Nguyen, Quynh, Ni, Nichols, Nishizawa, Nissanke, Nitoglia, Nocera, Norman, North, Nozaki, Siles, Nuttall, Oberling, O'Brien, Obuchi, O'Dell, Oelker, Ogaki, Oganesyan, Oh, Oh, Oh, Ohashi, Ohishi, Ohkawa, Ohme, Ohta, Okada, Okutani, Okutomi, Olivetto, Oohara, Ooi, Oram, O'Reilly, Ormiston, Ormsby, Ortega, O'Shaughnessy, O'Shea, Oshino, Ossokine, Osthelder, Otabe, Ottaway, Overmier, Pace, Pagano, Page, Pagliaroli, Pai, Pai, Palamos, Palashov, Palomba, Pan, Pan, Panda, Pang, Pang, Pankow, Pannarale, Pant, Panther, Paoletti, Paoli, Paolone, Parisi, Park, Park, Parker, Pascucci, Pasqualetti, Passaquieti, Passuello, Patel, Pathak, Patricelli, Patron, Paul, Payne, Pedraza, Pegoraro, Pele, Arellano, Penn, Perego, Pereira, Pereira, Perez, P\'erigois, Perkins, Perreca, Perri\`es, Petermann, Petterson, Pfeiffer, Pham, Phukon,
  Piccinni, Pichot, Piendibene, Piergiovanni, Pierini, Pierro, Pillant, Pillas, Pilo, Pinard, Pinto, Pinto, Piotrzkowski, Piotrzkowski, Pirello, Pitkin, Placidi, Planas, Plastino, Pluchar, Poggiani, Polini, Pong, Ponrathnam, Popolizio, Porter, Poulton, Powell, Pracchia, Pradier, Prajapati, Prasai, Prasanna, Pratten, Principe, Prodi, Prokhorov, Prosposito, Prudenzi, Puecher, Punturo, Puosi, Puppo, P\"urrer, Qi, Quetschke, Quitzow-James, Qutob, Raab, Raaijmakers, Radkins, Radulesco, Raffai, Rail, Raja, Rajan, Ramirez, Ramirez, Ramos-Buades, Rana, Rapagnani, Rapol, Ray, Raymond, Raza, Razzano, Read, Rees, Regimbau, Rei, Reid, Reid, Reitze, Relton, Renzini, Rettegno, Reza, Rezac, Ricci, Richards, Richardson, Richardson, Riemenschneider, Riles, Rinaldi, Rink, Rizzo, Robertson, Robie, Robinet, Rocchi, Rodriguez, Rolland, Rollins, Romanelli, Romano, Romel, Romero-Rodr\'{\i}guez, Romero-Shaw, Romie, Ronchini, Rosa, Rose, Rosi\ifmmode~\acute{n}\else \'{n}\fi{}ska, Ross, Rowan, Rowlinson, Roy, Roy, Roy, Rozza, Ruggi,
  Ruiz-Rocha, Ryan, Sachdev, Sadecki, Sadiq, Sago, Saito, Saito, Sakai, Sakai, Sakellariadou, Sakuno, Salafia, Salconi, Saleem, Salemi, Samajdar, Sanchez, Sanchez, Sanchez, Sanchis-Gual, Sanders, Sanuy, Saravanan, Sarin, Sassolas, Satari, Sathyaprakash, Sato, Sato, Sauter, Savage, Sawada, Sawant, Sawant, Sayah, Schaetzl, Scheel, Scheuer, Schiworski, Schmidt, Schmidt, Schnabel, Schneewind, Schofield, Sch\"onbeck, Schulte, Schutz, Schwartz, Scott, Scott, Seglar-Arroyo, Sekiguchi, Sekiguchi, Sellers, Sengupta, Sentenac, Seo, Sequino, Sergeev, Setyawati, Shaffer, Shahriar, Shams, Shao, Sharma, Sharma, Shawhan, Shcheblanov, Shibagaki, Shikauchi, Shimizu, Shimoda, Shimode, Shinkai, Shishido, Shoda, Shoemaker, Shoemaker, ShyamSundar, Sieniawska, Sigg, Singer, Singh, Singh, Singha, Sintes, Sipala, Skliris, Slagmolen, Slaven-Blair, Smetana, Smith, Smith, Soldateschi, Somala, Somiya, Son, Soni, Soni, Sordini, Sorrentino, Sorrentino, Sotani, Soulard, Souradeep, Sowell, Spagnuolo, Spencer, Spera, Srinivasan, Srivastava,
  Srivastava, Staats, Stachie, Steer, Steinhoff, Steinlechner, Steinlechner, Stevenson, Stops, Stover, Strain, Strang, Stratta, Strunk, Sturani, Stuver, Sudhagar, Sudhir, Sugimoto, Suh, Sullivan, Sullivan, Summerscales, Sun, Sun, Sunil, Sur, Suresh, Sutton, Suzuki, Suzuki, Swinkels, Szczepa\ifmmode~\acute{n}\else \'{n}\fi{}czyk, Szewczyk, Tacca, Tagoshi, Tait, Takahashi, Takahashi, Takamori, Takano, Takeda, Takeda, Talbot, Talbot, Tanaka, Tanaka, Tanaka, Tanaka, Tanaka, Tanasijczuk, Tanioka, Tanner, Tao, Tao, Mart\'{\i}n, Taranto, Tasson, Telada, Tenorio, Terhune, Terkowski, Thirugnanasambandam, Thomas, Thomas, Thomas, Thompson, Thondapu, Thorne, Thrane, Tiwari, Tiwari, Tiwari, Toivonen, Toland, Tolley, Tomaru, Tomigami, Tomura, Tonelli, Torres-Forn\'e, Torrie, e~Melo, T\"oyr\"a, Trapananti, Travasso, Traylor, Trevor, Tringali, Tripathee, Troiano, Trovato, Trozzo, Trudeau, Tsai, Tsai, Tsang, Tsang, Tsao, Tse, Tso, Tsubono, Tsuchida, Tsukada, Tsuna, Tsutsui, Tsuzuki, Turbang, Turconi, Tuyenbayev, Ubhi,
  Uchikata, Uchiyama, Udall, Ueda, Uehara, Ueno, Ueshima, Unnikrishnan, Uraguchi, Urban, Ushiba, Utina, Vahlbruch, Vajente, Vajpeyi, Valdes, Valentini, Valsan, van Bakel, van Beuzekom, van~den Brand, Van Den~Broeck, Vander-Hyde, van~der Schaaf, van Heijningen, Vanosky, van Putten, van Remortel, Vardaro, Vargas, Varma, Vas\'uth, Vecchio, Vedovato, Veitch, Veitch, Venneberg, Venugopalan, Verkindt, Verma, Verma, Veske, Vetrano, Vicer\'e, Vidyant, Viets, Vijaykumar, Villa-Ortega, Vinet, Virtuoso, Vitale, Vo, Vocca, von Reis, von Wrangel, Vorvick, Vyatchanin, Wade, Wade, Wagner, Walet, Walker, Wallace, Wallace, Walsh, Wang, Wang, Wang, Ward, Warner, Was, Washimi, Washington, Watchi, Weaver, Webster, Weinert, Weinstein, Weiss, Weller, Weller, Wellmann, Wen, We\ss{}els, Wette, Whelan, White, Whiting, Whittle, Wilken, Williams, Williams, Williams, Williamson, Willis, Willke, Wilson, Winkler, Wipf, Wlodarczyk, Woan, Woehler, Wofford, Wong, Wu, Wu, Wu, Wu, Wysocki, Xiao, Xu, Yamada, Yamamoto, Yamamoto, Yamamoto,
  Yamamoto, Yamashita, Yamazaki, Yang, Yang, Yang, Yang, Yang, Yap, Yeeles, Yelikar, Ying, Yokogawa, Yokoyama, Yokozawa, Yoo, Yoshioka, Yu, Yu, Yuzurihara, Zadro\ifmmode~\dot{z}\else \.{z}\fi{}ny, Zanolin, Zeidler, Zelenova, Zendri, Zevin, Zhan, Zhang, Zhang, Zhang, Zhang, Zhang, Zhao, Zhao, Zhao, Zhao, Zheng, Zhou, Zhou, Zhu, Zhu, Zimmerman, Zlochower, Zucker, \& Zweizig}]{LIGO_2021c}
Abbott, R., Abbott, T.~D., Acernese, F., {et~al.} 2023{\natexlab{a}}, Phys. Rev. X, 13, 041039, \dodoi{10.1103/PhysRevX.13.041039}

\bibitem[{Abbott {et~al.}(2023{\natexlab{b}})Abbott, Abbott, Acernese, Ackley, Adams, Adhikari, Adhikari, Adya, Affeldt, Agarwal, Agathos, Agatsuma, Aggarwal, Aguiar, Aiello, Ain, Ajith, Akutsu, de~Alarc\'on, Akcay, Albanesi, Allocca, Altin, Amato, Anand, Anand, Ananyeva, Anderson, Anderson, Ando, Andrade, Andres, Andri\ifmmode~\acute{c}\else \'{c}\fi{}, Angelova, Ansoldi, Antelis, Antier, Antonini, Appert, Arai, Arai, Arai, Araki, Araya, Araya, Areeda, Ar\`ene, Aritomi, Arnaud, Arogeti, Aronson, Arun, Asada, Asali, Ashton, Aso, Assiduo, Aston, Astone, Aubin, Austin, Babak, Badaracco, Bader, Badger, Bae, Bae, Baer, Bagnasco, Bai, Baiotti, Baird, Bajpai, Ball, Ballardin, Ballmer, Balsamo, Baltus, Banagiri, Bankar, Barayoga, Barbieri, Barish, Barker, Barneo, Barone, Barr, Barsotti, Barsuglia, Barta, Bartlett, Barton, Bartos, Bassiri, Basti, Bawaj, Bayley, Baylor, Bazzan, B\'ecsy, Bedakihale, Bejger, Belahcene, Benedetto, Beniwal, Bennett, Bentley, BenYaala, Bergamin, Berger, Bernuzzi, Berry, Bersanetti,
  Bertolini, Betzwieser, Beveridge, Bhandare, Bhardwaj, Bhattacharjee, Bhaumik, Bilenko, Billingsley, Bini, Birney, Birnholtz, Biscans, Bischi, Biscoveanu, Bisht, Biswas, Bitossi, Bizouard, Blackburn, Blair, Blair, Blair, Bobba, Bode, Boer, Bogaert, Boldrini, Bonavena, Bondu, Bonilla, Bonnand, Booker, Boom, Bork, Boschi, Bose, Bose, Bossilkov, Boudart, Bouffanais, Bozzi, Bradaschia, Brady, Bramley, Branch, Branchesi, Brandt, Brau, Breschi, Briant, Briggs, Brillet, Brinkmann, Brockill, Brooks, Brooks, Brown, Brunett, Bruno, Bruntz, Bryant, Bulik, Bulten, Buonanno, Buscicchio, Buskulic, Buy, Byer, Cadonati, Cagnoli, Cahillane, Bustillo, Callaghan, Callister, Calloni, Cameron, Camp, Canepa, Canevarolo, Cannavacciuolo, Cannon, Cao, Cao, Capocasa, Capote, Carapella, Carbognani, Carlin, Carney, Carpinelli, Carrillo, Carullo, Carver, Diaz, Casentini, Castaldi, Caudill, Cavagli\`a, Cavalier, Cavalieri, Ceasar, Cella, Cerd\'a-Dur\'an, Cesarini, Chaibi, Chakravarti, Subrahmanya, Champion, Chan, Chan, Chan, Chan, Chan,
  Chandra, Chanial, Chao, Chapman-Bird, Charlton, Chase, Chassande-Mottin, Chatterjee, Chatterjee, Chatterjee, Chaturvedi, Chaty, Chatziioannou, Chen, Chen, Chen, Chen, Chen, Chen, Chen, Chen, Cheng, Cheong, Cheung, Chia, Chiadini, Chiang, Chiarini, Chierici, Chincarini, Chiofalo, Chiummo, Cho, Cho, Choudhary, Choudhary, Christensen, Chu, Chu, Chu, Chua, Chung, Ciani, Ciecielag, Cie\ifmmode~\acute{s}\else \'{s}\fi{}lar, Cifaldi, Ciobanu, Ciolfi, Cipriano, Cirone, Clara, Clark, Clark, Clarke, Clearwater, Clesse, Cleva, Coccia, Codazzo, Cohadon, Cohen, Cohen, Colleoni, Collette, Colombo, Colpi, Compton, Constancio, Conti, Cooper, Corban, Corbitt, Cordero-Carri\'on, Corezzi, Corley, Cornish, Corre, Corsi, Cortese, Costa, Cotesta, Coughlin, Coulon, Countryman, Cousins, Couvares, Coward, Cowart, Coyne, Coyne, Creighton, Creighton, Criswell, Croquette, Crowder, Cudell, Cullen, Cumming, Cummings, Cunningham, Cuoco, Cury\l{}o, Dabadie, Canton, Dall'Osso, D\'alya, Dana, DaneshgaranBajastani, D'Angelo, Danila,
  Danilishin, D'Antonio, Danzmann, Darsow-Fromm, Dasgupta, Datrier, Datta, Dattilo, Dave, Davier, Davies, Davis, Davis, Daw, Dean, DeBra, Deenadayalan, Degallaix, De~Laurentis, Del\'eglise, Del~Favero, De~Lillo, De~Lillo, Del~Pozzo, DeMarchi, De~Matteis, D'Emilio, Demos, Dent, Depasse, De~Pietri, De~Rosa, De~Rossi, DeSalvo, De~Simone, Dhurandhar, D\'{\i}az, Diaz-Ortiz, Didio, Dietrich, Di~Fiore, Di~Fronzo, Di~Giorgio, Di~Giovanni, Di~Giovanni, Di~Girolamo, Di~Lieto, Ding, Di~Pace, Di~Palma, Di~Renzo, Divakarla, Dmitriev, Doctor, D'Onofrio, Donovan, Dooley, Doravari, Dorrington, Drago, Driggers, Drori, Ducoin, Dupej, Durante, D'Urso, Duverne, Dwyer, Eassa, Easter, Ebersold, Eckhardt, Eddolls, Edelman, Edo, Edy, Effler, Eguchi, Eichholz, Eikenberry, Eisenmann, Eisenstein, Ejlli, Engelby, Enomoto, Errico, Essick, Estell\'es, Estevez, Etienne, Etzel, Evans, Evans, Ewing, Fafone, Fair, Fairhurst, Farah, Farinon, Farr, Farr, Farrow, Fauchon-Jones, Favaro, Favata, Fays, Fazio, Feicht, Fejer, Fenyvesi, Ferguson,
  Fernandez-Galiana, Ferrante, Ferreira, Fidecaro, Figura, Fiori, Fishbach, Fisher, Fittipaldi, Fiumara, Flaminio, Floden, Fong, Font, Fornal, Forsyth, Franke, Frasca, Frasconi, Frederick, Freed, Frei, Freise, Frey, Fritschel, Frolov, Fronz\'e, Fujii, Fujikawa, Fukunaga, Fukushima, Fulda, Fyffe, Gabbard, Gadre, Gair, Gais, Galaudage, Gamba, Ganapathy, Ganguly, Gao, Gaonkar, Garaventa, Garc\'{\i}a, Garc\'{\i}a-N\'u\~nez, Garc\'{\i}a-Quir\'os, Garufi, Gateley, Gaudio, Gayathri, Ge, Gemme, Gennai, George, George, Gerberding, Gergely, Gewecke, Ghonge, Ghosh, Ghosh, Ghosh, Ghosh, Giacomazzo, Giacoppo, Giaime, Giardina, Gibson, Gier, Giesler, Giri, Gissi, Glanzer, Gleckl, Godwin, Golomb, Goetz, Goetz, Gohlke, Goncharov, Gonz\'alez, Gopakumar, Gosselin, Gouaty, Gould, Grace, Grado, Granata, Granata, Grant, Gras, Grassia, Gray, Gray, Greco, Green, Green, Gretarsson, Gretarsson, Griffith, Griffiths, Griggs, Grignani, Grimaldi, Grimm, Grote, Grunewald, Gruning, Guerra, Guidi, Guimaraes, Guix\'e, Gulati, Guo, Guo,
  Gupta, Gupta, Gupta, Gustafson, Gustafson, Guzman, Ha, Haegel, Hagiwara, Haino, Halim, Hall, Hamilton, Hammond, Han, Haney, Hanks, Hanna, Hannam, Hannuksela, Hansen, Hansen, Hanson, Harder, Hardwick, Haris, Harms, Harry, Harry, Hartwig, Hasegawa, Haskell, Hasskew, Haster, Hattori, Haughian, Hayakawa, Hayama, Hayes, Healy, Heidmann, Heidt, Heintze, Heinze, Heinzel, Heitmann, Hellman, Hello, Helmling-Cornell, Hemming, Hendry, Heng, Hennes, Hennig, Hennig, Hernandez, Vivanco, Heurs, Hild, Hill, Himemoto, Hines, Hiranuma, Hirata, Hirose, Hochheim, Hofman, Hohmann, Holcomb, Holland, Hollows, Holmes, Holt, Holz, Hong, Hopkins, Hough, Hourihane, Howell, Hoy, Hoyland, Hreibi, Hsieh, Hsu, Huang, Huang, Huang, Huang, Huang, Huang, H\"ubner, Huddart, Hughey, Hui, Hui, Husa, Huttner, Huxford, Huynh-Dinh, Ide, Idzkowski, Iess, Ikenoue, Imam, Inayoshi, Ingram, Inoue, Ioka, Isi, Isleif, Ito, Itoh, Iyer, Izumi, JaberianHamedan, Jacqmin, Jadhav, Jadhav, James, Jan, Jani, Janquart, Janssens, Janthalur, Jaranowski, Jariwala,
  Jaume, Jenkins, Jenner, Jeon, Jeunon, Jia, Jin, Johns, Jones, Jones, Jones, Jones, Jones, Jonker, Ju, Jung, Jung, Junker, Juste, Kaihotsu, Kajita, Kakizaki, Kalaghatgi, Kalogera, Kamai, Kamiizumi, Kanda, Kandhasamy, Kang, Kanner, Kao, Kapadia, Kapasi, Karat, Karathanasis, Karki, Kashyap, Kasprzack, Kastaun, Katsanevas, Katsavounidis, Katzman, Kaur, Kawabe, Kawaguchi, Kawai, Kawasaki, K\'ef\'elian, Keitel, Key, Khadka, Khalili, Khan, Khazanov, Khetan, Khursheed, Kijbunchoo, Kim, Kim, Kim, Kim, Kim, Kim, Kimball, Kimura, Kinley-Hanlon, Kirchhoff, Kissel, Kita, Kitazawa, Kleybolte, Klimenko, Knee, Knowles, Knyazev, Koch, Koekoek, Kojima, Kokeyama, Koley, Kolitsidou, Kolstein, Komori, Kondrashov, Kong, Kontos, Koper, Korobko, Kotake, Kovalam, Kozak, Kozakai, Kozu, Kringel, Krishnendu, Kr\'olak, Kuehn, Kuei, Kuijer, Kulkarni, Kumar, Kumar, Kumar, Kumar, Kume, Kuns, Kuo, Kuo, Kuromiya, Kuroyanagi, Kusayanagi, Kuwahara, Kwak, Lagabbe, Laghi, Lalande, Lam, Lamberts, Landry, Landry, Lane, Lang, Lange, Lantz,
  La~Rosa, Lartaux-Vollard, Lasky, Laxen, Lazzarini, Lazzaro, Leaci, Leavey, Lecoeuche, Lee, Lee, Lee, Lee, Lee, Lee, Lehmann, Lema\^{\i}tre, Leonardi, Leroy, Letendre, Levesque, Levin, Leviton, Leyde, Li, Li, Li, Li, Li, Li, Lin, Lin, Lin, Lin, Lin, Linde, Linker, Linley, Littenberg, Liu, Liu, Liu, Liu, Llamas, Llorens-Monteagudo, Lo, Lockwood, Loh, London, Longo, Lopez, Portilla, Lorenzini, Loriette, Lormand, Losurdo, Lott, Lough, Lousto, Lovelace, Lucaccioni, L\"uck, Lumaca, Lundgren, Luo, Lynam, Macas, MacInnis, Macleod, MacMillan, Macquet, Hernandez, Magazz\`u, Magee, Maggiore, Magnozzi, Mahesh, Majorana, Makarem, Maksimovic, Maliakal, Malik, Man, Mandic, Mangano, Mango, Mansell, Manske, Mantovani, Mapelli, Marchesoni, Marchio, Marion, Mark, M\'arka, M\'arka, Markakis, Markosyan, Markowitz, Maros, Marquina, Marsat, Martelli, Martin, Martin, Martinez, Martinez, Martinez, Martinovic, Martynov, Marx, Masalehdan, Mason, Massera, Masserot, Massinger, Masso-Reid, Mastrogiovanni, Matas, Mateu-Lucena, Matichard,
  Matiushechkina, Mavalvala, McCann, McCarthy, McClelland, McClincy, McCormick, McCuller, McGhee, McGuire, McIsaac, McIver, McRae, McWilliams, Meacher, Mehmet, Mehta, Meijer, Melatos, Melchor, Mendell, Menendez-Vazquez, Menoni, Mercer, Mereni, Merfeld, Merilh, Merritt, Merzougui, Meshkov, Messenger, Messick, Meyers, Meylahn, Mhaske, Miani, Miao, Michaloliakos, Michel, Michimura, Middleton, Milano, Miller, Miller, Miller, Miller, Millhouse, Mills, Milotti, Minazzoli, Minenkov, Mio, Mir, Miravet-Ten\'es, Mishra, Mishra, Mistry, Mitra, Mitrofanov, Mitselmakher, Mittleman, Miyakawa, Miyamoto, Miyazaki, Miyo, Miyoki, Mo, Modafferi, Moguel, Mogushi, Mohapatra, Mohite, Molina, Molina-Ruiz, Mondin, Montani, Moore, Moraru, Morawski, More, Moreno, Moreno, Mori, Morisaki, Moriwaki, Morr\'as, Mours, Mow-Lowry, Mozzon, Muciaccia, Mukherjee, Mukherjee, Mukherjee, Mukherjee, Mukherjee, Mukund, Mullavey, Munch, Mu\~niz, Murray, Musenich, Muusse, Nadji, Nagano, Nagano, Nagar, Nakamura, Nakano, Nakano, Nakashima, Nakayama,
  Napolano, Nardecchia, Narikawa, Naticchioni, Nayak, Nayak, Negishi, Neil, Neilson, Nelemans, Nelson, Nery, Neubauer, Neunzert, Ng, Ng, Nguyen, Nguyen, Nguyen, Quynh, Ni, Nichols, Nishizawa, Nissanke, Nitoglia, Nocera, Norman, North, Nozaki, Siles, Nuttall, Oberling, O'Brien, Obuchi, O'Dell, Oelker, Ogaki, Oganesyan, Oh, Oh, Oh, Ohashi, Ohishi, Ohkawa, Ohme, Ohta, Okada, Okutani, Okutomi, Olivetto, Oohara, Ooi, Oram, O'Reilly, Ormiston, Ormsby, Ortega, O'Shaughnessy, O'Shea, Oshino, Ossokine, Osthelder, Otabe, Ottaway, Overmier, Pace, Pagano, Page, Pagliaroli, Pai, Pai, Palamos, Palashov, Palomba, Pan, Pan, Panda, Pang, Pang, Pankow, Pannarale, Pant, Panther, Paoletti, Paoli, Paolone, Parisi, Park, Park, Parker, Pascucci, Pasqualetti, Passaquieti, Passuello, Patel, Pathak, Patricelli, Patron, Paul, Payne, Pedraza, Pegoraro, Pele, Arellano, Penn, Perego, Pereira, Pereira, Perez, P\'erigois, Perkins, Perreca, Perri\`es, Petermann, Petterson, Pfeiffer, Pham, Phukon, Piccinni, Pichot, Piendibene, Piergiovanni,
  Pierini, Pierro, Pillant, Pillas, Pilo, Pinard, Pinto, Pinto, Piotrzkowski, Piotrzkowski, Pirello, Pitkin, Placidi, Planas, Plastino, Pluchar, Poggiani, Polini, Pong, Ponrathnam, Popolizio, Porter, Poulton, Powell, Pracchia, Pradier, Prajapati, Prasai, Prasanna, Pratten, Principe, Prodi, Prokhorov, Prosposito, Prudenzi, Puecher, Punturo, Puosi, Puppo, P\"urrer, Qi, Quetschke, Quitzow-James, Raab, Raaijmakers, Radkins, Radulesco, Raffai, Rail, Raja, Rajan, Ramirez, Ramirez, Ramos-Buades, Rana, Rapagnani, Rapol, Ray, Raymond, Raza, Razzano, Read, Rees, Regimbau, Rei, Reid, Reid, Reitze, Relton, Renzini, Rettegno, Reza, Rezac, Ricci, Richards, Richardson, Richardson, Riemenschneider, Riles, Rinaldi, Rink, Rizzo, Robertson, Robie, Robinet, Rocchi, Rodriguez, Rolland, Rollins, Romanelli, Romano, Romel, Romero-Rodr\'{\i}guez, Romero-Shaw, Romie, Ronchini, Rosa, Rose, Rosi\ifmmode~\acute{n}\else \'{n}\fi{}ska, Ross, Rowan, Rowlinson, Roy, Roy, Roy, Rozza, Ruggi, Ryan, Sachdev, Sadecki, Sadiq, Sago, Saito, Saito,
  Sakai, Sakai, Sakellariadou, Sakuno, Salafia, Salconi, Saleem, Salemi, Samajdar, Sanchez, Sanchez, Sanchez, Sanchis-Gual, Sanders, Sanuy, Saravanan, Sarin, Sassolas, Satari, Sathyaprakash, Sato, Sato, Sauter, Savage, Sawada, Sawant, Sawant, Sayah, Schaetzl, Scheel, Scheuer, Schiworski, Schmidt, Schmidt, Schnabel, Schneewind, Schofield, Sch\"onbeck, Schulte, Schutz, Schwartz, Scott, Scott, Seglar-Arroyo, Sekiguchi, Sekiguchi, Sellers, Sengupta, Sentenac, Seo, Sequino, Sergeev, Setyawati, Shaffer, Shahriar, Shams, Shao, Sharma, Sharma, Shawhan, Shcheblanov, Shibagaki, Shikauchi, Shimizu, Shimoda, Shimode, Shinkai, Shishido, Shoda, Shoemaker, Shoemaker, ShyamSundar, Sieniawska, Sigg, Singer, Singh, Singh, Singha, Sintes, Sipala, Skliris, Slagmolen, Slaven-Blair, Smetana, Smith, Smith, Soldateschi, Somala, Somiya, Son, Soni, Soni, Sordini, Sorrentino, Sorrentino, Sotani, Soulard, Souradeep, Sowell, Spagnuolo, Spencer, Spera, Srinivasan, Srivastava, Srivastava, Staats, Stachie, Steer, Steinhoff, Steinlechner,
  Steinlechner, Stevenson, Stops, Stover, Strain, Strang, Stratta, Strunk, Sturani, Stuver, Sudhagar, Sudhir, Sugimoto, Suh, Sullivan, Summerscales, Sun, Sun, Sunil, Sur, Suresh, Sutton, Suzuki, Suzuki, Swinkels, Szczepa\ifmmode~\acute{n}\else \'{n}\fi{}czyk, Szewczyk, Tacca, Tagoshi, Tait, Takahashi, Takahashi, Takamori, Takano, Takeda, Takeda, Talbot, Talbot, Tanaka, Tanaka, Tanaka, Tanaka, Tanaka, Tanasijczuk, Tanioka, Tanner, Tao, Tao, Mart\'{\i}n, Taranto, Tasson, Telada, Tenorio, Terhune, Terkowski, Thirugnanasambandam, Thomas, Thomas, Thomas, Thompson, Thondapu, Thorne, Thrane, Tiwari, Tiwari, Tiwari, Toivonen, Toland, Tolley, Tomaru, Tomigami, Tomura, Tonelli, Torres-Forn\'e, Torrie, e~Melo, T\"oyr\"a, Trapananti, Travasso, Traylor, Trevor, Tringali, Tripathee, Troiano, Trovato, Trozzo, Trudeau, Tsai, Tsai, Tsang, Tsang, Tsao, Tse, Tso, Tsubono, Tsuchida, Tsukada, Tsuna, Tsutsui, Tsuzuki, Turbang, Turconi, Tuyenbayev, Ubhi, Uchikata, Uchiyama, Udall, Ueda, Uehara, Ueno, Ueshima, Unnikrishnan,
  Uraguchi, Urban, Ushiba, Utina, Vahlbruch, Vajente, Vajpeyi, Valdes, Valentini, Valsan, van Bakel, van Beuzekom, van~den Brand, Van Den~Broeck, Vander-Hyde, van~der Schaaf, van Heijningen, Vanosky, van Putten, van Remortel, Vardaro, Vargas, Varma, Vas\'uth, Vecchio, Vedovato, Veitch, Veitch, Venneberg, Venugopalan, Verkindt, Verma, Verma, Veske, Vetrano, Vicer\'e, Vidyant, Viets, Vijaykumar, Villa-Ortega, Vinet, Virtuoso, Vitale, Vo, Vocca, von Reis, von Wrangel, Vorvick, Vyatchanin, Wade, Wade, Wagner, Walet, Walker, Wallace, Wallace, Walsh, Wang, Wang, Wang, Ward, Warner, Was, Washimi, Washington, Watchi, Weaver, Webster, Weinert, Weinstein, Weiss, Weller, Wellmann, Wen, We\ss{}els, Wette, Whelan, White, Whiting, Whittle, Wilken, Williams, Williams, Williamson, Willis, Willke, Wilson, Winkler, Wipf, Wlodarczyk, Woan, Woehler, Wofford, Wong, Wu, Wu, Wu, Wu, Wysocki, Xiao, Xu, Yamada, Yamamoto, Yamamoto, Yamamoto, Yamamoto, Yamashita, Yamazaki, Yang, Yang, Yang, Yang, Yang, Yap, Yeeles, Yelikar, Ying,
  Yokogawa, Yokoyama, Yokozawa, Yoo, Yoshioka, Yu, Yu, Yuzurihara, Zadro\ifmmode~\dot{z}\else \.{z}\fi{}ny, Zanolin, Zeidler, Zelenova, Zendri, Zevin, Zhan, Zhang, Zhang, Zhang, Zhang, Zhang, Zhao, Zhao, Zhao, Zhao, Zheng, Zhou, Zhou, Zhu, Zhu, Zimmerman, Zlochower, Zucker, \& Zweizig}]{Abbott2023}
---. 2023{\natexlab{b}}, Phys. Rev. X, 13, 011048, \dodoi{10.1103/PhysRevX.13.011048}

\bibitem[{{Antonini} \& {Perets}(2012)}]{Antonini_Perets12}
{Antonini}, F., \& {Perets}, H.~B. 2012, \apj, 757, 27, \dodoi{10.1088/0004-637X/757/1/27}

\bibitem[{{Astropy Collaboration} {et~al.}(2018){Astropy Collaboration}, {Price-Whelan}, {Sip{\H{o}}cz}, {G{\"u}nther}, {Lim}, {Crawford}, {Conseil}, {Shupe}, {Craig}, {Dencheva}, {Ginsburg}, {VanderPlas}, {Bradley}, {P{\'e}rez-Su{\'a}rez}, {de Val-Borro}, {Aldcroft}, {Cruz}, {Robitaille}, {Tollerud}, {Ardelean}, {Babej}, {Bach}, {Bachetti}, {Bakanov}, {Bamford}, {Barentsen}, {Barmby}, {Baumbach}, {Berry}, {Biscani}, {Boquien}, {Bostroem}, {Bouma}, {Brammer}, {Bray}, {Breytenbach}, {Buddelmeijer}, {Burke}, {Calderone}, {Cano Rodr{\'\i}guez}, {Cara}, {Cardoso}, {Cheedella}, {Copin}, {Corrales}, {Crichton}, {D'Avella}, {Deil}, {Depagne}, {Dietrich}, {Donath}, {Droettboom}, {Earl}, {Erben}, {Fabbro}, {Ferreira}, {Finethy}, {Fox}, {Garrison}, {Gibbons}, {Goldstein}, {Gommers}, {Greco}, {Greenfield}, {Groener}, {Grollier}, {Hagen}, {Hirst}, {Homeier}, {Horton}, {Hosseinzadeh}, {Hu}, {Hunkeler}, {Ivezi{\'c}}, {Jain}, {Jenness}, {Kanarek}, {Kendrew}, {Kern}, {Kerzendorf}, {Khvalko}, {King}, {Kirkby}, {Kulkarni},
  {Kumar}, {Lee}, {Lenz}, {Littlefair}, {Ma}, {Macleod}, {Mastropietro}, {McCully}, {Montagnac}, {Morris}, {Mueller}, {Mumford}, {Muna}, {Murphy}, {Nelson}, {Nguyen}, {Ninan}, {N{\"o}the}, {Ogaz}, {Oh}, {Parejko}, {Parley}, {Pascual}, {Patil}, {Patil}, {Plunkett}, {Prochaska}, {Rastogi}, {Reddy Janga}, {Sabater}, {Sakurikar}, {Seifert}, {Sherbert}, {Sherwood-Taylor}, {Shih}, {Sick}, {Silbiger}, {Singanamalla}, {Singer}, {Sladen}, {Sooley}, {Sornarajah}, {Streicher}, {Teuben}, {Thomas}, {Tremblay}, {Turner}, {Terr{\'o}n}, {van Kerkwijk}, {de la Vega}, {Watkins}, {Weaver}, {Whitmore}, {Woillez}, {Zabalza}, \& {Astropy Contributors}}]{Astropy+18}
{Astropy Collaboration}, {Price-Whelan}, A.~M., {Sip{\H{o}}cz}, B.~M., {et~al.} 2018, \aj, 156, 123, \dodoi{10.3847/1538-3881/aabc4f}

\bibitem[{Beloborodov(2003)}]{Beloborodov:2002af}
Beloborodov, A.~M. 2003, Astrophys. J., 588, 931, \dodoi{10.1086/374217}

\bibitem[{{Blinnikov} {et~al.}(1990){Blinnikov}, {Imshennik}, {Nadezhin}, {Novikov}, {Perevodchikova}, \& {Polnarev}}]{Blinnikov+90}
{Blinnikov}, S.~I., {Imshennik}, V.~S., {Nadezhin}, D.~K., {et~al.} 1990, \sovast, 34, 595

\bibitem[{{Blinnikov} {et~al.}(1984){Blinnikov}, {Novikov}, {Perevodchikova}, \& {Polnarev}}]{Blinnikov+84}
{Blinnikov}, S.~I., {Novikov}, I.~D., {Perevodchikova}, T.~V., \& {Polnarev}, A.~G. 1984, Soviet Astronomy Letters, 10, 177, \dodoi{10.48550/arXiv.1808.05287}

\bibitem[{{Camilletti} {et~al.}(2024){Camilletti}, {Perego}, {Guercilena}, {Bernuzzi}, \& {Radice}}]{Camilletti+2024}
{Camilletti}, A., {Perego}, A., {Guercilena}, F.~M., {Bernuzzi}, S., \& {Radice}, D. 2024, arXiv e-prints, arXiv:2401.04102, \dodoi{10.48550/arXiv.2401.04102}

\bibitem[{{Chiaramello} \& {Nagar}(2020)}]{Chiaramello_Nagar20}
{Chiaramello}, D., \& {Nagar}, A. 2020, \prd, 101, 101501, \dodoi{10.1103/PhysRevD.101.101501}

\bibitem[{Chirenti {et~al.}(2017)Chirenti, Gold, \& Miller}]{Chirenti:2016xys}
Chirenti, C., Gold, R., \& Miller, M.~C. 2017, Astrophys. J., 837, 67, \dodoi{10.3847/1538-4357/aa5ebb}

\bibitem[{{Clark} \& {Eardley}(1977)}]{ClarkEardley1977}
{Clark}, J.~P.~A., \& {Eardley}, D.~M. 1977, \apj, 215, 311, \dodoi{10.1086/155360}

\bibitem[{{Davies} {et~al.}(2005){Davies}, {Levan}, \& {King}}]{Davies+05_BHNS}
{Davies}, M.~B., {Levan}, A.~J., \& {King}, A.~R. 2005, \mnras, 356, 54, \dodoi{10.1111/j.1365-2966.2004.08423.x}

\bibitem[{De \& Siegel(2021)}]{De:2020jdt}
De, S., \& Siegel, D.~M. 2021, Astrophys. J., 921, 94, \dodoi{10.3847/1538-4357/ac110b}

\bibitem[{{Drozda} {et~al.}(2022){Drozda}, {Belczynski}, {O'Shaughnessy}, {Bulik}, \& {Fryer}}]{Drozda+22_FMG}
{Drozda}, P., {Belczynski}, K., {O'Shaughnessy}, R., {Bulik}, T., \& {Fryer}, C.~L. 2022, \aap, 667, A126, \dodoi{10.1051/0004-6361/202039418}

\bibitem[{East {et~al.}(2012)East, Pretorius, \& Stephens}]{East:2011xa}
East, W.~E., Pretorius, F., \& Stephens, B.~C. 2012, Phys. Rev. D, 85, 124009, \dodoi{10.1103/PhysRevD.85.124009}

\bibitem[{{Eggleton}(1983)}]{Eggleton1983}
{Eggleton}, P.~P. 1983, \apj, 268, 368, \dodoi{10.1086/160960}

\bibitem[{{Etienne} {et~al.}(2009){Etienne}, {Liu}, {Shapiro}, \& {Baumgarte}}]{Etienne+09}
{Etienne}, Z.~B., {Liu}, Y.~T., {Shapiro}, S.~L., \& {Baumgarte}, T.~W. 2009, \prd, 79, 044024, \dodoi{10.1103/PhysRevD.79.044024}

\bibitem[{Fern\'andez {et~al.}(2017)Fern\'andez, Foucart, Kasen, Lippuner, Desai, \& Roberts}]{Fernandez:2016sbf}
Fern\'andez, R., Foucart, F., Kasen, D., {et~al.} 2017, Class. Quant. Grav., 34, 154001, \dodoi{10.1088/1361-6382/aa7a77}

\bibitem[{Fern\'andez \& Metzger(2013)}]{Fernandez:2013tya}
Fern\'andez, R., \& Metzger, B.~D. 2013, Mon. Not. Roy. Astron. Soc., 435, 502, \dodoi{10.1093/mnras/stt1312}

\bibitem[{{Fern{\'a}ndez} \& {Metzger}(2013)}]{Fernandez&Metzger13}
{Fern{\'a}ndez}, R., \& {Metzger}, B.~D. 2013, \mnras, 435, 502, \dodoi{10.1093/mnras/stt1312}

\bibitem[{Foucart(2012)}]{Foucart:2012nc}
Foucart, F. 2012, Phys. Rev. D, 86, 124007, \dodoi{10.1103/PhysRevD.86.124007}

\bibitem[{{Foucart} {et~al.}(2018){Foucart}, {Hinderer}, \& {Nissanke}}]{Foucart+18}
{Foucart}, F., {Hinderer}, T., \& {Nissanke}, S. 2018, \prd, 98, 081501, \dodoi{10.1103/PhysRevD.98.081501}

\bibitem[{{Foucart} {et~al.}(2014){Foucart}, {Deaton}, {Duez}, {O'Connor}, {Ott}, {Haas}, {Kidder}, {Pfeiffer}, {Scheel}, \& {Szilagyi}}]{Foucart+14}
{Foucart}, F., {Deaton}, M.~B., {Duez}, M.~D., {et~al.} 2014, \prd, 90, 024026, \dodoi{10.1103/PhysRevD.90.024026}

\bibitem[{{Godzieba} {et~al.}(2021){Godzieba}, {Radice}, \& {Bernuzzi}}]{Godzieba+21_GW190814}
{Godzieba}, D.~A., {Radice}, D., \& {Bernuzzi}, S. 2021, \apj, 908, 122, \dodoi{10.3847/1538-4357/abd4dd}

\bibitem[{Gottlieb {et~al.}(2023)}]{Gottlieb:2023est}
Gottlieb, O., {et~al.} 2023, Astrophys. J. Lett., 954, L21, \dodoi{10.3847/2041-8213/aceeff}

\bibitem[{{Hamers} \& {Dosopoulou}(2019)}]{Hamers&Dosopoulou19}
{Hamers}, A.~S., \& {Dosopoulou}, F. 2019, \apj, 872, 119, \dodoi{10.3847/1538-4357/ab001d}

\bibitem[{{Hamers} {et~al.}(2021){Hamers}, {Rantala}, {Neunteufel}, {Preece}, \& {Vynatheya}}]{Hamers+21}
{Hamers}, A.~S., {Rantala}, A., {Neunteufel}, P., {Preece}, H., \& {Vynatheya}, P. 2021, \mnras, 502, 4479, \dodoi{10.1093/mnras/stab287}

\bibitem[{Hamilton \& Rafikov(2019)}]{Hamilton:2019yij}
Hamilton, C., \& Rafikov, R.~R. 2019, Astrophys. J. Lett., 881, L13, \dodoi{10.3847/2041-8213/ab3468}

\bibitem[{Harris {et~al.}(2020)Harris, Millman, van~der Walt, Gommers, Virtanen, Cournapeau, Wieser, Taylor, Berg, Smith, Kern, Picus, Hoyer, van Kerkwijk, Brett, Haldane, Fernández~del Río, Wiebe, Peterson, Gérard-Marchant, Sheppard, Reddy, Weckesser, Abbasi, Gohlke, \& Oliphant}]{2020NumPy-Array}
Harris, C.~R., Millman, K.~J., van~der Walt, S.~J., {et~al.} 2020, Nature, 585, 357–362, \dodoi{10.1038/s41586-020-2649-2}

\bibitem[{{Hayashi} {et~al.}(2024){Hayashi}, {Kiuchi}, {Kyutoku}, {Sekiguchi}, \& {Shibata}}]{Hayashi+24}
{Hayashi}, K., {Kiuchi}, K., {Kyutoku}, K., {Sekiguchi}, Y., \& {Shibata}, M. 2024, arXiv e-prints, arXiv:2410.10958, \dodoi{10.48550/arXiv.2410.10958}

\bibitem[{{Hoang} {et~al.}(2018){Hoang}, {Naoz}, {Kocsis}, {Rasio}, \& {Dosopoulou}}]{Hoang+2018}
{Hoang}, B.-M., {Naoz}, S., {Kocsis}, B., {Rasio}, F.~A., \& {Dosopoulou}, F. 2018, \apj, 856, 140, \dodoi{10.3847/1538-4357/aaafce}

\bibitem[{{Hunter}(2007)}]{Hunter07_matplotlib}
{Hunter}, J.~D. 2007, Computing in Science and Engineering, 9, 90, \dodoi{10.1109/MCSE.2007.55}

\bibitem[{{Kasen} {et~al.}(2013){Kasen}, {Badnell}, \& {Barnes}}]{Kasen13}
{Kasen}, D., {Badnell}, N.~R., \& {Barnes}, J. 2013, \apj, 774, 25, \dodoi{10.1088/0004-637X/774/1/25}

\bibitem[{{Khalil} {et~al.}(2021){Khalil}, {Buonanno}, {Steinhoff}, \& {Vines}}]{KhalilBuonanno+21}
{Khalil}, M., {Buonanno}, A., {Steinhoff}, J., \& {Vines}, J. 2021, \prd, 104, 024046, \dodoi{10.1103/PhysRevD.104.024046}

\bibitem[{{Kramarev} {et~al.}(2024){Kramarev}, {Kuranov}, {Yudin}, \& {Postnov}}]{Kramarev+24}
{Kramarev}, N.~I., {Kuranov}, A.~G., {Yudin}, A.~V., \& {Postnov}, K.~A. 2024, Astronomy Letters, 50, 302, \dodoi{10.1134/S1063773724700166}

\bibitem[{{Kyutoku} {et~al.}(2013){Kyutoku}, {Ioka}, \& {Shibata}}]{Kyutoku+13}
{Kyutoku}, K., {Ioka}, K., \& {Shibata}, M. 2013, \prd, 88, 041503, \dodoi{10.1103/PhysRevD.88.041503}

\bibitem[{{Kyutoku} {et~al.}(2018){Kyutoku}, {Kiuchi}, {Sekiguchi}, {Shibata}, \& {Taniguchi}}]{Kyutoku+18}
{Kyutoku}, K., {Kiuchi}, K., {Sekiguchi}, Y., {Shibata}, M., \& {Taniguchi}, K. 2018, \prd, 97, 023009, \dodoi{10.1103/PhysRevD.97.023009}

\bibitem[{{Lattimer} \& {Schramm}(1976)}]{Lattimer_Schramm76}
{Lattimer}, J.~M., \& {Schramm}, D.~N. 1976, \apj, 210, 549, \dodoi{10.1086/154860}

\bibitem[{{Lippuner} \& {Roberts}(2015)}]{Lippuner&Luke15}
{Lippuner}, J., \& {Roberts}, L.~F. 2015, \apj, 815, 82, \dodoi{10.1088/0004-637X/815/2/82}

\bibitem[{{Martineau} {et~al.}(2024){Martineau}, {Foucart}, {Scheel}, {Duez}, {Kidder}, \& {Pfeiffer}}]{Martineau_GW230529}
{Martineau}, T., {Foucart}, F., {Scheel}, M., {et~al.} 2024, arXiv e-prints, arXiv:2405.06819, \dodoi{10.48550/arXiv.2405.06819}

\bibitem[{Most {et~al.}(2024)Most, Kim, Chatziioannou, \& Legred}]{Most:2024eig}
Most, E.~R., Kim, Y., Chatziioannou, K., \& Legred, I. 2024, Astrophys. J. Lett., 973, L37, \dodoi{10.3847/2041-8213/ad785c}

\bibitem[{Most {et~al.}(2021)Most, Papenfort, Tootle, \& Rezzolla}]{Most:2021ytn}
Most, E.~R., Papenfort, L.~J., Tootle, S.~D., \& Rezzolla, L. 2021, Mon. Not. Roy. Astron. Soc., 506, 3511, \dodoi{10.1093/mnras/stab1824}

\bibitem[{{Most} {et~al.}(2020){Most}, {Papenfort}, {Weih}, \& {Rezzolla}}]{Most+20_GW190814}
{Most}, E.~R., {Papenfort}, L.~J., {Weih}, L.~R., \& {Rezzolla}, L. 2020, \mnras, 499, L82, \dodoi{10.1093/mnrasl/slaa168}

\bibitem[{{Naoz}(2016)}]{Naoz16}
{Naoz}, S. 2016, \araa, 54, 441, \dodoi{10.1146/annurev-astro-081915-023315}

\bibitem[{{Paczynski}(1977)}]{Paczynski1977}
{Paczynski}, B. 1977, \apj, 216, 822, \dodoi{10.1086/155526}

\bibitem[{{Pannarale} {et~al.}(2011){Pannarale}, {Tonita}, \& {Rezzolla}}]{Pannarale&Tonita&Rezzolla+11}
{Pannarale}, F., {Tonita}, A., \& {Rezzolla}, L. 2011, \apj, 727, 95, \dodoi{10.1088/0004-637X/727/2/95}

\bibitem[{{Paschalidis} {et~al.}(2015){Paschalidis}, {Ruiz}, \& {Shapiro}}]{Paschalidis+15}
{Paschalidis}, V., {Ruiz}, M., \& {Shapiro}, S.~L. 2015, \apjl, 806, L14, \dodoi{10.1088/2041-8205/806/1/L14}

\bibitem[{Penner {et~al.}(2012)Penner, Andersson, Jones, Samuelsson, \& Hawke}]{Penner:2011br}
Penner, A.~J., Andersson, N., Jones, D.~I., Samuelsson, L., \& Hawke, I. 2012, Astrophys. J. Lett., 749, L36, \dodoi{10.1088/2041-8205/749/2/L36}

\bibitem[{{Perego} {et~al.}(2014){Perego}, {Rosswog}, {Cabez{\'o}n}, {Korobkin}, {K{\"a}ppeli}, {Arcones}, \& {Liebend{\"o}rfer}}]{Prego+14}
{Perego}, A., {Rosswog}, S., {Cabez{\'o}n}, R.~M., {et~al.} 2014, \mnras, 443, 3134, \dodoi{10.1093/mnras/stu1352}

\bibitem[{{Peters}(1964)}]{Peters1964}
{Peters}, P.~C. 1964, Physical Review, 136, 1224, \dodoi{10.1103/PhysRev.136.B1224}

\bibitem[{{Radice} {et~al.}(2020){Radice}, {Bernuzzi}, \& {Perego}}]{Radice+20_AnnRev}
{Radice}, D., {Bernuzzi}, S., \& {Perego}, A. 2020, Annual Review of Nuclear and Particle Science, 70, 95, \dodoi{10.1146/annurev-nucl-013120-114541}

\bibitem[{{Ramos-Buades} {et~al.}(2022{\natexlab{a}}){Ramos-Buades}, {Buonanno}, {Khalil}, \& {Ossokine}}]{Ramos-Buades+22a}
{Ramos-Buades}, A., {Buonanno}, A., {Khalil}, M., \& {Ossokine}, S. 2022{\natexlab{a}}, \prd, 105, 044035, \dodoi{10.1103/PhysRevD.105.044035}

\bibitem[{{Ramos-Buades} {et~al.}(2022{\natexlab{b}}){Ramos-Buades}, {van de Meent}, {Pfeiffer}, {R{\"u}ter}, {Scheel}, {Boyle}, \& {Kidder}}]{Ramos-Buades+22b}
{Ramos-Buades}, A., {van de Meent}, M., {Pfeiffer}, H.~P., {et~al.} 2022{\natexlab{b}}, \prd, 106, 124040, \dodoi{10.1103/PhysRevD.106.124040}

\bibitem[{Rosofsky {et~al.}(2019)Rosofsky, Gold, Chirenti, Huerta, \& Miller}]{Rosofsky:2018vyg}
Rosofsky, S., Gold, R., Chirenti, C., Huerta, E.~A., \& Miller, M.~C. 2019, Phys. Rev. D, 99, 084024, \dodoi{10.1103/PhysRevD.99.084024}

\bibitem[{{Shariat} {et~al.}(2024){Shariat}, {Naoz}, {El-Badry}, {Rodriguez}, {Hansen}, {Angelo}, \& {Stephan}}]{Shartiat+24}
{Shariat}, C., {Naoz}, S., {El-Badry}, K., {et~al.} 2024, arXiv e-prints, arXiv:2407.06257, \dodoi{10.48550/arXiv.2407.06257}

\bibitem[{{Shibata} {et~al.}(2021){Shibata}, {Fujibayashi}, \& {Sekiguchi}}]{Shibata+21}
{Shibata}, M., {Fujibayashi}, S., \& {Sekiguchi}, Y. 2021, \prd, 104, 063026, \dodoi{10.1103/PhysRevD.104.063026}

\bibitem[{{Shibata} \& {Hotokezaka}(2019)}]{Shibata_Kenta19}
{Shibata}, M., \& {Hotokezaka}, K. 2019, Annual Review of Nuclear and Particle Science, 69, 41, \dodoi{10.1146/annurev-nucl-101918-023625}

\bibitem[{{Shibata} \& {Taniguchi}(2006)}]{Shibata_Taniguchi06}
{Shibata}, M., \& {Taniguchi}, K. 2006, \prd, 73, 064027, \dodoi{10.1103/PhysRevD.73.064027}

\bibitem[{{Shibata} \& {Ury{\={u}}}(2006)}]{Shibata&Uryu06}
{Shibata}, M., \& {Ury{\={u}}}, K. 2006, \prd, 74, 121503, \dodoi{10.1103/PhysRevD.74.121503}

\bibitem[{{Siegel} \& {Metzger}(2017)}]{SiegelMetzger17PRL}
{Siegel}, D.~M., \& {Metzger}, B.~D. 2017, \prl, 119, 231102, \dodoi{10.1103/PhysRevLett.119.231102}

\bibitem[{{Steiner} {et~al.}(2013){Steiner}, {Hempel}, \& {Fischer}}]{Steiner+13_SHFo}
{Steiner}, A.~W., {Hempel}, M., \& {Fischer}, T. 2013, \apj, 774, 17, \dodoi{10.1088/0004-637X/774/1/17}

\bibitem[{{Tanaka} \& {Hotokezaka}(2013)}]{TankaHotokezaka2013}
{Tanaka}, M., \& {Hotokezaka}, K. 2013, \apj, 775, 113, \dodoi{10.1088/0004-637X/775/2/113}

\bibitem[{{The LIGO Scientific Collaboration} {et~al.}(2024){The LIGO Scientific Collaboration}, {the Virgo Collaboration}, \& {the KAGRA Collaboration}}]{GW230529_2024}
{The LIGO Scientific Collaboration}, {the Virgo Collaboration}, \& {the KAGRA Collaboration}. 2024, arXiv e-prints, arXiv:2404.04248, \dodoi{10.48550/arXiv.2404.04248}

\bibitem[{Tsang {et~al.}(2012)Tsang, Read, Hinderer, Piro, \& Bondarescu}]{Tsang:2011ad}
Tsang, D., Read, J.~S., Hinderer, T., Piro, A.~L., \& Bondarescu, R. 2012, Phys. Rev. Lett., 108, 011102, \dodoi{10.1103/PhysRevLett.108.011102}

\bibitem[{{Wanajo} {et~al.}(2014){Wanajo}, {Sekiguchi}, {Nishimura}, {Kiuchi}, {Kyutoku}, \& {Shibata}}]{Wanajo+14}
{Wanajo}, S., {Sekiguchi}, Y., {Nishimura}, N., {et~al.} 2014, \apjl, 789, L39, \dodoi{10.1088/2041-8205/789/2/L39}

\bibitem[{{Xuan} {et~al.}(2024){Xuan}, {Naoz}, {Li}, {Kocsis}, {Petigura}, {Knee}, {McIver}, {Kremer}, \& {Farr}}]{XuanZeyuan+24}
{Xuan}, Z., {Naoz}, S., {Li}, A. K.~Y., {et~al.} 2024, arXiv e-prints, arXiv:2409.15413, \dodoi{10.48550/arXiv.2409.15413}

\bibitem[{{Yudin} {et~al.}(2020){Yudin}, {Razinkova}, \& {Blinnikov}}]{Yudin+20}
{Yudin}, A.~V., {Razinkova}, T.~L., \& {Blinnikov}, S.~I. 2020, Astronomy Letters, 45, 847, \dodoi{10.1134/S1063773719120077}

\bibitem[{{Zenati} {et~al.}(2024){Zenati}, {Krolik}, {Werneck}, {Etienne}, {Noble}, {Murguia-Berthier}, \& {Schnittman}}]{Zenati+24BNS}
{Zenati}, Y., {Krolik}, J., {Werneck}, L., {et~al.} 2024, arXiv e-prints, arXiv:2404.03156, \dodoi{10.48550/arXiv.2404.03156}

\bibitem[{{Zenati} {et~al.}(2023){Zenati}, {Krolik}, {Werneck}, {Murguia-Berthier}, {Etienne}, {Noble}, \& {Piran}}]{zenati+23BNS}
{Zenati}, Y., {Krolik}, J.~H., {Werneck}, L.~R., {et~al.} 2023, \apj, 958, 161, \dodoi{10.3847/1538-4357/acf714}

\bibitem[{{Zevin} {et~al.}(2020){Zevin}, {Spera}, {Berry}, \& {Kalogera}}]{Zevin+20}
{Zevin}, M., {Spera}, M., {Berry}, C. P.~L., \& {Kalogera}, V. 2020, \apjl, 899, L1, \dodoi{10.3847/2041-8213/aba74e}

\end{thebibliography}
\bibliographystyle{aasjournal}

\end{document}